\documentclass[reprint,superscriptaddress,twocolumn,preprintnumbers,aps,prd]{revtex4-1}

\usepackage[utf8]{inputenc}
\usepackage{amssymb,amsmath,amsthm,dsfont}
\usepackage{xcolor}
\usepackage{bm,bbm}
\usepackage{bbold}

\usepackage[colorlinks,bookmarks=true,linktocpage=true]{hyperref}

\hypersetup{
    colorlinks,
    citecolor=blue,
    filecolor=blue,
    linkcolor=blue,
   urlcolor=blue,
   linktoc=page
}

\definecolor{darkgreen}{rgb}{0,0.6,0}
\definecolor{gray}{rgb}{.7,.7,.7}

\usepackage{graphicx}
 \graphicspath{{./figures/}}

\def\Eq#1{Eq.\,(\ref{#1})}

\def\eq#1{(\ref{#1})}

\newcommand{\Tr}{\mathrm{Tr}}

\newcommand{\tr}{\mathrm{tr}}

\def\CA{{\mathcal A}}

\def\CD{{\mathcal D}}

\def\CG{{\mathcal G}}

\def\CL{{\mathcal L}}

\def\CN{{\mathcal N}}
\def\CO{{\mathcal O}}

\def\CT{{\mathcal T}}

\def\CV{{\mathcal V}}
\def\CW{{\mathcal W}}

\def\llangle{\left\langle}
\def\rrangle{\right\rangle}

\begin{document}

\title{Quantum gravity: a fluctuating point of view}

\author{Jan M.~Pawlowski}
\affiliation{Institut f\"ur Theoretische Physik, Universit\"at Heidelberg,
Philosophenweg 16, 69120 Heidelberg, Germany}
\affiliation{ExtreMe Matter Institute EMMI, GSI Helmholtzzentrum f\"ur
Schwerionenforschung mbH, Planckstr.\ 1, 64291 Darmstadt, Germany}

\author{Manuel Reichert}
\affiliation{CP3-Origins, University of Southern Denmark, Campusvej 55, 5230 Odense M, Denmark}
\affiliation{Department  of  Physics  and  Astronomy,  University  of  Sussex,  Brighton,  BN1  9QH,  U.K.}

\begin{abstract}
In this contribution, we discuss the asymptotic safety scenario for quantum gravity with a functional renormalisation group approach that disentangles dynamical metric fluctuations from the background metric. We review the state of the art in pure gravity and general gravity-matter systems. This includes the discussion of results on the existence and properties of the asymptotically safe ultraviolet fixed point, full ultraviolet-infrared trajectories with classical gravity in the infrared, and the curvature dependence of couplings also in gravity-matter systems. The results in gravity-matter systems concern the ultraviolet stability of the fixed point and the dominance of gravity fluctuations in minimally coupled gravity-matter systems. Furthermore, we discuss important physics properties such as locality of the theory, diffeomorphism invariance, background independence, unitarity, and access to observables, as well as open challenges.
\end{abstract}

\maketitle

\section{Introduction}
One of the major challenges in theoretical physics is the unification of the Standard Model of particle physics (SM) with quantum gravity. Based on the classical Einstein-Hilbert action, gravity is perturbatively non-renormalisable and hence cannot be expanded about a vanishing gravitational coupling, the Newton coupling. A very promising way out has been proposed by Weinberg \cite{Weinberg:1980gg}, the \textit{asymptotic safety scenario}. It draws from the theory of critical phenomena developed for investigating the phase structure of condensed matter and statistical systems. In the language of critical phenomena, standard perturbation theory about a vanishing Newton coupling is an expansion about the free, Gau\ss ian fixed point of the theory and fails since this fixed point is ultraviolet (UV) repulsive in the relevant couplings. In turn, the asymptotic safety scenario builds upon the conjecture that quantum gravity also exhibits a nontrivial UV fixed point, the Reuter fixed point. This asymptotically safe fixed point should exhibit a finite-dimensional critical hypersurface, which renders the theory finite and predictive even beyond the Planck scale. 

The method of choice for respective investigations is the renormalisation group. Most investigations of asymptotically safe gravity have been performed with the functional renormalisation group (fRG) in its form for the effective action \cite{Wetterich:1992yh}. The fRG-approach to quantum gravity has been initiated by the seminal paper \cite{Reuter:1996cp}, where the UV fixed point has been studied in the Einstein-Hilbert truncation. In this approximation, one retains only two couplings, the Newton coupling $G_\text{N}$ and the cosmological constant $\Lambda$. Already this basic truncation exhibits a UV fixed point in four dimensions, see \cite{Reuter:1996cp, Souma:1999at}. This exciting finding has triggered a plethora of works for asymptotically safe gravity with and without matter, and we refer the reader to the textbooks \cite{Percacci:2017fkn, Reuter:2019byg} and reviews \cite{Niedermaier:2006wt, Litim:2011cp, Reuter:2012id, Ashtekar:2014kba, Eichhorn:2017egq, Bonanno:2017pkg, Eichhorn:2018yfc, Pereira:2019dbn, Reichert:2020mja}. For very recent accounts of the challenges for asymptotically safe gravity see \cite{Bonanno:2020bil, Donoghue:2019clr}. For generic reviews on the fRG we refer to~\cite{Berges:2000ew, Aoki:2000wm, Polonyi:2001se, Pawlowski:2005xe, Gies:2006wv, Delamotte:2007pf, Kopietz:2010zz, Rosten:2010vm, Braun:2011pp, Dupuis:2020fhh}.

The fRG-approach to gravity centres around the quantum effective action of the theory $\Gamma[\bar g_{\mu\nu}, h_{\mu\nu}]$, the quantum analogue of the classical action. Here, $\bar g_{\mu\nu}$ is a generic metric background and the graviton field $h_{\mu\nu}$ accounts for quantum fluctuations about this background. The computation of the effective action $\Gamma[\bar g_{\mu\nu}, h_{\mu\nu}]$ is tantamount to that of the path integral: the $n$-point correlation functions of the dynamical fluctuation field $h$ are given by $n$ derivatives of the effective action with respect to the correlation field, evaluated on the equations of motion, $h=0$ and $\bar g= \bar g_\text{EoM}$, i.e.\ on-shell. These correlation functions are nothing but the moments of the path integral and carry the dynamics of the quantum theory. 

This seemingly introduces a background dependence of the approach. However, the approach has inherent on-shell background independence, also related to physical diffeomorphism invariance. Indeed, the background effective action $\Gamma[g_{\mu\nu}]= \Gamma[g_{\mu\nu}, 0]$ is diffeomorphism invariant. The latter properties are the backbones of any quantum gravity approach and their realisation even within approximations is chiefly important. 

The present review outlines the properties and results of the fRG-approach to asymptotically safe quantum gravity in terms of \textit{background} and \textit{fluctuation} correlation functions of gravitons, shortly baptised the \textit{fluctuation approach} to gravity. This approach is based on the observation, that the dynamics of quantum gravity is encoded in the correlation functions of the fluctuation field~$h$. Reliable computations of observables can only be done from these correlation functions. This situation calls for a systematic improvement of the standard background-field approximation. In this approximation, the correlation functions of the background metric and the fluctuation field are identified. We refrain from going into more details here, the underlying assumptions and challenges are discussed in \autoref{sec:BackApprox} and \autoref{sec:SymId}. 

The fluctuation approach resolves these differences and by now it has matured enough to host a large number of results: this includes investigations of the Reuter fixed point in pure gravity in a rather elaborate truncation within a vertex expansion with momentum-dependent two-, three-, and four-point functions; the computation of the background-effective action for backgrounds with constant curvature; investigations of the stability of general gravity-matter systems; investigation of convergence properties of the expansion (\textit{apparent convergence}) as well as a potential close-perturbativeness of the asymptotically safe UV regime (\textit{effective universality}). We refer the reader to \autoref{sec:StateArt} for an explanation of the terminology and respective results.

In \autoref{sec:GaugeBack}, we discuss the general quantum field theory setting of quantum gravity, which we use for the fluctuation approach. This includes a discussion of the necessary gauge fixing and background independence of the approach. In \autoref{sec:FieldPara}, we discuss general parametrisations of the full metric in terms of a metric background and a fluctuation field. The preparation in \autoref{sec:GaugeBack} and \autoref{sec:FieldPara} allows us to introduce the fRG-approach to quantum gravity in \autoref{sec:fRG} as well as discussing the standard approximation used in the field, the background-field approximation, in \autoref{sec:BackApprox}. The symmetry identities that relate the dynamical correlation functions of the fluctuation field and that of the background metric are discussed in \autoref{sec:SymId}. These symmetry identities imply the necessity to go beyond the background-field approximation and thus we detail the fluctuation approach in \autoref{sec:Fluctuation-flows}. With the preparation of the sections before we discuss the results of the fluctuation approach in \autoref{sec:StateArt}, and close with a short conclusion and outlook in \autoref{sec:Summary}.

\section{Quantum field theory approach to quantum gravity}
\label{sec:GaugeBack}
The present contribution discusses the advances and open problems of a quantum field theory approach to quantum gravity that is based on the computation of metric correlation functions or, more generally, correlation functions of operators in quantum gravity. Formally, such an approach is based on the existence of a path integral for quantum gravity, for example defined by the integration over the space of all metrics $\{g_{\mu\nu}\}$ with a specific \textit{classical} action for gravity $S_\text{grav}$, a standard choice being the Einstein-Hilbert action, 
\begin{align} 
 \label{eq:EH-Action} 
 S_\text{EH}[g_{\mu\nu}]= \frac{1}{16 \pi G} \int\! \mathrm d^4 x \sqrt{g} \left(2 \Lambda - R\right),
\end{align}
with the abbreviation $g= \text{det}\,g_{\mu\nu}$. In \eq{eq:EH-Action} we have introduced the Newton coupling $G$ and the cosmological constant $\Lambda$. $R$ stands for the Ricci scalar. In most works in the fRG-approach, the theory is considered in its Euclidean version, which is indicated here by the missing minus sign in the square root of the determinant. The expectation value of a diffeomorphism invariant operator $\CO[g_{\mu\nu}]$ is formally given by 
\begin{align}
 \label{eq:ExpO}
 \langle \CO[\hat g_{\mu\nu}]\rangle =
 \frac{\int\!\CD \hat g_{\mu\nu}\, {\CO}[\hat g_{\mu\nu}]\,e^{-S_\text{grav}[\hat g_{\mu\nu}]}}{\int\!\CD \hat g_{\mu\nu}\, e^{-S_\text{grav}[\hat g_{\mu\nu}]}}\,.
\end{align}
Here and in the following, the $\hat{}$ indicates the fields that are integrated over. The formal definition \eq{eq:ExpO} faces several problems. Some of them are standard problems of the quantisation of gauge theories and some of them are specific to quantum gravity. The latter problems include e.g.\ the lack of perturbative renormalisability of gravity for $S_\text{EH}$ \cite{tHooft:1974toh, Goroff:1985sz, Goroff:1985th, vandeVen:1991gw}, the apparent unitarity problems for higher-derivative gravity a la Stelle \cite{Stelle:1976gc, Stelle:1977ry, Antoniadis:1986tu}, and the question, whether the integration measure $\CD\hat g$ includes a sum over all topologies \cite{Houthoff:2017oam}. The latter question is also an eminent one in lattice gravity, see e.g.~\cite{Ambjorn:1991pq, Bilke:1996zd, Ambjorn:1998xu, Ambjorn:2004qm, Ambjorn:2005qt, Laiho:2011ya, Ambjorn:2016fbd}. Note in this context that a general measure $\CD\mu(\hat g)$ can always be absorbed with a change of the gravity action in \eq{eq:ExpO}, 
\begin{align}
 \CD\mu(\hat g) = \CD \hat g_{\mu\nu}\, e^{-\Delta S_\text{grav}[\hat g_{\mu\nu}]}\,, 
\end{align}
with a potentially non-local action $\Delta S_\text{grav}$. In the fRG-approach, the task of a finite definition of \eq{eq:ExpO} and its computation is turned into the task of solving a flow equation for the quantum effective action $\Gamma[\bar g,\phi]$. Here $\bar g_{\mu\nu}$ is the background or reference metric, and $\phi$ are fluctuation fields, the expectation values of the fluctuation field operators $\hat{\phi}$. The latter includes the fluctuation field $\hat h_{\mu\nu}$ of the metric, $g_{\mu\nu}=g_{\mu\nu}(\bar g, h)$, as well as potential matter fields $\phi_\text{mat}$ and auxiliary fields such as the ghosts $c_\mu$ of the gauge fixing in gravity,  
\begin{align}
 \label{eq:def-phi}
 \phi&=\langle \hat\phi\rangle \,,
 &
 \phi&=(h_{\mu\nu},c_\mu, \bar c_\mu,\,\phi_\text{mat} )\,.
\end{align}
In the case of further gauge fields, one may also use background fields for the gauge fields, which are suppressed here for the sake of convenience. A reparameterisation $g_{\mu\nu}(\bar g, h)$ seemingly introduces a background-metric dependence of the formulation. This is common to many approaches to quantum gravity due to the necessity of defining metric fluctuations. Accordingly, the question of background independence of the present approach is an eminent one and is discussed later. Here we only want to mention the most common split between the background metric and the fluctuation field, the \textit{linear} split, 
\begin{align}
 \label{eq:LinSplitIntro} 
g_{\mu\nu} =\bar g_{\mu\nu} + h_{\mu\nu}\,. 
\end{align} 
This split also underlies most of the results discussed in \autoref{sec:StateArt}. Note that from now on the lowering and raising of indices is done with the background metric $\bar g$ if not specified otherwise. \Eq{eq:LinSplitIntro} emphasises one specific problem with the background field approach in quantum gravity: while $g_{\mu\nu}$ and $\bar g_{\mu\nu}$ are metrics, their difference $h_{\mu\nu} = g_{\mu\nu}- \bar g_{\mu\nu}$ is not. Indeed $h_{\mu\nu}$ has no geometrical meaning at all. This is discussed in more detail in \autoref{sec:FieldPara}. 

\subsection{Gauge fixing}
\label{sec:GaugeFixing}
In gauge theories such as gravity with the diffeomorphism (gauge) group, or the simpler case of non-Abelian gauge theories, the practical computation of observables \eq{eq:ExpO} faces the gauge group redundancy in the path integral measure. While this redundancy is a finite-dimensional one within discrete lattice formulations, it is an infinite-dimensional one in functional approaches based on graviton correlation functions. In particular, it prohibits the straightforward definition of the propagator, which is key in most functional approaches. 

Therefore, most of the latter approaches require a gauge fixing, for a brief discussion of gauge-invariant functional approaches see \autoref{sec:DiffInv} . Put differently, we have to choose a parametrisation of the theory. Typically this is done with a linear gauge fixing for the fluctuation field $h_{\mu\nu}$ that carries the metric degrees of freedom,
\begin{align}
 \label{eq:gf} 
 S_{\text{gf}}[\bar g, h]=\frac{1}{2 \alpha} \int \!\mathrm{d}^4x
 \sqrt{\bar{g}}\; \bar{g}^{\mu \nu} F_\mu F_\nu \,.
\end{align}
A common gauge-fixing condition $F_\mu$ is given by
\begin{align}
 \label{eq:gf-condition}
 F_\mu[\bar g, h] =
 \bar{\nabla}^\nu h_{\mu \nu} -\frac{1+ \beta}{4} \bar{\nabla}_\mu h^{\nu}_{~\nu} \,, 
\end{align}
where $\bar\nabla$ is the covariant derivative with the background metric $\bar g_{\mu\nu}$. The gauge fixing \eq{eq:gf-condition} is introduced in the path integral with the Faddeev-Popov trick and the Jacobi determinant of the respective reparameterisation. The Faddeev-Popov determinant $\Delta_\text{FP}$ can be rewritten in terms of a fermionic path integral with the ghost fields $c_\mu$ and $\bar c_\mu$. The ghost action related to \eq{eq:gf-condition} reads 
\begin{align}
 \label{eq:Sghost}
 S_{\text{gh}}[\bar g, \phi]=\int \!\mathrm{d}^4x
 \sqrt{\bar{g}}\; \bar c^\mu M_{\mu\nu} c^\nu\,, 
\end{align}
with the Faddeev-Popov operator 
\begin{align}
 \label{eq:OpFP}
 M_{\mu\nu}= \bar\nabla^\rho\! \left(g_{\mu\nu} \nabla_\rho +g_{\rho\nu} \nabla_\mu\right) -\frac{1+\beta}{2} \bar g^{\sigma\rho} \bar\nabla_\mu g_{\nu\sigma} \nabla_\rho\,.
\end{align}
Again, $\bar\nabla$ is the covariant derivative with the background metric $\bar g_{\mu\nu}$ while $\nabla$ is that with the full metric $g_{\mu\nu}$. Note that $M_{\mu\nu}$ is linear in the fluctuation field $h$. The background metric $\bar g_{\mu\nu}$ cannot be avoided and both, gauge-fixing and ghost action, depend on it. This implies that also the quantum effective action depends on both metrics, the background metric $\bar g_{\mu\nu}$ and the full metric $g_{\mu\nu}(\bar g,h)$, as we shall see later. Note, however, that the correlation functions of diffeomorphism-invariant operators as well as the solutions to the quantum equations of motion do not depend on the gauge fixing. Hence, they are background-independent as explained below.

\subsection{Background independence}
\label{sec:BackInd}
Background independence of the construction is more than a formal property to aim for. We briefly recollect the standard arguments for background independence in the background-field approach to quantum gauge field theories. We first restrict ourselves to pure gravity. Seemingly, background dependence of the path integral is introduced by a gauge fixing such as \eq{eq:gf} and the respective Faddeev-Popov determinant $\Delta_\text{FP}$. The latter is the Jacobian of the reparameterisation of the path integral in terms of gauge-fixed fields. We emphasise that the gauge fixing should be rather understood as a specific choice of field coordinates in the configuration space that facilitates the integration. The Faddeev-Popov trick is nothing but a convenient way to introduce these coordinates. In any case it leads us to the expectation values of diffeomorphism invariant operators defined in \eq{eq:ExpO} for pure gravity with a path integral with the gauge-fixed action, 
\begin{align} 
 \label{eq:backpath}
 \langle \CO[\hat g]\rangle = \frac{\int\!\CD \hat \phi\, 
 e^{-S_\text{gf}[\bar g, \hat h]-S_\text{gh}[\bar g, \hat \phi]}\, \CO [\hat g]\,e^{-S_\text{grav}[\hat g]} }{\int\!\CD \hat \phi\, 
 e^{-S_\text{gf}[\bar g, \hat h]-S_\text{gh}[\bar g, \hat \phi]-S_\text{grav}[\hat g]} }\,.
\end{align}
Note that an integration over the diffeomorphism group (from the Faddeev-Popov trick) has been factored out in the numerator and denominator. This relies on the diffeomorphism-invariance of $S_\text{grav}$, $\CO [\hat g]$, and $\CD \hat g$. The full integration measure $\CD\hat\phi$ in \eq{eq:backpath} now also includes the ghost fields, $\hat\phi=(\hat h_{\mu\nu}, \hat c_\mu, \hat{\bar{c}}_\mu)$. Naturally, the right-hand side in \eq{eq:backpath} is independent of the background field as the left-hand side trivially is, see \eq{eq:ExpO}. This background-metric independence is captured in the Nielsen or split-Ward identity derived from taking a $\bar g_{\mu\nu}$-derivative of \eq{eq:backpath}. Typically one also subtracts the Dyson-Schwinger equation for $\langle \CO \rangle$, which reads schematically 
\begin{align} 
 \int\!\CD \hat h\,\frac{\delta}{\delta \hat h}\left[ e^{-S_{\text{gf}}-S_{\text{gh}}}\,
 {\CO}\,e^{-S_\text{grav}}\right]=0 \,. 
\end{align}
This leads us to the Nielsen identity for general diffeomorphism invariant operators $\CO[g]$ with 
\begin{align}
 \label{eq:NI-path}
 \llangle \CO[\hat g] \left(\frac{\delta}{\delta \bar g}-\frac{\delta}{\delta \hat h}\right) 
 \Bigl( S_\text{gh}+S_\text{gf}\Bigr)[\bar g,\hat h]\rrangle=0\,. 
\end{align}
If solving the path integral within approximations, the check of the Nielsen identity \eq{eq:NI-path} is crucial as it carries the physical background independence. 

The identity \eq{eq:NI-path} constitutes infinitely many relations for diffeomorphism-invariant correlation functions and can be rephrased in terms of derivatives of the effective action. Correlation functions are conveniently derived from the generating functional $Z[\bar g,J]$ obtained by adding source terms for the fluctuation fields to the exponent in the path integral, 
\begin{align}
 \label{eq:ZpathI}
 Z[\bar g,J] = \frac{1}{\CN} \int\!\CD \hat \phi\,
 e^{-S-S_\text{gf}-S_\text{gh}+\int\!\mathrm{d}^4x \sqrt{\bar g} \, J^a \hat\phi_a}\,, 
\end{align}
where $J=(J_{h_{\mu\nu}},J_{c_\mu},J_{\bar c_\mu},J_\text{mat})$ and the normalisation $\CN$ is the denominator in \eq{eq:backpath}. Lowering and rising the field indices is done with the metric $\gamma^{ab}$ in field space, for details see App.~\ref{app:Notation}. 

In \eq{eq:ZpathI}, the action $S=S[\bar g, \hat \phi]$ is the 'classical' action of the gravity-matter system under consideration. The gauge-fixing action $S_\text{gf}$ and the ghost action $S_\text{gh}$ of the full gravity-matter system may include further gauge fixings of gauge fields. Note that for gravity-matter systems the 'classical' action may not be based on the Einstein-Hilbert action of general relativity as discussed before. More generally also the matter part may not simply be that of a standard renormalisable QFT in the presence of a dynamical metric background. 

The generating functional $Z[\bar g,J]$, or rather the Schwinger functional $\log [\bar g,J]$, generates connected $n$-point correlation functions of the fluctuation field with
\begin{align}
 \label{eq:Zh}
 \frac{ \delta ^n \log Z[\bar g,J]}{\delta J^{a_1}\cdots \delta J^{a_n}} =\llangle \hat\phi_{a_1} \cdots \hat\phi_{a_n}\rrangle_{\!\text{con}}\,,
\end{align}
where the indices $a_i$ stand for Lorentz and internal indices as well as species of fields. The subscript ${}_\text{con}$ in \eq{eq:Zh} indicates the connected part of the correlation function. We have included a factor of $1/\sqrt{\bar g}$ in the definition of the functional derivative, see App.~\ref{app:Notation}. This cancels the $\sqrt{\bar g}$ factor in the spacetime integral in the source term of \eq{eq:ZpathI}. If instead, we had used $\sqrt{\hat g}$ in the source term, derivatives with respect to the current $J$ would generate infinite order correlation functions.

Note that the generating functional \eq{eq:ZpathI} can be expressed with the right-hand side of \eq{eq:backpath} with the operator $\CO=e^{\int\! \mathrm d^4 x \sqrt{\bar g} \, J^a \hat\phi_a}$. However, this operator is neither diffeomorphism invariant nor background-independent. For that reason it cannot be mapped into a manifestly background-independent form such as \eq{eq:backpath}. For $J=0$, we have $\CO=1$ and $Z=1$, which is trivially background-independent. Accordingly, for $J\neq0$ the gauge-fixed generating functional $Z[\bar g, J]$ is background-\textit{dependent}, as is the effective action $\Gamma[\bar g, \phi]$, 
\begin{align}
 \label{eq:GammaCon}
 e^{-\Gamma[\bar g, \phi]}= \frac{1}{\CN} \int\!\CD \hat \phi\,
 e^{-\left(S+S_\text{gf}+S_\text{gh}\right)+ \int\!\mathrm{d}^4x\sqrt{\bar g} \,(\hat\phi_a-\phi_a)\frac{\delta \Gamma}{\delta\phi_a}}\,.
\end{align}
For the relation \eq{eq:GammaCon}, we have used that the effective action $\Gamma$ is the Legendre transformation of the Schwinger functional $W[\bar g, J]=\log Z[\bar g,J]$. This leads to  
\begin{align}
 \label{eq:EoMJ}
 J^a[\bar g, \phi] = (-1)^{s_a}\frac{\delta \Gamma[\bar g, \phi]}{\delta \phi_a}\,,
\end{align}
with the fermion number $s_a=1$ for fermions and $s_a=0$ for bosons. Then, background-independence is achieved on the \textit{fluctuation} field equations of motion (EoM) for $J[\bar g, \phi] =0$. The on-shell vanishing of the currents entails, that all diffeomorphism-invariant quantities are background-independent on-shell, and this independence is carried by \eq{eq:NI-path}. 

An important consequence of background-independence is the equivalence of the solutions $\bar g_\text{EoM}[\phi]$ to the \textit{fluctuation} field EoM and the \textit{background} field EoM, 
\begin{align}
 \label{eq:FlucEoM=backEoM} 
 \frac{\delta\Gamma[\bar g^\text{fluc}_\text{EoM},\phi]}{\delta h_{\mu\nu}} =0 \quad \longleftrightarrow\quad 
	\frac{\delta\Gamma[\bar g^\text{back}_\text{EoM},\phi]}{\delta \bar g_{\mu\nu}} =0\,,
\end{align}
with 
\begin{align}
 \bar g_\text{EoM}^\text{fluc}=\bar g_\text{EoM}^\text{back}=\bar g^{\,}_\text{EoM}\,.
\end{align}
If the fluctuation EoM holds, the current $J$ is vanishing and hence the background EoM is nothing but the Nielsen identity \eq{eq:NI-path}. In turn, if the background EoM holds, the current $J$ necessarily vanishes.

\section{Field parametrisations}
\label{sec:FieldPara}
So far we have not specified the relation between the background metric $\bar g$ and the full metric $g(\bar g,h)$, which defines the r\^ole of the fluctuation field $h$. While most of the computations are done within the linear split, \eq{eq:LinSplitIntro}, it is worth discussing the general case. This not only allows us to achieve a better understanding of the linear split but also to discuss the challenges for manifestly diffeomorphism-invariant formulations.  

The importance of the different splits for the path integral has been already mentioned in the context of the path integral measure, see the introduction of \autoref{sec:GaugeBack} around \eq{eq:ExpO}. In the flow equation approach to quantum gravity detailed in the next section, \autoref{sec:fRG}, the discussion of the path integral measure translates into that of the ordering of fluctuations: the fRG-approach to quantum gravity is based on a Wilsonian successive integrating-out of quantum fluctuations. In its form of a flow equation for the quantum effective action $\Gamma[\bar g, \phi]$ is has a simple form in terms of the full field-dependent fluctuation field propagator $G[\bar g,\phi]$ of the theory, see \eq{eq:flow}. This is the connected part of the two-point function of the fluctuation field, 
\begin{align}
 \label{eq:FullProp}
 G[\bar g,\phi](x,y) = \langle \hat\phi(x) \hat\phi(y) \rangle - \phi(x) \phi(y)\,.
\end{align}
The definition \eq{eq:FullProp} requires a gauge fixing (or reparameterisation) as discussed in the previous section. Moreover, the Wilsonian cutoff regularises the spectrum of the propagator. Consequently, the fRG-approach crucially depends on the split of the full metric $g$ into the background metric $\bar g$ and the fluctuation field $h$ for two reasons: 
\begin{itemize} 
\item[(i)] \textit{Ordering of fluctuations}: The quantum fluctuations of the fluctuation field $h$ are successively integrated out and are ordered in terms of the background covariant Laplacian. Therefore, the meaning of this ordering depends on the chosen split.
\item[(ii)] \textit{Relevance of higher-order correlations:} The physics included with higher-order correlation functions crucially depends on the chosen split. Thus, a different split orders quantum fluctuations differently. This leads to potentially qualitative differences for the convergence of a given approximation scheme. 
\end{itemize}
In this section, we briefly introduce and discuss the different splits considered so far in the fRG-approach to asymptotically safe quantum gravity. 

\subsection{Linear split}
\label{sec:LinSplit}
We begin with the standard and simplest split, the linear split, see also \eq{eq:LinSplitIntro}. It is given by 
\begin{align}
 \label{eq:LinSplit}
 g=\bar g+ h\,,
 \qquad 
 \text{with}\qquad \CD \hat g = \CD \hat h\,.
\end{align}
The Jacobian of this transformation is unity and the path integral measures agree. As mentioned before, with such a definition, the fluctuation field $h=g-\bar g$ is not a metric and has no geometrical meaning in the configuration space of metrics. Still, it is the natural choice as it facilitates explicit computations as well as the implementation of the quantisation of the theory for a given classical action $S_\text{grav}$ on the space of metrics. Still, its lack of a geometrical interpretation makes it difficult to discuss the reparameterisation invariance of the theory as well as the consequences of background independence. For more details see \autoref{sec:SymId}. These intricacies have led to more elaborated splits based on the fibre bundle structure of the configuration space of metrics. 

\subsection{Exponential split}
\label{sec:ExpSplit}
In recent years the exponential split has attracted some attention, \cite{Kawai:1992np, Nink:2014yya, Falls:2015qga, Demmel:2015zfa, Percacci:2015wwa, Gies:2015tca, Ohta:2015efa, Labus:2015ska, Ohta:2015fcu, Ohta:2015zwa, Dona:2015tnf, Falls:2016msz, Ohta:2016npm, Ohta:2016jvw, Ohta:2017dsq, Alkofer:2018fxj, deBrito:2018jxt}. It is given by  
\begin{align}
 \label{eq:ExpSplit}
 g&=\bar g \exp h \,, 
\qquad 
\text{with}\qquad \CD \hat g = J_\text{exp}\CD \hat h\,. 
\end{align}
The full metric is proportional to the exponential of the fluctuation field $h$ indicating a Lie-algebra nature of the fluctuation field $h$. Note that the parametrisation \eq{eq:ExpSplit} restricts the metric $g$, and in particular it does not allow for signature changes. Therefore, it is potentially not a reparameterisation of the path integral in terms of an integration over all metrics but a definition of another candidate for quantum gravity. Moreover, the assumption $J_\text{exp}=\mathbb{1}$ may change the integration. In summary, it is unclear whether a path integral with the exponential split and the measure $\CD \hat h$ describes the same quantum theory as that with the measure $\CD \hat g$. This parametrisation is also linked to unimodular gravity, see e.g.~\cite{Eichhorn:2013xr, Eichhorn:2015bna, Benedetti:2015zsw, Ardon:2017atk, Percacci:2017fsy, deBrito:2019umw, deBrito:2020rwu}. 

\subsection{Geometrical split}
\label{sec:GeoSplit}
We briefly describe the geometrical approach to quantum gravity pioneered by Vilkovisky and DeWitt, see e.g.~\cite{DeWitt:1980jv, Fradkin:1983nw, Vilkovisky:1984st, DeWitt:2003pm}. In the fRG-approach to gravity it has been discussed in \cite{Branchina:2003ek, Pawlowski:2003sk, Pawlowski:2005xe, Donkin:2012ud, Demmel:2014hla, Falls:2020tmj}. It is a general framework, and all parametrisations used in the literature can be understood as different choices for the geometrical structure of the configuration space of metrics $g_{\mu\nu}$. This also allows for a better understanding of the Wilsonian integrating-out of quantum fluctuations underlying the different splits. 

\begin{figure}[t]
 \includegraphics[width=.85\linewidth]{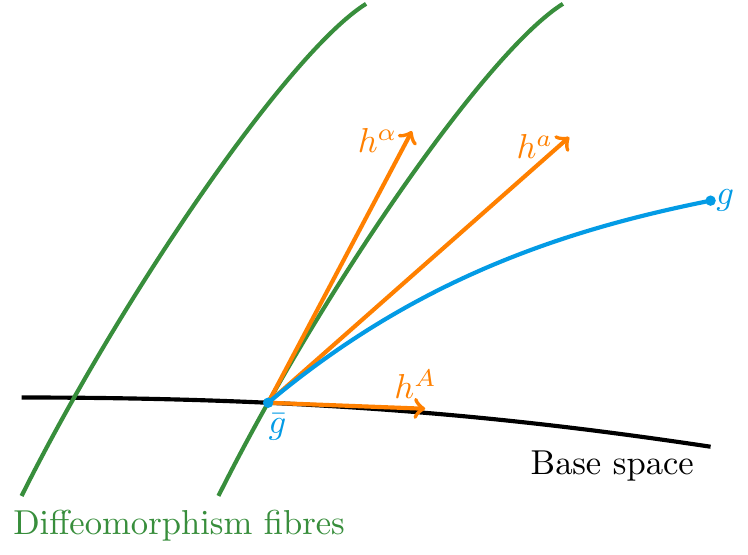}
 \caption{
 Illustration of the configuration space of metrics with the Vilkovisky connection. The background metric $\bar g$ and the full metric $g$ are connected by geodesics. The fluctuation field $h^a$ is a tangent vector of these geodesics at the background metric. $h^A$ is the projection on the base space while $h^\alpha$ is the projection on the diffeomorphism fibre. The effective action depends only on $h^A$ and not on $h^\alpha$.
 }
 \label{fig:geo-split}
\end{figure}

In the linear split, as discussed in \autoref{sec:LinSplit}, the fluctuation field $h$ neither is a metric nor does it have a geometrical interpretation in the configuration space $\Phi$. In turn, in the geometrical approach, the fluctuation field is constructed such that it has a geometrical meaning. The background metric and the full metric are linked by geodesics with respect to a given connection in the configuration space. The Vilkovisky connection $\Gamma_{\text{V}}$ is a specifically useful one: it is constructed with the demand of maximal orthogonality between the diffeomorphism fibre in the configuration space and the base space. If such a disentanglement is achieved, the path integral as well as the effective action only depends on the propagating degrees of freedom and the gauge redundancies are completely removed. This leads to the following conditions, 
\begin{align}
 \label{eq:Vilkovisky}
 \Gamma_{\text{V}}^A{}_{BC} &= \Gamma_{\text{g}}^A{}_{BC} \,,
 &
 \Gamma_{\text{V}}^A{}_{B\gamma}&= 0 \,,
 &
 \Gamma_{\text{V}}^A{}_{\beta\gamma}&=0\,,
\end{align}
where 
\begin{align}
 \Gamma_{\text{g}}^A{}_{BC}=\frac12 g^{AD}\!\left(g_{DB,C}+g_{DC,B}+g_{BC,D}\right),
\end{align}
is the Riemannian metric on the quotient space $\Phi/\CG$, where $\CG$ is the group of diffeomorphisms. This quotient space is labelled with capital latin letters $A,B,C,....$, while the diffeomorphism fibre is labelled with greek letters $\alpha,\beta, \gamma,...$. The full space is labelled with small latin letters $a,b,c,...$. For further details on the notation and the setup in the context of RG-gravity see e.g.~\cite{Pawlowski:2003sk}. 

The background metric $\bar g$ and the full metric $g$ are connected by a geodesic. With the Vilkovisky connection, the fluctuation field is a tangent vector on this geodesic at the background metric (Gau\ss ian or geodesic normal coordinates). This is illustrated in \autoref{fig:geo-split} and leads to
\begin{align}
 \label{eq:GeoSplit}
 g&=\bar g+ h -\frac12 \Gamma_V h^2 + \CO(h^3)\,,
 &
 \text{with}\quad \CD \hat g &\simeq \CD \hat h\,.
\end{align}
The relation between $g$ and $\bar g$ is non-polynomial. Still, the Jacobian does not depend on the fluctuation field and we have dropped it in \eq{eq:GeoSplit}. In this setting, it can be shown that the effective action $\Gamma$ only depends on the projection $h^A$ of the tangent vector $h^a$ onto the base space of the fibre bundle: $\Gamma=\Gamma[\bar g, h^A]$. In turn, the projection of $h^a$ onto the diffeomorphism fibre, $h^\alpha$, drops out. Hence, the effective action is diffeomorphism-invariant as $h^A$ is a diffeomorphism scalar. Trivially, an infrared (IR) regularisation of the $h^A$-path integral is diffeomorphism-invariant. 

We close this section with some remarks on the implications of such a geometrical setup for 'physical' gauge fixings, linear and exponential splits, and locality.

The geometrical construction comes as close as possible to the definition of the configuration space of a gauge theory in terms of 'physical' gauge-invariant fields and correlation functions. Such a parameterisation is tantamount to a specific gauge fixing as already mentioned in \autoref{sec:BackInd}. We may call such a gauge fixing 'physical', having in mind that it removes most of the redundancies related to the gauge group, in gravity that related to the diffeomorphism group. Note, however, that the terminology 'physical gauge fixing' is not well-defined and also used differently in other contexts. In non-Abelian gauge theories, the projection is unique and singles out the Landau-DeWitt gauge as the 'physical' one. In gravity, one is left with a one-parameter family of gauges with the gauge-fixing parameter $\beta$, see \eq{eq:gf-condition}. 

It is worth emphasising that a gauge-fixing condition for the geometrical field (or Gau\ss ian normal field) $h^a$ is different from that for the fluctuation field $h$ in the linear split. Only for specific choices of the latter, the maximal disentanglement of the geometrical construction is manifestly obtained. We also remark that the linear split is obtained by using a vanishing connection and hence entirely ignoring the geometrical structure of the configuration space. The exponential split simply uses the Riemannian part $\Gamma_g$ of the configuration space, hence ignoring the diffeomorphism group. 

Finally, the geometrical construction with the Vilkovisky connection is highly non-local in configuration space, one of the ensuing problems being caustics as well as Gribov copies. This also raises the question of locality in the configuration space, and that of momentum locality of the correlation functions of the geometrical fluctuation field $h$. The latter is discussed in detail in \autoref{sec:MomentumLocality}. Both locality issues highlight the challenges for manifest gauge- or diffeomorphism-invariant functional approaches to quantum gravity.

\section{Flow equation for gravity}
\label{sec:fRG}
With the quantum field theory approach to quantum gravity outlined in the last sections, we are now in the position to discuss the flow-equation approach to gravity, for reviews see \cite{Niedermaier:2006wt, Litim:2011cp, Reuter:2012id, Ashtekar:2014kba, Eichhorn:2017egq, Bonanno:2017pkg, Eichhorn:2018yfc, Pereira:2019dbn, Reichert:2020mja, Bonanno:2020bil, Donoghue:2019clr} and for generic fRG reviews see \cite{Berges:2000ew, Aoki:2000wm, Polonyi:2001se, Pawlowski:2005xe, Gies:2006wv, Delamotte:2007pf, Kopietz:2010zz, Rosten:2010vm, Braun:2011pp, Dupuis:2020fhh}. As already mentioned in the introduction of \autoref{sec:FieldPara}, the fRG-approach to gravity is based on a successive integrating out of quantum fluctuations. Typically this is done with an ordering of quantum fluctuations in momentum space: the regulator introduces a suppression of low momentum fluctuations below an IR cutoff scale $p^2\lesssim k^2$, and one RG step with $k\to k-\Delta k$ relates to the integration of momentum modes $p^2 \approx k^2$. In gravity, the implementation of such a momentum cutoff necessitates the choice of a background metric $\bar g$, and the (covariant) momenta are those related to the covariant Laplacian in the background metric, $\Delta_{\bar g}$, with the spectral values $p^2_{\bar g}$. 

Remarkably, the flow equation is insensitive to field reparameterisations of quantum gravity discussed in the last section or even physically different formulations: For the derivation let us assume that a finite generating functional for correlation functions of the fluctuation field is given. In terms of a path integral, this is given by \eq{eq:ZpathI} with an assumed diffeomorphism invariant regularisation and renormalisation procedure. More generally such a \textit{finite} generating functional is given by its defining property \eq{eq:Zh} under the assumption that these correlation functions are \textit{finite}. Then, the flow equation can be readily derived without the necessity of referring to a specific representation of $Z[\bar g, J]$ such as the path integral, for a detailed discussion see \cite{Pawlowski:2003hq}. The correlation functions of $h$ depend on $\bar g$, as does the generating functional for $J\neq 0$ via the gauge fixing, see \autoref{sec:GaugeBack}. 

The flow equation for the effective action is derived from the IR regularised generating functional, 
\begin{align}
 \label{eq:Zhk} 
 Z_k[\bar g, J]= \exp\!\left(-\!\int \!\mathrm d^4 x \sqrt{\bar g} \frac{\delta}{\delta J^a} R^{ab}_k \frac{\delta}{\delta J^b} \right) Z[\bar g, J]\,,
\end{align}
with a $\bar g$-dependent IR regulator $R_k$. Typically the background dependence enters the regulator via a background Laplacian and background covariant derivatives. In flat space, the eigenvalues of the Laplacian are just momentum squared, $p^2$. As already discussed above, the regulator suppresses then IR-momentum modes with $p^2 \lesssim k^2$. In turn, UV-momentum modes with $p^2 \gtrsim k^2$ propagate freely and the generating functional includes all quantum contributions generated by these modes.

It is convenient to write the regulator $R_k$ in terms of the classical or quantum dispersion of the field at hand, 
\begin{align}
 \label{eq:RegSplits} 
 R_k^{ab}(p) &= T^{ab}_k(p)\, r_k(x)\,,
 &
 \text{with} \quad x&=\frac{p^2}{k^2}\,, 
\end{align}
where momentum-squared is counted in cutoff units. In these units the IR regime is given by $x\lesssim 1$ and the UV regime by $x\gtrsim 1$. The tensor part $T^{ab}_k$ of the regulator is proportional to the classical or quantum dispersion of the field. Classically it is the second derivative of the action with respect to the fields $\phi^a$ and $\phi^b$, i.e., $(S^{(2)})^{ab}(p)$. It carries the kinetic information about the field whose propagation is regularised. In turn, the dimensionless \textit{shape function} $r_k$ specifies how the propagation is regularised. In most cases the latter part is chosen such that the physical cutoff scales agree for all fields. This is typically achieved with identical (e.g.\ for several scalar or bosonic fields) or related shape functions (e.g.\ for scalars and Dirac fermions). It can be shown that such a choice also improves the convergence of generic expansion schemes, see \cite{Pawlowski:2005xe, Pawlowski:2015mlf}. Moreover, $r_k$ has to be chosen such that the IR suppression of momentum modes as well as the UV decay of the regularisation is guaranteed. These properties lead to the following asymptotics of the regulator shape function, 
\begin{subequations} \label{eq:rkLimits}
\begin{align}
 \label{eq:rkLimitsIR}
 \lim_{x\to 0} r_k(x) \to&\, \infty \,, \\
 \label{eq:rkLimitsUV}
 \lim_{x\to \infty} r_k(x)=&\, 0 \,. 
\end{align} 
\end{subequations}
The first limit, \eq{eq:rkLimitsIR}, guarantees the IR suppression of momentum modes. For example, for a scalar field in $d$ dimensions with a quadratic dispersion $\propto p^2$, a regulator shape function $r_k(x\to 0) = 1/x$ introduces a low momentum mass $k^2$ for this field. Indeed this is the common choice for the IR limit, but more singular choices work as well. \Eq{eq:rkLimitsIR} also entails that for $k\to\infty$ all momentum modes are suppressed and the theory approaches the UV-scaling regime. For asymptotically free theories this is the classical theory, for asymptotically safe theories this is the non-trivial quantum UV theory.

The second limit, \eq{eq:rkLimitsUV}, guarantees that the UV behaviour of the theory is unchanged by the IR regularisation. We shall see below that the limit in \eq{eq:rkLimitsUV} has to be approached sufficiently fast. In our example of a scalar field in $d$ dimensions, the regulator shape function has to decay with at least $\min(1/x^\frac{d}{2},1/x)$ for rendering the IR flows finite. This is also discussed later in more details below \eq{eq:backflowfreer}. Note that the latter limit is that of a mass or Callan-Symanzik cutoff. Then, changing $k$ changes a relevant parameter of the theory, and hence changes the theory at all scales. Accordingly, the Callan-Symanzik cutoff is not a \textit{local} momentum cutoff. The limit \eq{eq:rkLimitsUV} also has another implication: for $k\to 0$ the limit \eq{eq:rkLimitsUV} holds for all momenta and the cutoff is removed from the theory. We remark that it is precisely this property, which is at stake for the Callan-Symanzik cutoff and similar ones.

Subject to the existence of a finite full generating functional $Z[\bar g,J]$, the regularised generating functional $Z_k[\bar g,J]$ is also finite (and smaller than $Z[\bar g,J]$). The flow equation for the Schwinger functional $W_k[\bar g, J] = \log Z_k[\bar g, J]$ is derived by taking the logarithmic $k$-derivative of \eq{eq:Zhk}. Schematically this leads us to 
\begin{align}
 \label{eq:FlowWk} 
 \partial_t W_k[\bar g, J] =
 - \frac12 \Tr\! \left( \frac{\delta^2 W_k}{\delta J^2} + \frac{\delta W_k}{\delta J} \, \frac{\delta W_k}{\delta J}\right)\! \partial_t R_k\,, 
\end{align}
where the RG-'time' $t$ is defined with $t=\ln k/\Lambda$, and $\Lambda$ is some reference scale $\Lambda$. The trace sums over position space, Lorentz and internal indices as well as species of fields. For the sake of a concise presentation, we have suppressed all space-time and internal indices including species of fields. We emphasise that for the explicit form of \eq{eq:FlowWk} the order of derivatives is important as $J$ contains fermionic currents. 

The term in parenthesis in \eq{eq:FlowWk} is nothing but the full two-point correlation functions of the theory: the first term is the connected part, i.e.\ the scale-dependent propagator $G_k[\bar g, \phi]$ of the theory, see \eq{eq:FullProp}. The second term is simply $\phi^2$, the disconnected part. The scale-dependent effective action $\Gamma_k[\bar g,\phi]$ is defined as the modified Legendre transformation of the Schwinger functional, 
\begin{align}  
 \label{eq:Gammak}
 \Gamma_k[\bar g,\phi] &= \int \! \mathrm d^4 x\sqrt{\bar g}\, J^a\phi_a - W_k[\bar g,J] \notag \\
 &\quad-\frac12 \int \! \mathrm d^4 x\sqrt{\bar g} \,\phi_a R^{ab}_k\phi_b , 
\end{align}
where $J=J[\bar g,\phi]$ is given by \eq{eq:EoMJ}. The source term in \eq{eq:Gammak} depends on $\sqrt{\bar g}$, just as the source term in \eq{eq:ZpathI}. Otherwise, the Legendre transform would not be linear in the mean field $\phi$. Note also that the classical action of gravity may be unbounded, e.g.\ in the case of the Einstein-Hilbert action. Then the Legendre transformation is defined on a saddle point. 

The flow equation for the effective action \cite{Wetterich:1992yh,Ellwanger:1993mw,Morris:1993qb} follows straightforwardly from \eq{eq:FlowWk}. The part proportional to $\phi^2$ is cancelled by the flow of the last term in \eq{eq:Gammak}, and the flow of $\Gamma_k[\bar g,\phi]$ is given by  
\begin{align}
 \label{eq:flow} 
 \partial_t \Gamma_k[\bar g,\phi]= \frac12 \Tr\, G_k[\bar g,\phi]\,\partial_t R_k\,,
\end{align}
where $G_k[\bar g, \phi]$ is the full field-dependent propagator $\delta^2 W_k/\delta J^2$ and the trace has been defined below \eq{eq:FlowWk}. It now contains a relative minus sign for Gra\ss mann-valued fields. With the definition of the Legendre transformation in \eq{eq:Gammak}, the full propagator is given by 
\begin{align}
 \label{eq:Gk}
 G_k[\bar g,\phi] = \frac{1}{\Gamma_k^{(0,2)}[\bar g,\phi]+ R_k} \,.
\end{align}
The flow equation for the effective action depends on the second derivative of the effective action with respect to the fluctuation fields, $\Gamma_k^{(0,2)}[\bar g,\phi]$. The flow of the latter is derived from \eq{eq:flow} with two derivatives w.r.t.\ the fluctuation field $\phi$. This flow depends on itself as well as the vertices $\Gamma_k^{(0,3)}[\bar g,\phi]$ and $\Gamma_k^{(0,4)}[\bar g,\phi]$. This leads to a tower of coupled differential equations for the $n$-point vertices $\Gamma^{(n,m)}_k[\bar g,\phi]$, which is discussed in more detail in \autoref{sec:Hierarchy}. We use the following notation for derivatives, 
\begin{align}
 \label{eq:def-derivates}
 \Gamma_k^{(n,m)}[\bar g,\phi] =\frac{\delta^{n+m}\Gamma_k[\bar g,\phi] }{\delta \bar g^n\delta\phi^m}\,, 
\end{align}
for general functionals of $\bar g$ and $\phi$. The functional derivative in \eq{eq:def-derivates} includes a factor of $1/\sqrt{\bar g}$, see App.~\ref{app:Notation}. 

The different parameterisations of the metric field, discussed in \autoref{sec:FieldPara}, do not influence the flow equation for the effective action \eq{eq:Gk} and they only differ by their corresponding expansion schemes induced by the relations between metric and fluctuations \eq{eq:LinSplit}, \eq{eq:ExpSplit}, \eq{eq:GeoSplit}. Still, from the viewpoint of diffeomorphism invariance, the different parameterisations differ qualitatively. While the geometrical approach with the fluctuation field \eq{eq:GeoSplit} by construction leads to a diffeomorphism-invariant effective action at all cutoff scales, diffeomorphism invariance is broken in the linear split \eq{eq:LinSplit} and the exponential split \eq{eq:ExpSplit} at a finite cutoff scale. 

For all field parameterisations, a diffeomorphism-invariant effective action with one metric $g$ is obtained at vanishing fluctuation graviton field $h=0$, 
\begin{align}
 \label{eq:backG}
 \Gamma_k[g,\varphi] &= \Gamma_k[g,\phi]\big|_{h=0}\,,
 &
 \text{with}\quad \varphi&=\phi(h=0)\,, 
\end{align}
the \textit{background-effective action}. Its flow equation is given by \eq{eq:flow}, evaluated at vanishing fluctuation field $h=0$,
\begin{align}
 \label{eq:backflow} 
 \partial_t \Gamma_k[g,\varphi]= \frac12 \Tr\, G_k[g,\varphi]\,\partial_t R_k\,. 
\end{align}
Importantly, \eq{eq:backflow} is not closed: the right-hand side depends on $\Gamma_k^{(0,2)}$, the two-point function of the fluctuation fields including the fluctuation graviton field $h$, while the left-hand side knows nothing about $h$. Hence the information about $\delta^2\Gamma_k/\delta h^2$ has to be obtained separately. 

\section{Background field approximation}
\label{sec:BackApprox}
The background-field approximation, introduced in \cite{Reuter:1993kw, Reuter:1996cp} for Yang-Mills theory and gravity respectively, is the most commonly used approximation in the fRG-approach to quantum gravity, see the reviews \cite{Niedermaier:2006wt, Litim:2011cp, Reuter:2012id, Ashtekar:2014kba, Eichhorn:2017egq, Bonanno:2017pkg, Eichhorn:2018yfc, Pereira:2019dbn, Reichert:2020mja, Bonanno:2020bil}. It elevates the diffeomorphism invariance of the background-effective action to that of the full effective action. To that end we write the full effective action in an expansion about the background-effective action in \eq{eq:backG}, 
\begin{subequations} \label{eq:BackApprox}
\begin{align}
 \label{eq:BackApproxG}
 \Gamma_k[\bar g,\phi]=
 \Gamma_k[g,\varphi]+S_{\text{gf}}[\bar g,h]+\Delta\Gamma_k[\bar g , \phi]\,.
\end{align}
The gauge-fixing term $S_{\text{gf}}$ is defined in \eq{eq:gf} and $\Delta\Gamma_k[\bar g, \phi=0]=0$. In the background-field approximation the last term in \eq{eq:BackApproxG} is assumed to be negligible, 
\begin{align}
 \label{eq:DeltaG=0} 
 \Delta\Gamma_k[\bar g , \phi] \approx 0\,.
\end{align}
\end{subequations}
The underlying assumption is that the dynamics of a gauge theory is carried by gauge-invariant fluctuations, while $\Delta\Gamma_k$ carries quantum deformations of the gauge-fixing procedure and should not drive the dynamics. Then, derivatives w.r.t.\ $\bar g$ and w.r.t.\ $h$ agree in the linear split and are related in a simple way in the other parameterisations via \eq{eq:ExpSplit} and \eq{eq:GeoSplit}. 

In the approximation \eq{eq:BackApprox} and with the linear split \eq{eq:LinSplit}, the second derivatives of the effective action w.r.t.\ the background metric and the fluctuation field agree at $\phi=0$ up to the gauge-fixing term: 
\begin{align}
 \label{eq:BackGamma2s} 
 \Gamma_k^{(0,2)}[\bar g,\varphi] = \Gamma_k^{(2,0)}[\bar g,\varphi]+S_\text{gf}^{(0,2)}[\bar g,0]\,.
\end{align}
Inserting \eq{eq:BackApprox} into \eq{eq:flow} leads us to a closed and diffeomorphism-invariant flow for the background-effective action $\Gamma_k[g,\varphi]$. 

\subsection{Properties of the background approximation}
\label{sec:PropBack}
It is the simple relation \eq{eq:BackGamma2s} and the manifest diffeomorphism invariance of the approximation at all cutoff scales that make the background-field approximation so attractive. A large amount of the results in asymptotically safe quantum gravity has been obtained in this approximation and it is still the commonly used approximation in the field. This asks for independent checks of these results as well as its embedding in systematic expansion schemes that go beyond it. In the present work, we review the \textit{fluctuation approach}, see \autoref{sec:Fluctuation-flows}, which includes the correlation functions of the fluctuation graviton field $h$. The results in the background-field approximation are qualitatively in line with the results in the fluctuation approach discussed in \autoref{sec:StateArt}. This confirms -in most cases- the underlying assumption \eq{eq:DeltaG=0}. Nonetheless, some words of caution are needed.

Despite its seeming manifest diffeomorphism invariance, the background-field approximation is at odds with diffeomorphism invariance and background independence. To understand this counterintuitive remark, we recall some features of the background-field formalism to standard quantum field theories, e.g.\ the SM and QCD. The introduction of the background field to the gauge fixing allows defining a gauge-invariant background-effective action. It is evident from its introduction that it is an auxiliary symmetry. The background field can even generate gauge-invariant background-effective actions in theories that explicitly break gauge invariance. This is clear from the construction of diffeomorphism-invariant background-effective actions in gravity in the presence of a background-covariant momentum regulator. In a gauge-invariant theory without a cutoff, it can be shown that the physical gauge invariance of the theory is carried by the fluctuation field in terms of non-trivial Ward- or Slavnov-Taylor identities. The underlying transformations are called quantum gauge/diffeomorphism transformations. This physical symmetry carries over to the auxiliary background gauge invariance via non-trivial Nielsen or split-Ward identities. The latter encode background independence of the theory and have been introduced in \autoref{sec:BackInd}. The Slavnov-Taylor and Nielsen identities for gravity are discussed in detail in \autoref{sec:SymId}.

In summary, \textit{only} if the fluctuation correlation functions satisfy the non-trivial symmetry relations \textit{and} the Nielsen identities, the auxiliary background gauge-invariance is physical. Then it carries the underlying symmetry and we have background independence.

\subsection{Regulator dependence of the background-effective action}
\label{sec:RegDep}
In this section, we first argue that regulator choices within the general class defined with \eq{eq:RegSplits} and \eq{eq:rkLimits} can be used within the background-field approximation to even change the (non-)existence or the nature of an asymptotically-safe UV fixed point. This seems to casts some doubts on the reliability of results obtained in the background-field approximation. However, we then show that the comparison with fluctuation results and the proper use of Nielsen identities (see \autoref{sec:SymId}) suffices to further restrict the general class of regulators such that it is adapted to the background-field approximation.

The regulator term is the origin of the reliability problems of a na\"ive use of the background-field approximation within the fRG-approach: it generates additional terms in $\Delta\Gamma_k[\bar g,h]$ in \eq{eq:BackApproxG} via the background-metric dependence of the regulator. In the background-field approximation \eq{eq:DeltaG=0}, this background-metric dependence is elevated to a dynamical one: in the approximation~\eq{eq:BackGamma2s}, the fluctuation two-point function $\Gamma_k^{(0,2)}$ is computed from background-metric derivatives of the (integrated) flow with the exception of the gauge-fixing term. These derivatives also hit the regulator. Accordingly, we have added dynamics via the choice of the regulator and it remains to be proven in each application that this does not change the results qualitatively.

This has been discussed early on at the example of scalar theories and Yang-Mills theory in \cite{Litim:2002hj, Litim:2002ce}. In particular, it has been shown that the one-loop $\beta$-function in Yang-Mills theory can be changed from its universal result with regulator choices in the background-field approximation. More precisely, it has been shown, that for regulators $R_k(\bar \Delta_s)$ with spin $s=1$ and spin $s=0$ covariant Laplacians $\bar \Delta_s=\Delta_s(\bar A)$ the coefficient $Z_F$ of the $\tr F^2$-term in the effective action runs at one-loop as
\begin{align} 
 \label{eq:nbeta}
 \left. \frac{\partial_t Z_F}{Z_F}\right|_{\text{1-loop}}\!&= n\,\beta_{\alpha_s,\text{1-loop}}\,,
 & 
 \text{for}\;\; R_k(x\to 0)&\propto \frac{1}{x^{n-1}}\,, 
\end{align}
for details we refer to \cite{Litim:2002ce}. This spoils the universality of the one-loop $\beta$-function in Yang-Mills theory. If one does not resort to the background-field approximation, the correct one-loop $\beta$-function is obtained. 

We now discuss the origin of this peculiar behaviour. We follow the argument in \cite{Folkerts:2011jz} and for the general case including gravity, we refer to \cite{Pawlowski:2003sk, Pawlowski:2005xe, Folkerts:2011jz, Donkin:2012ud}. Simply put, we would like to show that the background-effective action at a finite cutoff scale $k$ and in particular in the limit $k\to\infty$ carries no physics without further restrictions of the regulator. We parameterise the regulator with
\begin{align}
 \label{eq:reg}
 R_k=\Gamma_k^{(0,2)} r_k(\bar\nabla)\,,
\end{align}
see also \eq{eq:RegSplits}. Note that in \eq{eq:reg} we have introduced a $\bar{\nabla}$-dependent shape function, which is more general than the $x=\bar{\Delta}/k^2$-dependent one defined in \eq{eq:RegSplits}. Still we use $x$ in a slight abuse of notation for identifying the UV-and IR-limits. As already explained around \eq{eq:RegSplits}, the shape function $r_k$ is a free function of the covariant derivative with the limits \eq{eq:rkLimits}. In particular it has to decay in the UV. With the parameterisation \eq{eq:reg}, the flow equation \eq{eq:backflow} for the background-effective action with $\bar g=g$ reads 
\begin{align}
 \label{eq:backflowfreer} 
 \partial_t \Gamma_k 
 = \frac12 \text{Tr}\,
 \frac{1}{1+r_k}\,\partial_t r_k + \frac12 \Tr\,\frac{1}{\Gamma_k^{(0,2)}}\partial_t\Gamma_k^{(0,2)} 
 \frac{1}{1+r_k}\, r_k \,. 
\end{align}
From the first term on the right-hand side of the flow \eq{eq:backflowfreer}, we deduce that the UV limit of the shape function is constrained: $r_k(x\to \infty) \leq 1/x^{d+\epsilon}$ with $x=\Delta/k^2$, as discussed below \eq{eq:rkLimits}. In turn, the IR limit $x\to 0$ of $r_k$ can be singular without spoiling the finiteness of \eq{eq:backflowfreer}. In order to obtain a general background-effective action, we simply demand that $r_k$ solves the differential equation,
\begin{align}
 \label{eq:rkSol}
 \partial_t r_k = - r_k \,
 \frac{1}{\Gamma_k^{(0,2)}}\partial_t\Gamma_k^{(0,2)} +2 \partial_t Y_k \,. 
\end{align}
This is a simple differential equation that admits a solution at least locally (in the flow time $t$). Note that the UV-decay of $r_k$ also constrains the UV-limit of $Y_k$ with $\partial_t Y_k(x\to\infty) \leq 1/x^{d+\epsilon}$. Inserting a shape function $r_k$ of \eq{eq:rkSol} into \eq{eq:backflowfreer} we arrive at 
\begin{align}
 \label{eq:backflowfreer1} 
 \partial_t \Gamma_k 
 = \text{Tr}\, \partial_t Y_k(\nabla) \,.
\end{align}
\Eq{eq:backflowfreer1} constrains the IR-limit of the function $Y_k$: its flow $\partial_t Y_k$ has to be trace-class for rendering the flow of the background-effective action finite. If we also assume the trace-class property for $Y_k$, the order of $t$-derivative and trace can be swapped.

Apart from these trivial constraints, the choice of $Y_k(\nabla)$ is at our disposal. Integrating the flow \eq{eq:backflowfreer1} from some scale $\Lambda<k$, and taking the UV limit with $k\to\infty$ we arrive at 
\begin{align}
\label{eq:UVlimitGen} 
\lim_{k\to\infty} \Gamma_k[g,\varphi ] = \Bigl[\Gamma_\Lambda - \Tr\, Y_\Lambda\Bigr]+ \Tr \,Y_k (\nabla)
\to \Tr\, Y_k(\nabla)\,. 
\end{align} 
The term $\Gamma_\Lambda - \text{Tr}\,Y_\Lambda$ is $k$- and $\Lambda$-independent, the latter property follows from RG-consistency: $\partial_\Lambda \Gamma_k\equiv 0$ for $k\neq \Lambda$, see e.g.~\cite{Braun:2018svj}. In the last relation in \eq{eq:UVlimitGen} we have assumed that the effective action is dominated by the UV-term $\text{Tr} \,Y_k$. This assumption underlies most fixed-point analyses. 

We emphasise that the result \eq{eq:UVlimitGen} is exact and no approximation has been applied. \Eq{eq:UVlimitGen} implies that without suitable restrictions on the regulator function $r_k$ the flow of the background-effective action $\Gamma_k[g,\varphi] $ (for large cutoff scales) has no physics content at all. Even at one- and two-loop order in perturbatively renormalisable theories, it does not reproduce universal results without further restrictions on the regulator.

The IR limit with $r_{k\to 0} = 0$ puts a severe restriction onto $r_k$, which constrains the integrated flow together with the RG-consistency at the initial cutoff scale $\Lambda$, $\partial_\Lambda\Gamma_{k=0}=0$. However, in the UV-limit the restriction
\begin{align}
 \label{eq:UVlimit}
 r_{k\to\infty} \to \infty\,,
\end{align}
does effectively not restrict the UV-scaling. The latter is dominated by the UV-relevant operators that all satisfy \eq{eq:UVlimit} by definition. Note that so far we have discussed the flow of the background-effective action $\Gamma_k[g,\varphi]$ without resorting to approximations. 

The above issues are already present for the full flow and emphasise the auxiliary nature of the background-effective action at $k\neq 0$. In particular, no conclusion can be drawn from its regularity or singular behaviour in the UV-limit with $k\to\infty$. This situation is further complicated by the background-field approximation. Then, the field dependence that originates from the regulator term is fed back into the flow equation as dynamical contributions. As we have discussed above, these contributions are ambiguous in particular in the UV-limit. In conclusion, the background-field approximation, while having the appeal of simplicity and seeming diffeomorphism invariance has to be applied with great care. To that end we split the problems discussed above into their physics origin:
\begin{itemize}
 \item[(1)] Physical diffeomorphism invariance and background independence is carried by non-trivial Slavnov-Taylor and Nielsen identities of the fluctuation field.
 \item[(2)] The background-field dependence of the regulator term is potentially dangerous in the UV and has to be separated. 
\end{itemize}
A first step in the resolution of the issues of the background-field dependence is to monitor the field-dependence that originates in the regulator. The related equation and discussion in Yang-Mills theory and gravity can be found in \cite{Pawlowski:2001df, Pawlowski:2002eb, Litim:2002ce, Pawlowski:2003sk, Pawlowski:2005xe, Folkerts:2011jz, Donkin:2012ud}, for applications to gravity see also \cite{Morris:2016spn, Percacci:2016arh, Labus:2016lkh, Ohta:2017dsq, Nieto:2017ddk}. The equation that monitors this dependence is given by
\begin{align}
\label{eq:backrdep}
 \text{Tr}\!\left[ \frac{\delta \sqrt{\bar g} R_k }{\delta\bar g_{\mu\nu}} \frac{\delta}{\delta R_k}\right]\Gamma_k[\bar g,\phi]=
 \frac12 \text{Tr}\, \frac{\delta \sqrt{\bar g} R_k }{\delta\bar g_{\mu\nu}} G_k[\bar g,\phi] \,.
\end{align}
\Eq{eq:backrdep} allows to disentangle the background-metric dependence stemming from the regulator from the rest. In the Yang-Mills example from \eq{eq:nbeta}, it can be shown that the regulator-field dependence is responsible for a contribution $(1-n) \beta_{\alpha_s,\text{1-loop}} $. Subtracting the contribution from the regulator-field dependence, the universal result is obtained. Indeed, even without an explicit computation, we can already infer from \eq{eq:backrdep} that the universal 1-loop $\beta$-function of the dimensionless Yang-Mills coupling is achieved for IR-regular regulators: the projection of the right-hand side of \eq{eq:backrdep} on the dimensionless term proportional to $\tr F_{\mu\nu}^2$ can only depend on the cutoff scale $k$ in the presence of an additional scale. For IR-regular regulators such a scale is absent and the $k$-derivative of \eq{eq:backrdep} vanishes. In turn, IR-singular regulators implicitly introduce a further IR scale and the $k$-derivative of \eq{eq:backrdep} does not vanish. This explains the structure of the result in \eq{eq:nbeta}. We emphasise that the modification of the dynamics in the background-field approximation via the regulator term is not restricted to IR-singular regulators. The latter fact is a peculiarity of the \textit{universal} one-loop running of the \textit{dimensionless} Yang-Mills coupling. In particular, we emphasise that for non-universal couplings as well as theories with dimensionful couplings such as gravity the flow of \eq{eq:backrdep} does not vanish for IR-regular regulators.

Based on this analysis it has been suggested in \cite{Litim:2002hj,Litim:2002ce}, that within the background-field approximation the corresponding field-dependence should be subtracted before applying the approximation $\Gamma_k^{(0,2)}\simeq \Gamma_k^{(2,0)}$ for the right-hand side of the flow. This idea has been picked up by \cite{Bridle:2013sra, Dietz:2015owa, Eichhorn:2018akn} for scalar theories, $f(R)$-gravity and gravity-matter systems, for more details see \autoref{sec:SymId}. These works are based on the relation \eq{eq:backrdep}, where one derivative with respect to the background is taken. To fully resolve $\Delta \Gamma_k$ in \eq{eq:BackApproxG}, a further field derivative of \eq{eq:backrdep} is needed. Furthermore, \eq{eq:backrdep} does not comprise the full difference between $h$- and $\bar g$-derivatives. While the background-field correlation functions are diffeomorphism-covariant due to background diffeomorphism invariance, the fluctuation correlation functions satisfy difficult Slavnov-Taylor identities. This is well-known and well-studied (though not fully conclusively) in non-Abelian gauge theories where one also has access to respective lattice results, for a recent review and related references, see \cite{Dupuis:2020fhh}. In turn, the related analysis, while in high demand, is less advanced in quantum gravity, see also \cite{Bonanno:2020bil, Dupuis:2020fhh}. This is detailed in the next section.

\section{Symmetry identities}
\label{sec:SymId}
Physical observables are diffeomorphism invariant and background-independent. The underlying symmetry is dynamical and is solely carried by the dynamical fluctuation fields. It is called \textit{quantum} diffeomorphism invariance and reads 
\begin{align}
 \label{eq:quantum-diff}
 h_{\mu\nu} &\longrightarrow h_{\mu\nu} + \CL_{\omega} ( \bar g_{\mu\nu} + h_{\mu\nu} ) \,,
 &
 \bar g_{\mu\nu} &\longrightarrow \bar g_{\mu\nu} \,.
\end{align}
The background metric triggers an a priori \textit{auxiliary} symmetry, the \textit{background} diffeomorphism invariance. It is given by the transformation
\begin{align}
 \label{eq:backgr-diff}
 h_{\mu\nu} &\longrightarrow h_{\mu\nu} + \CL_{\omega} h_{\mu\nu} \,,
 &
 \bar g_{\mu\nu} &\longrightarrow \bar g_{\mu\nu} + \CL_{\omega} \bar g_{\mu\nu} \,.
\end{align}
Here $\CL_{\omega}$ is the Lie derivative with respect to some vector field $\omega_\mu$, which reads for a rank-two tensor 
\begin{align}\label{eq:Lie}
 \CL_{\omega} T_{\mu\nu} = \omega_\rho \bar \nabla ^\rho T_{\mu\nu} + T_{\mu\rho} \bar \nabla^\rho \omega_\nu + T_{\nu\rho} \bar \nabla^\rho \omega_\mu \,.
\end{align} 
Both tranformations, \eq{eq:quantum-diff} and \eq{eq:backgr-diff}, generated diffeomorphism transformations on the full metric $g_{\mu\nu}$, so they do not differ on the functional of $g_{\mu\nu}$. Moreover, while \eq{eq:backgr-diff} is an auxiliary symmetry, it still comprises the information of the dynamical quantum-diffeomorphism symmetry \eq{eq:quantum-diff} via the Nielsen identities. The latter carry the background independence of the theory. 

Any fRG-computation needs to introduce a gauge fixing and a regularisation, which both apparently break diffeomorphism invariance and (on-shell) background independence. Thus, it is an important issue in the fRG-approach to quantum gravity to discuss how these properties can be preserved in a non-perturbative computation. For each symmetry broken by the cutoff term, we can formulate a non-trivial modified symmetry identity, which captures the cutoff-deformation of the underlying symmetry and smoothly approaches the unbroken symmetry identity at vanishing cutoff scale, $k=0$. We now first discuss how the Nielsen identities take care of background independence and afterwards discuss quantum diffeomorphism invariance due to the Slavnov-Taylor identities. Note that also in discrete gravity models the Ward identities play a crucial r\^ole, see \cite{Baloitcha:2020idd} for a review of tensor models.

\subsection{Background independence}
\label{sec:BackIndep}
As discussed in \autoref{sec:GaugeFixing}, we always need to split the full metric into a background metric $\bar g$ and a fluctuation field $h$. This split introduces an additional symmetry given by all transformations of the background metric and of the fluctuation field that leave the full metric invariant.
\begin{align}
 \label{eq:metric-trafo}
 g( \bar g, h) \longrightarrow g( \bar g + \delta \bar g, h+ \delta h) = g( \bar g, h)\,.
\end{align}
For example in the linear split \eq{eq:LinSplit}, we have $\delta \bar g = -\delta h$. This symmetry is guaranteeing background independence since we can always find a transformation that changes the background according to our choice. This symmetry is broken off-shell by the gauge-fixing and ghost action and further broken by the regularisation on- and off-shell. The breaking of the symmetry is described by the Nielsen (or split-Ward) identities\cite{Nielsen:1975fs, Fukuda:1975di}. They encode the background independence of the physical observables and allow us to restore the symmetry at vanishing cutoff.

Let us first discuss the Nielsen identities without the regulator. The Ward identity for the effective action for any symmetry transformation $\CG$ is given by
\begin{align}
 \label{eq:Ward}
 \CW = \CG \Gamma - \langle \CG (S_\text{gf} + S_\text{gh}) \rangle= 0\,,
\end{align}
where $S_\text{gf}$ and $S_\text{gh}$ are defined as in \eq{eq:gf} and \eq{eq:Sghost}. We apply this to the transformation of the metric split \eq{eq:metric-trafo} and obtain the Nielsen identity $\text{NI} =0$, with
\begin{align} 
 \label{eq:NI}
 \text{NI}&= \frac{\delta\Gamma}{\delta \bar g_{\mu\nu}} -\int \!\llangle \frac{\delta \hat h}{\delta \bar g_{\mu\nu}} \rrangle \cdot\frac{\delta\Gamma}{\delta h} \notag \\
 &\quad -\llangle \left[\frac{\delta}{\delta \bar g_{\mu\nu}}- \int \! \frac{\delta \hat h}{\delta \bar g_{\mu\nu}}\cdot \frac{\delta}{\delta \hat h} \right] ( S_\text{gf}+S_\text{gh} )\rrangle\,,
\end{align}
where $h_{\mu\nu}=\langle \hat h_{\mu\nu}\rangle$, and the fluctuation field is understood as function of the full metric and the background metric $\hat h (g,\bar g)$. For the linear split \eq{eq:LinSplit}, we have $\frac{\delta \hat h_{\rho\sigma}(x)}{\delta \bar g_{\mu\nu}(y)} = \frac1{\sqrt{\bar g}} \delta(x-y) \frac12 (\delta^\mu_\rho \delta^\nu_\sigma+\delta^\mu_\sigma \delta^\nu_\rho)$, see App.~\ref{app:Notation}, and thus, 
\begin{align}
 \label{eq:linNielsen}
 \text{NI}_\text{lin}&=\frac{\delta\Gamma_\text{lin}}{\delta \bar g_{\mu\nu}} -\frac{\delta\Gamma_\text{lin}}{\delta h_{\mu\nu}}
 -\llangle \!\left[\frac{\delta}{\delta \bar g_{\mu\nu}}-\frac{\delta}{\delta \hat h_{\mu\nu}} \right] \!( S_\text{gf}+S_\text{gh})\!\rrangle. 
\end{align} 
The Nielsen identity for the exponential split \eq{eq:ExpSplit} resembles \eq{eq:linNielsen}: there is a non-trivial difference between the background-metric and fluctuation-field derivatives due to the gauge-fixing and ghost terms. In \eq{eq:FlucEoM=backEoM}, we have pointed out that at $k=0$, a solution of the background EoM is also a solution of the quantum EoM and vice versa. This implies together with \eq{eq:linNielsen} that the expectation value $\langle [\delta_{\bar g}-\delta_{\hat h}]( S_{\text{gf}}+S_{\text{gh}} )\rangle$ needs to vanish on-shell. This is indeed non-trivial and does not happen off-shell.

In comparison, for the fully diffeomorphism-invariant Vilkovisky-DeWitt or geometrical effective action with the split given by \eq{eq:Vilkovisky}, the dependence on the gauge-fixing action and the ghost action is vanishing and thus the Nielsen identity reads
\begin{align}
 \label{eq:geoNielsen} 
 \text{NI}_{\text{geo}}=
 \frac{\delta\Gamma_{\text{geo}}}{\delta \bar g_{\mu\nu}}
 - \int \!\llangle \frac{\delta \hat h}{\delta \bar g_{\mu\nu}} \rrangle\cdot
 \frac{\delta\Gamma_{\text{geo}}}{\delta h} \,. 
\end{align}
In contradistinction to the linear and exponential split, the $\bar g$ and $h$-derivatives are directly related.

\begin{figure*}[t]
 \includegraphics[width=.48\linewidth]{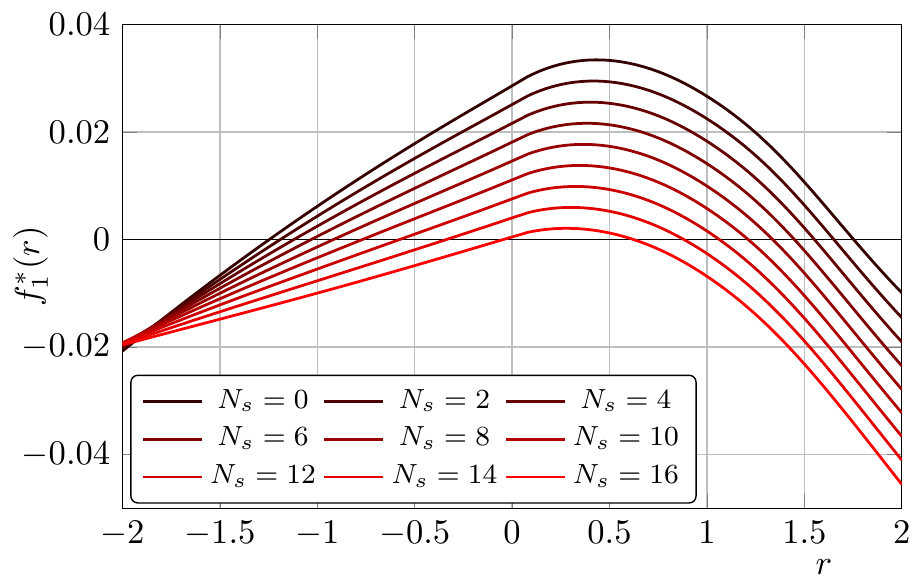} \hfill
 \includegraphics[width=.505\linewidth]{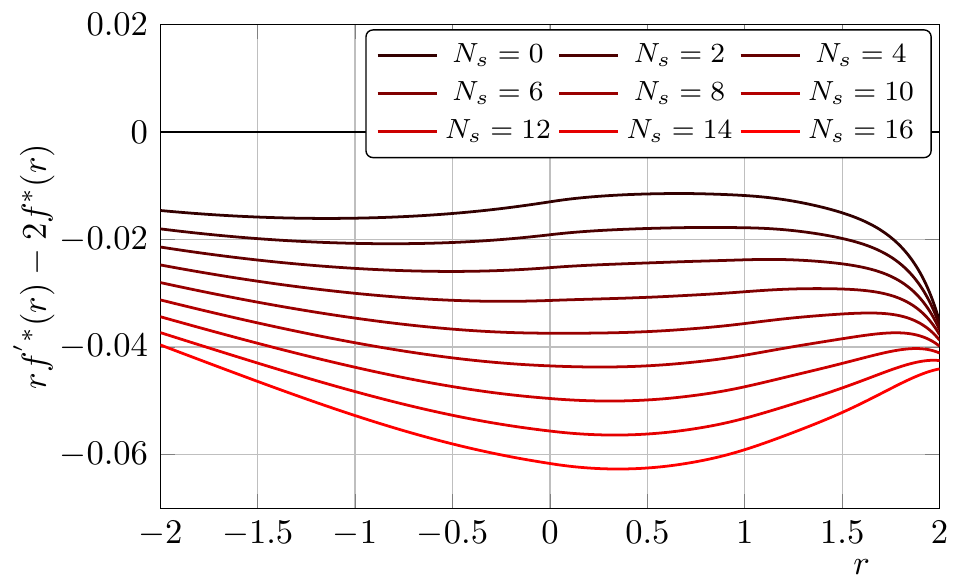} 
 \caption{
 Displayed are the potential of the one-point function and the derivate of the background potential for different numbers of scalar fields at the fixed point as defined in \eq{eq:explicit-beom} and \eq{eq:explicit-qeom}. A zero in these functions indicates a solution to the quantum and background EoM, respectively. While the former always has two solutions, a minimum at negative curvature and a maximum at positive curvature, the latter shows no solution at all. The figures are taken from \cite{Burger:2019upn}.
 }
 \label{fig:EoMs}
\end{figure*}

The Nielsen identities entail that for all metric-splits the effective action is not a function of the full metric $g$, but depends separately on the background metric $\bar g$ and the fluctuation field $h$. Consequently, the effective action has no simple expansion in terms of diffeomorphism invariant quantities in $g_{\mu\nu}$. Still, the Nielsen identities relate $\bar g$- and $h$-derivatives such, that on the solution of the Nielsen identities, the effective action carries background independence and only depends on one field.

So far, the analysis has been performed in the absence of the cutoff term, that is at $k=0$. At finite $k$, the regulator term introduces a further breaking of the split symmetry \eq{eq:metric-trafo}. The Nielsen identities turn into modified Nielsen identities, $\text{mNI}=0$, that read for a general split  
\begin{align}
 \label{eq:mNI}
 \text{mNI} = \text{NI} 
 &- \frac12 \text{Tr}\,\frac{\delta \sqrt{\bar g} R_k}{\sqrt{\bar g}\,\delta \bar g_{\mu\nu}} G_k 
 - \text{Tr}\,R_k G_k \!\left[\frac{\delta}{\delta \phi}\!\llangle \frac{\delta \hat \phi}{\delta \bar g_{\mu\nu}} \rrangle\!\right].
\end{align}
Note, that in the last term in \eq{eq:mNI}, only the metric fluctuation $h$ contributes as the other fluctuation fields do not depend on the background metric. Furthermore, in the linear split, the last term is vanishing and consequently, the mNI simplifies to
\begin{align}
 \label{eq:mNIlin}
 \text{mNI}_\text{lin} = \text{NI}_\text{lin} - \frac12 \text{Tr}\,\frac{\delta \sqrt{\bar g} R_k}{\sqrt{\bar g}\,\delta \bar g_{\mu\nu}} G_k \,.
\end{align}
While some of the properties and consequences of the mNI are theory-dependent, most of them are generic and much can be learned about applications in gravity from investigations in general theories: mNIs have been discussed in detail in gravity, gauge theories, and in scalar theories, see \cite{Reuter:1996cp, Reuter:1997gx, Freire:2000bq, Litim:2002ce, Litim:2002hj, Pawlowski:2003sk, Pawlowski:2005xe, Manrique:2009uh, Manrique:2010mq, Donkin:2012ud, Bridle:2013sra, Becker:2014qya, Dietz:2015owa, Safari:2015dva, Safari:2016dwj, Safari:2016gtj, Labus:2016lkh, Morris:2016spn, Percacci:2016arh, Nieto:2017ddk, Ohta:2017dsq, Eichhorn:2018akn, Lippoldt:2018wvi, Reichert:2020mja}.
 
There is an important qualitative difference between the breaking of the metric-split symmetry \eq{eq:metric-trafo} at finite $k$ and at $k=0$. We have already discussed in \autoref{sec:BackInd}, that the Nielsen identity at vanishing cutoff scale, $k=0$, encodes background independence, manifested in the equivalence of the solutions of the background and fluctuation EoMs, \eq{eq:FlucEoM=backEoM}. At finite cutoff scale, $k\neq 0$, we necessarily have background dependence, as the quantum fluctuations have to be ordered in a specific background. This is also manifest in the missing equivalence of the background and fluctuation EoMs, the respective solutions do not agree,  
\begin{align}
 \label{eq:FlucEoM-neq-backEoM} 
 \frac{\delta\Gamma_k[\bar g^\text{fluc}_\text{EoM},0]}{\delta h_{\mu\nu}} &=0 =  
 \frac{\delta\Gamma_k[\bar g^\text{back}_\text{EoM},0]}{\delta \bar g_{\mu\nu}}\,,
 &
 \bar g^\text{fluc}_\text{EoM}&\neq \bar g^\text{back}_\text{EoM}\,, 
\end{align}
for a detailed discussion see \cite{Christiansen:2017bsy, Lippoldt:2018wvi, Burger:2019upn}. The difference between the solutions can be parameterised by a term proportional to the regulator $R_k$, which is most easily seen in the modified Nielsen identity in the geometric approach, \eq{eq:geoNielsen} and \eq{eq:mNI}. 

The difference between $\bar g^\text{fluc}_\text{EoM}$ and $\bar g^\text{back}_\text{EoM}$ was explicitly computed in \cite{Christiansen:2017bsy, Burger:2019upn} for backgrounds with constant curvature. The ansatz for the background-effective action is
\begin{align}
\Gamma_k[\bar g] = \int \!\mathrm d^4x \sqrt{\bar g}\, k^4 f(r) = V \tilde f(r) \,,
\end{align}
where $V$ is the spacetime volume and $r = \bar R/k^2$ is the dimensionless background curvature. Thus the background EoM becomes
\begin{align}
 \label{eq:explicit-beom}
 \Gamma_k^{(\bar g)}[\bar g,0] \sim r f'(r) - 2 f(r) = 0 \,,
\end{align} 
which is displayed in the right panel of \autoref{fig:EoMs} at the UV fixed point for different numbers of scalar fields $N_s$. The ansatz for the fluctuation one-point function reads
\begin{align}
\label{eq:Gamma-f1}
\Gamma_k^{(h_\text{tr})}[ \bar g,0]=\int \! \mathrm d^4 x\sqrt{\bar g}\, k^3 f_1(r)=\frac{V}{k} f_1(r) \,,
\end{align}
and thus the quantum EoM is simply
\begin{align}
 \label{eq:explicit-qeom}
 \Gamma_k^{(h_\text{tr})}[\bar g,0] \sim f_1(r) = 0 \,.
\end{align}
This is shown in the left panel of \autoref{fig:EoMs} at the UV fixed point for different numbers of scalar fields $N_s$.

The background EoM does not display a solution in the whole investigated region, while the quantum EoM has two solutions, a minimum at negative curvature and a maximum at positive curvature. For a larger number of scalar fields, these two solutions merge. However, in this regime, the approximation lacks reliability due to large values of the graviton anomalous dimension. Importantly, \autoref{fig:EoMs} manifests in a explicit computation the difference between the background and quantum EoM \eq{eq:FlucEoM-neq-backEoM}. The background EoM was also extensively investigated in the background-field approximation with different choices of regulator and parameterisation. For example in \cite{Falls:2014tra}, the linear split was used and a solution at large negative curvature was found. However, in \cite{Falls:2017lst,Falls:2018ylp}, two further solutions at positive curvature were found due to a different choice of the regulator. A solution at positive curvature was also found in \cite{Demmel:2015oqa} and with the exponential parameterisation in \cite{Ohta:2015fcu}.

In \cite{Lippoldt:2018wvi}, a modification of the fRG equation was proposed. There, the effective action was defined as the Legendre transform of a normalised Schwinger functional, $\hat W_k[\bar g,J] = \log (Z_k[\bar g,J]/Z_k[\bar g,0])$. This modification implies that the solutions to the quantum and background EoMs agree even at a finite cutoff scale. This does not imply that the modified effective action is background-independent at finite $k$ since there are differences in the higher-order correlation function. However, it allows constructing improved background-field approximations, which might allow resolving some tensions between background and fluctuation results.

\subsection{From BRST to diffeomorphism invariance}
\label{sec:BRST+Diff}
While the auxiliary background diffeomorphism invariance \eq{eq:backgr-diff} remains unbroken, the physical quantum diffeomorphism invariance \eq{eq:quantum-diff} turns into a BRST symmetry due to the gauge fixing, which is then further broken by the regulator. The related symmetry identities are called (modified) Slavnov-Taylor identities ((m)STI) \cite{Taylor:1971ff,Slavnov:1972fg}. They encode physical diffeomorphism invariance. We sketch the main ideas of the derivation and apply them to gravity.

\begin{figure*}[t]
\includegraphics[width=\linewidth]{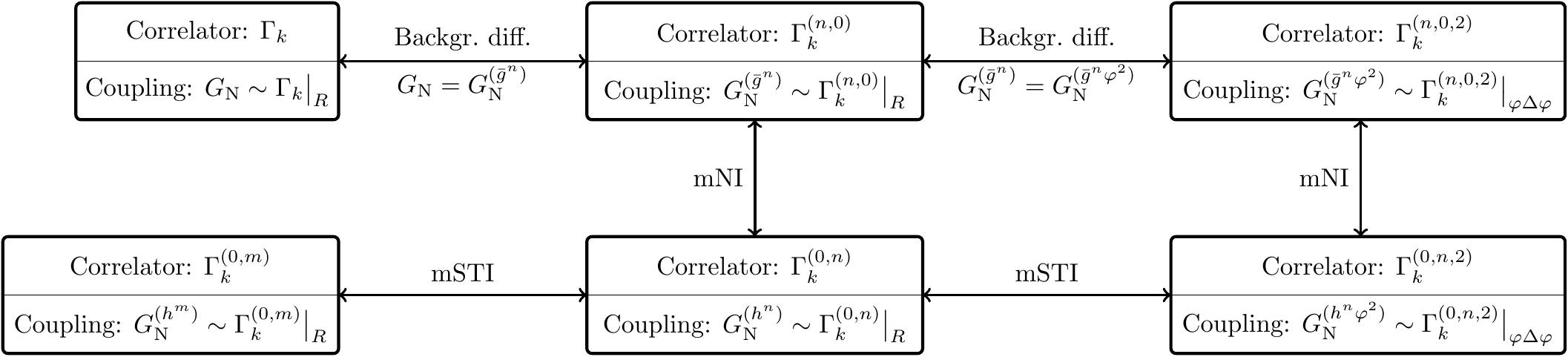}
\caption{Displayed are the relations between background and fluctuation correlation functions in terms of symmetry identities. The background diffeomorphism symmetry \eq{eq:backgr-diff} remains unbroken and trivially connects background correlation functions. The split symmetry \eq{eq:metric-trafo} is encoded in the modified Nielsen identity (mNI) \eq{eq:mNI} and relates background correlation functions with fluctuation ones. The quantum diffeomorphism symmetry \eq{eq:quantum-diff} is described by the modified Slavnov-Taylor identity (mSTI) \eq{eq:mSTI} and relates fluctuation correlation functions. For the purpose of illustration, we have assumed that $\Gamma_k=\Gamma_k[\bar g,h,\varphi]$ depends on the background metric $\bar g$, the metric fluctuation $h$ and a scalar field $\varphi$. The notation $\Gamma_k^{(n_1,n_2,n_3)}$ is then defined as in \eq{eq:def-derivates}.}
\label{fig:avatar-picture}
\end{figure*}

In case of the linear gauge-fixing condition \eq{eq:gf-condition} the generator of BRST-transformation (or BRST-operator) denoted by $\mathfrak{s}$, including the Nakanishi-Lautrup field $\lambda_\mu$, is given by 
\begin{align}
 \label{eq:brst}
 \mathfrak{s} (\bar g_{\mu\nu}&,h_{\mu\nu},c_\mu, \bar c_\mu, \lambda_\mu, \phi_\text{mat}) \notag \\[1ex]
 &= ( 0,\CL_{c}(\bar g_{\mu\nu} + h_{\mu\nu}),c_\rho \bar \nabla^\rho c_\mu, \lambda_\mu, 0, \mathfrak{s} \phi_\text{mat})\,. 
\end{align}
In \eq{eq:brst}, the vector field $\omega_\mu$ in the Lie derivative \eq{eq:Lie} is given by the ghost field, $\omega_\mu=c_\mu$, for more details on the setup and the condensed notation used below see \cite{Pawlowski:2005xe}. The Nakanishi-Lautrup field $\lambda_\mu$ transforms trivially under the BRST transformation, $\mathfrak{s}\lambda_\mu=0$. The classical gauge-fixed action incuding the gauge-fixing and the ghost action is invariant under this transformation, $\mathfrak{s}(S_\text{grav}+ S_\text{gf}+S_\text{gh})=0$. Furthermore, $\mathfrak{s}$ is a nilpotent operator with $\mathfrak{s}^2=0$.

For the derivation of the STI, we include a source term $Q^a \mathfrak{s} \hat\phi_a$ for the BRST variations of the fields in the generating functional. The Schwinger functional now reads
\begin{align}
 \label{eq:Schwinger}
 e^{W[\bar g, \phi, J, Q]} = \int \! \CD \hat\phi \, e^{-S_\text{tot}+ \int\!\mathrm d^4 x \sqrt{\bar g} ( J^a \hat\phi_a + Q^a \mathfrak{s} \hat\phi_a) }\,,
\end{align}
where $S_\text{tot} = S_\text{grav} + S_\text{gf} + S_\text{gh}$. The STI follows from the BRST-invariance of generating functional, 
\begin{align}
 \int \! \mathfrak{s}\left(\CD \hat\phi \, e^{-S_\text{tot} + \int\!\mathrm d^4 x \sqrt{\bar g} ( J^a \hat\phi_a + Q^a \mathfrak{s} \hat\phi_a) }\right) =0 \,.
\end{align}
The source term $J^a \hat\phi_a$ is the only BRST-variant term. The BRST-operator $\mathfrak{s}$ commutes with bosonic sources and anti-commutes with fermionic sources. This leads us to $\mathfrak{s}J^a\hat\phi_a = J^a \gamma^b{}_{a} \,\mathfrak{s}\hat\phi_b$, where the metric $\gamma^b{}_{a} $ carries the minus sign for the fermionic terms, see App.~\ref{app:Notation}. 

With these properties we obtain the STI for the Schwinger functional, 
\begin{align}
 \int \! \CD \hat\phi &\int\!\mathrm d^4 x \sqrt{\bar g}\, J^a \gamma^b_{\;\;a} (\mathfrak{s}\hat\phi_b)\, e^{-S_\text{tot} + \int\!\mathrm d^4 x \sqrt{\bar g} ( J^a \hat\phi_a + Q^a \mathfrak{s} \hat\phi_a)} \notag \\[1ex]
 =&\, \int\!\mathrm d^4 x \sqrt{\bar g}\, J^a \gamma^b_{\;\;a} \frac{\delta}{\delta Q^b} e^{W[\bar g, \phi, J, Q]} =0 \,.
\end{align}
This identity can be re-expressed in terms of the effective action, see \cite{Pawlowski:2005xe} for details. Here we just just state the result for the STI in the absence of the cutoff term, 
\begin{align}
 \label{eq:STI}
 \text{STI} =\int\!\mathrm d^4 x \sqrt{\bar g} \,\frac{\delta \Gamma}{\delta \phi_a} 
 \frac{\delta \Gamma}{\delta Q^a} = 0 \,.
\end{align}
This equation is known as the quantum-master equation. The BRST variation of the effective action is given by $\delta \Gamma/\delta Q^a = \langle \mathfrak{s} \hat \phi_a \rangle$. These variations can be interpreted as generalised vertices of the theory.

\Eq{eq:STI} encodes diffeomorphism invariance at $k=0$ where the regulator vanishes. At finite cutoff scale, an additional regulator contribution has to be taken into account, and we are led to the mSTI, 
\begin{align}
 \label{eq:mSTI}
 \text{mSTI} = \text{STI} - \,\text{Tr}\,R_k \,\frac{\delta^2 \Gamma_k}{\delta Q \delta \phi} \,G_k = 0 \,.
\end{align}
Some of the properties of the mSTI are theory-dependent, but most of them are generic: mSTIs in the presence and absence of background fields in gravity and gauge theories have been discussed in detail in \cite{Bonini:1993sj, Bonini:1994kp, Bonini:1994dz, Bonini:1995tx, Ellwanger:1994iz, DAttanasio:1996tzp, Reuter:1997gx, Litim:1998yc, Freire:2000bq, Igarashi:1999rm, Igarashi:2000vf, Igarashi:2001mf, Igarashi:2001ey, Igarashi:2001cv, Litim:2002ce, Pawlowski:2003sk, Pawlowski:2005xe, Gies:2006wv, Igarashi:2007fw, Igarashi:2008bb, Igarashi:2009tj, Donkin:2012ud, Sonoda:2013dwa, Safari:2015dva, Igarashi:2016gcf, Asnafi:2018pre, Igarashi:2019gkm, Reichert:2020mja}.

In summary, we have three symmetries:
\begin{itemize}
 \item[i)] The auxiliary background diffeomorphism invariance \eq{eq:backgr-diff}, which remains unbroken.
 \item[ii)] The quantum diffeomorphism invariance \eq{eq:quantum-diff}, which describes physical diffeomorphism invariance. It is broken and encoded in the mSTI \eq{eq:mSTI}.
 \item[iii)] The split symmetry \eq{eq:metric-trafo}, which guarantees background independence. It is broken as well and encoded in the mNI \eq{eq:mNI}.
\end{itemize} 
The relations between background and fluctuation correlation functions are summarised in \autoref{fig:avatar-picture}. The relation between two fluctuation correlation functions can be expressed either with an mSTI or with a combination of mNI and background diffeomorphism invariance. However, it should be noted that in a truncated non-perturbative computation these two possibilities of relating fluctuation correlation function do not agree with each other. Nonetheless, it can be used to check the error of the truncation, see \autoref{sec:Hierarchy} for more details. 

Last but not least, the flow of mNI and the mSTI is proportional to itself, respectively. This is conveniently expressed in terms of the flow equation for composite operators, derived in \cite{Pawlowski:2005xe, Igarashi:2009tj, Pagani:2016pad}. Schematically it reads 
\begin{align}
\label{eq:FlowComposite} 
\partial_t \CO_k[\bar g, \phi] = -\frac12 \text{Tr} \, G_k\,\partial_t R_k\,G_k \, \CO^{(0,2)}_k[\bar g,\phi]\,. 
\end{align}
The operator $\CO^{(0,2)}_k$ is contracted with $G_k\,\partial_t R_k\,G_k$ in the trace. The set of composite operators $\CO_k[\bar g, \phi]$ with the flow \eqref{eq:FlowComposite} includes general correlation functions $Z^{(0,n)}[\bar g, J[\bar g,\phi]]$ with their disconnected parts as well as more general functions of the field-dependent source such as $J[\bar g, \phi]=\Gamma_k^{(0,1)}[\bar g, \phi]$. In the most general case of a functional with an explicit cutoff-dependence, further terms enter \eqref{eq:FlowComposite}, see \cite{Pawlowski:2005xe}. An educative example is $\Gamma_k^{(0,1)}$: inserting it into \eqref{eq:FlowComposite} leads to the fluctuation-field derivative of the flow equation \eqref{eq:flow}. An instructive example for the case of general correlation functions and the necessity of including the disconnected terms is the full two-point function $G_{\phi_1\phi_2} +\phi_1 \phi_2$. \Eq{eq:FlowComposite} has been used in Yang-Mills theories for the traced Polyakov loop observables \cite{Herbst:2015ona} and in gravity for the study of the renormalisation and scaling of composite operators \cite{Pagani:2016dof, Becker:2019fhi, Houthoff:2020zqy, Kurov:2020csd}. 

Importantly, the set of composite operators $\{ \CO_k \}$ includes modified symmetry identities, i.e., $\text{Sym}_k=\text{mSTI},\text{mNI},\ldots$, see \cite{Pawlowski:2005xe} and also \cite{Ellwanger:1994iz, DAttanasio:1996tzp, Litim:1998qi, Litim:2002ce}. Hence the flow of the symmetry identities reads schematically 
\begin{align}
\label{eq:FlowSym} 
\partial_t \text{Sym}_k[\bar g,\phi]= -\frac12 \text{Tr} \, G_k\,\partial_t R_k\,G_k\, 
\text{Sym}_k^{(0,2)}[\bar g,\phi]\,. 
\end{align}
\Eq{eq:FlowSym} implies that once we have solved these identities at a scale $k$, then the identities are satisfied at all scales. However, this only holds for untruncated flows or truncations that are compatible with \eqref{eq:FlowSym}. More details can be found in \autoref{sec:Hierarchy}.

\subsection{Challenges for diffeomorphism-invariant flows}
\label{sec:DiffInv} 
Gauge-invariant approaches to quantum field theories have received much attention over the decades both in perturbation theory and beyond. Such formulations also have met considerable challenges except for lattice gauge theories that are based on link variables formulated in the gauge group. In turn, perturbation theory and non-perturbative functional approaches are based on correlation functions and in particular on the propagator of the algebra-valued gauge field. For reviews on lattice approaches to quantum gravity see e.g.~\cite{Ambjorn:1997yvj, Ambjorn:2012jv, Loll:2019rdj, Hamber:2009mt, Eichhorn:2018phj}.

Gauge-invariant functional formulations are based either on gauge-invariant or gauge-covariant variables such as the geometrical formulation, the field-strength formulation, or Wilson-line formulations similar to lattice gauge theories. Implementations in the flow equation approach range from generalised Polchinski equations with gauge-covariant kernels for the Wilson effective action \cite{Morris:1999px, Morris:2000fs, Arnone:2001iy, Arnone:2002cs, Arnone:2005fb, Arnone:2005vd, Morris:2005tv, Morris:2006in, Rosten:2006qx, Rosten:2006tk, Rosten:2006pd, Rosten:2008zp} and its recent manifestations \cite{Morris:2016nda, deAlwis:2017ysy, Bonanno:2019ukb}, over the geometrical or Vilkovisky-DeWitt flows for the effective action \cite{Branchina:2003ek, Pawlowski:2003sk, Pawlowski:2005xe, Donkin:2012ud, Demmel:2014hla}, to a recent suggestion for a gauge invariant flow for the effective action \cite{Wetterich:2016qee, Wetterich:2017aoy, Pawlowski:2018ixd, Wetterich:2019qzx, Wetterich:2019zdo}. 
 
Most of these approaches rely explicitly or implicitly on the definition of projection operators on the subspace of the dynamical degrees of freedom. Typically this is achieved by a gauge fixing but the notation of a projection is far more versatile. The appropriate definition of this projection and the respective geometrical structure of the configuration space is at the root of the geometrical construction. This has been discussed in \autoref{sec:GeoSplit}, \autoref{sec:BackIndep}, and \autoref{sec:BRST+Diff}, and we refer to the discussions there. The notable non-locality of the projections both in field space as well as momentum space is an \textit{inherent} property of the construction of gauge-invariant subspaces. Consequently, it should be considered an \textit{inherent} feature of such a construction. This inherent non-locality may be buried in functional self-consistency relations but it is present explicitly or implicitly without any doubt. 

In any case, the situation calls for self-consistency checks of the final formulations of gauge-invariant or diffeomorphism-invariant flows. This necessity has been discussed already in \cite{Litim:2002xm}: there the terminology of \textit{complete} and \textit{consistent} flows was introduced. The former flows generate all quantum fluctuations from a given classical action while the latter flows generate a well-defined subset of quantum fluctuations from a given -partial- effective action. A well-known example for the latter are thermal flows, which only generate thermal fluctuations from the full quantum effective action at vanishing temperature. In \cite{Litim:2002xm, Litim:2001hk}, an important and simple consistency check for flow equations has been suggested: any \textit{complete} flow equation must generate the complete perturbation theory upon iteration from the given classical action. While one-loop perturbation theory in the fluctuation field is trivially achieved within one-loop exact flow equations, two-loop perturbation theory provides a non-trivial necessary, while not sufficient, consistency check.  

These checks for diffeomorphism-invariant fRG-approaches have been passed for the Wilsonian approach \cite{Morris:1999px, Morris:2000fs, Arnone:2001iy, Arnone:2002cs, Arnone:2005fb, Arnone:2005vd, Morris:2005tv, Morris:2006in, Rosten:2006qx, Rosten:2006tk, Rosten:2006pd, Rosten:2008zp}, or are trivial for the geometrical effective action approach \cite{Branchina:2003ek, Pawlowski:2003sk, Pawlowski:2005xe, Donkin:2012ud, Demmel:2014hla}. It is a highly relevant and interesting question how the more recent proposals \cite{deAlwis:2017ysy, Bonanno:2019ukb, Wetterich:2016qee, Wetterich:2017aoy, Pawlowski:2018ixd, Wetterich:2019qzx, Wetterich:2019zdo} fare in such a self-consistency check. Respective investigations either confirm the \textit{completeness} of the approaches or may show their \textit{consistency}, i.e., they may integrate-out a well-defined subset of quantum fluctuations. Finally, for potentially \textit{consistent} flows such an investigation may enable the construction of non-trivial two-loop consistent extensions. We emphasise that such an extension does not simply pass a two-loop test, but more importantly allows for two-loop resummed non-perturbative approximations. The latter set of approximations certainly live-up to the self-consistency of other state-of-the-art computations in asymptotically safe quantum gravity while having the benefit of inherent diffeomorphism invariance. 

\section{Fluctuation approach}
\label{sec:Fluctuation-flows}
In the last sections, we have detailed the need for an fRG approach to quantum gravity that goes beyond the background-field approximation and that allows satisfying the non-trivial symmetry identities, the mSTI \eq{eq:mSTI} and the mNI \eq{eq:mNI}. For general metrics $g_{\mu\nu}$, this requires to solve the flow equation \eq{eq:flow} for the two-field action $\Gamma_k[\bar g,h]$. It is already a formidable task for the one-field flow in the background-field approximation discussed in \autoref{sec:BackApprox}. Indeed, already in scalar theories, one has to resort to approximations such as the derivative expansion or the vertex expansion and this is no different in gravity. As already discussed, while the quantum dynamics of asymptotically safe gravity is generated and carried by the fluctuation correlation functions, it is the diffeomorphism-invariant background-effective action $\Gamma[g]$ that allows for a more direct physics interpretation. The latter is extracted from the flow \eq{eq:backflow} that solely depends on the fluctuation two-point function $\Gamma^{(0,2)}[g,0]$. The flow of the latter depends on higher-order fluctuation correlation functions, see \autoref{sec:Hierarchy}. 

This suggests the expansion of the effective action $\Gamma_k[\bar g,h]$ in a vertex expansion of the fluctuation field $h$. Importantly, the vertex expansion in the fluctuation approach is a systematic approximation scheme, the strength and convergence of which has been shown in many non-perturbative approaches, and most notably in the fRG-approach to QCD, \cite{Mitter:2014wpa, Cyrol:2016tym, Cyrol:2016zqb, Cyrol:2017ewj, Corell:2018yil}. In the spirit of 'toy' theories that can teach us something about technical properties and convergence, we consider non-Abelian gauge theories as one of those standard quantum field theories that are as close as it gets to gravity. The vertex expansions fully disentangles the contributions from the background metric $\bar g$ and the fluctuation field $h$, and reads for the effective action, 
\begin{align}
 \label{eq:vertex-expansion}
 \Gamma_k [\bar g,\phi ] =
 \sum_{n=0}^{\infty} \frac{1}{n !}
 \int\!\Gamma_k^{(0,\phi_{a_1} \ldots \phi_{a_n})} [\bar g,0] \cdot \phi_{a_1} \dots \phi_{a_n} \,.
\end{align}
Evidently, if the expansion coefficients $\Gamma_k^{(0,\phi_{a_1} \ldots \phi_{a_n})}$ are evaluated for general $\bar g$, we have a simple access to the full effective action. For example, if we choose $\bar g=\bar g^\text{fluc}_\text{EoM}$, the solution of the fluctuation EoM in \eq{eq:FlucEoM-neq-backEoM}, we have chosen an on-shell expansion point. Accordingly, if we are interested in on-shell physics only small fluctuations $h$ should be relevant. In turn, if we choose another expansion point, e.g., for technical reasons, it is very important to assess whether on-shell physics is in the radius of convergence of the expansion. This will be discussed in more detail in \autoref{sec:Expbarg}.

\subsection{Hierarchy of flow equations}
\label{sec:Hierarchy}
The background-field approach leads to an extended hierarchy of flow equations. We first note that the background flow equation $\partial_t \Gamma_k[\bar g]$, \eq{eq:backflow}, depends on the fluctuation two-point function $\Gamma_k^{(0,2)}[\bar g, 0]$ in a general background. The knowledge of the latter allows us to determine $\Gamma[\bar g]$ and is tantamount to the determination of the full propagator of the theory in a general background. However, the flow of the two-point function $\partial_t \Gamma_k^{(0,2)}[\bar g, 0]$ depends on $\Gamma_k^{(0,m)}[\bar g, 0]$ with $m=2,3,4$. This continues for higher $n$-point functions and leads to an infinite tower of coupled equations, 
\begin{align}
 \label{eq:funfluc}
 \Gamma_k^{(0,m)}[\bar g,0]= \text{fRG}_{0,m}[\bar g, \{\Gamma_k^{(0\,,\,2\leq j\leq m+2)} [\bar g,0]\}]\,.
\end{align}
In other words, we need $\Gamma_k^{(0,2)}[\bar g,\phi]$ for general fluctuation fields for solving the flow equation of the background-effective action. For most interacting quantum field theories, the task of resolving the full field-dependence of the effective action is beyond reach. Already in scalar theories, one typically resorts to the computation of the full effective potential as well as additional vertices or momentum dependencies. In gravity, the full potential of the background curvature $R$ has been investigated: $f(R)$ as well as potentials of tensor invariants \cite{Falls:2013bv, Falls:2016wsa, Codello:2007bd, Machado:2007ea, Codello:2008vh, Benedetti:2012dx, Dietz:2012ic, Benedetti:2013jk, Dietz:2013sba, Demmel:2014sga, Falls:2014tra, Demmel:2015oqa, Ohta:2015efa, Ohta:2015fcu, Falls:2016msz, Falls:2017lst, Gonzalez-Martin:2017gza, Christiansen:2017bsy, deBrito:2018jxt, Alkofer:2018baq, Falls:2018ylp, Burger:2019upn}. Apart from this, as in other theories, explicitly or implicitly a vertex expansion has been used. This entails a further expansion of \eq{eq:funfluc} in powers of the background field and leads us to the hierarchy 
\begin{align}
 \label{eq:fun}
 \Gamma_k^{(n,m)}[\bar g,0]= \text{fRG}_{n,m}[\bar g, \{\Gamma_k^{(i\leq n\,,\,2\leq j\leq m+2)} [\bar g,0]\}]\,.
\end{align}
\Eq{eq:fun} is the full hierarchy of \textit{integrated} flow equations to solve for quantum gravity. While its solution in terms of the vertex expansion has been baptised the \textit{fluctuation approach}, it simply is the full problem at hand. 

Apparently, \eq{eq:fun} constitutes a system of equations for a two-field effective action. However, as discussed in \autoref{sec:SymId}, background-independence at vanishing cutoff, $k=0$, encoded in the Nielsen identities and carried over to the mNIs at finite cutoff scale $k$, turn the effective action into a one-field effective action. In terms of the vertex expansion this information is given by the mNI \eq{eq:mNI} for $(n,m)$-point functions,
\begin{align} 
 \label{eq:mNI-rel}
 \Gamma_k^{(n,m)}[\bar g,h]&= \Gamma_k^{(n-1,m+1)}[\bar g,h] \\ 
 &\quad + \text{mNI}_{n,m}[\bar g, \{\Gamma^{(i\leq n-1,j\leq m+1)}[\bar g,h]\}]\,.\notag
\end{align}
This leaves us with two towers of functional relations. While the first one, \eq{eq:fun}, describes the full set of correlation functions, the second one, \eq{eq:mNI-rel}, can be used to iteratively solve the tower of mixed fluctuation-background correlations on the basis of the fluctuating correlation functions $\{\Gamma^{(0,m)}\}$. In both cases, we can solve the system for the higher-order correlations of the background on the basis of the lower-order correlations. If we use \eq{eq:mNI-rel} with an iteration starting with the results from the flow equation for $\{\Gamma^{(0,m)}[\bar g_{\text{sp}}, h]\}$ for a specific background $\bar g_{\text{sp}}$, this closure of the system automatically satisfies the NI. Accordingly, {\it any} set of fluctuation correlation functions $\{\Gamma^{(0,m)}[\bar g_{\text{sp}}, h]\}$ can be iteratively extended to a full set of fluctuation-background correlation functions in an iterative procedure.

\begin{figure}[t]
\includegraphics[width=\linewidth]{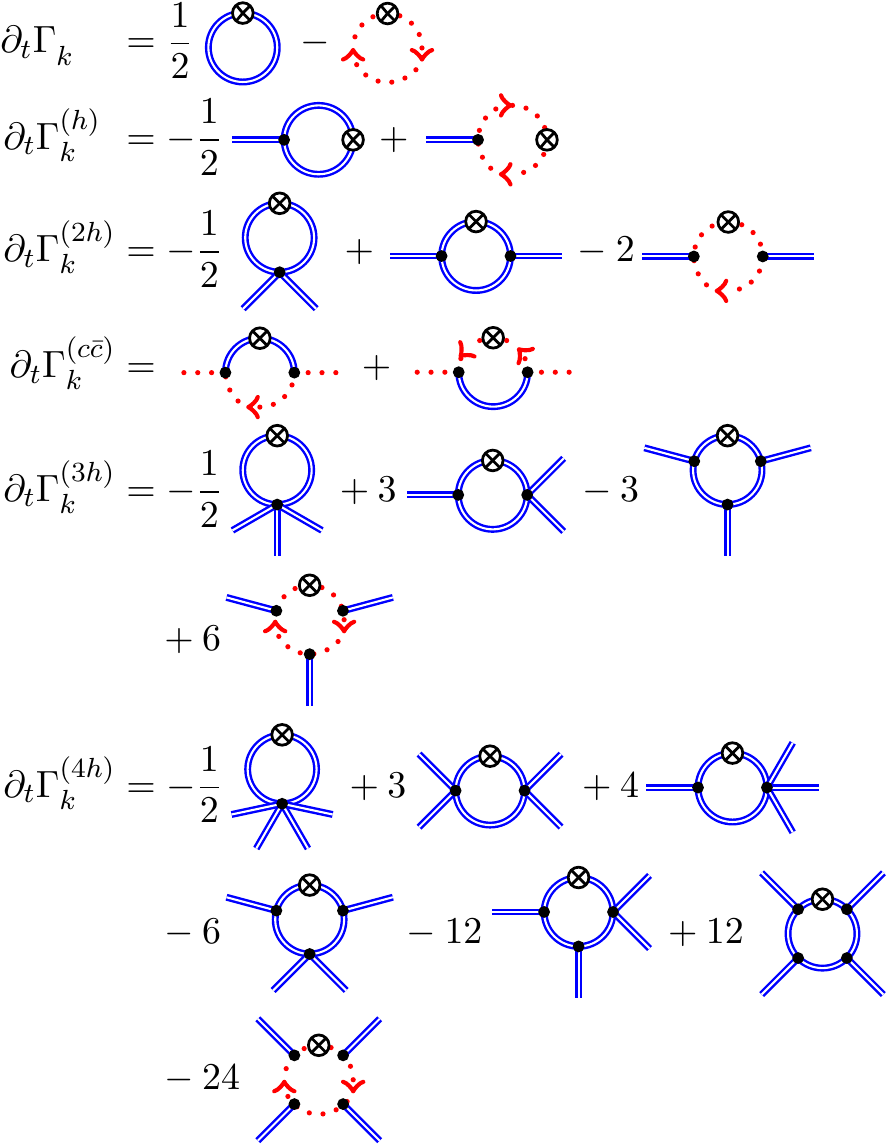}
\caption{
 Diagrammatic representation of the flow equations of the fluctuation $n$-point functions up to $n=4$. Graviton propagators are depicted with a blue double line and ghosts with a red dotted line. The crossed circle represents a regulator insertion. The flows can be augmented straightforwardly with contributions from matter fields. The figure is taken from \cite{Denz:2016qks}.
}
\label{fig:diagrams}
\end{figure}

An important feature of the fRG equations is that in the Landau limit of the gauge parameter $\alpha\to0$ in \eq{eq:gf}, the flow equations for the transverse vertices $\Gamma^{(0,n)}_{k,\bot}$ are closed: the external legs of the vertices in the flow are transverse due to the transverse projection of the flow, the internal legs are transverse as they are contracted with the transverse propagator. Schematically this reads for the integrated flows \eq{eq:fun},
\begin{align}\label{eq:4pt:flowbot}
 \Gamma^{(0,n)}_{k,\bot}=\text{fRG}^{(0,n)}_\bot[\{\Gamma^{(0,m)}_{k,\bot}\}]\,. 
\end{align}
In other words, the flow-equation system of transverse fluctuation correlation functions is closed and determines the dynamics of the system. In the fluctuation approach, the transverse system of graviton correlation function has been solved up to the four-graviton vertex \cite{Denz:2016qks}. A diagrammatic depiction of the system of flow equations is given in \autoref{fig:diagrams} and a description of the respective results can be found in \autoref{sec:StateArt}. 

In turn, the flow equation system for longitudinal fluctuation correlation functions is not closed, and the transverse correlation functions $\Gamma^{(0,n)}_{k,\bot}$ feed into it, 
\begin{align}
 \label{eq:4pt:flowlong}
 \Gamma^{(0,n)}_{k,L}=\text{fRG}^{(0,n)}_L[\{\Gamma^{(0,m)}_{k,\bot}\}\,, \,\{\Gamma^{(0,m)}_{k,L}\}]\,. 
\end{align}
Note that $\{\Gamma^{(0,n)}_{k,L}\}$ is the complement of the set of purely transverse correlation functions, so it consists of correlation functions with at least one longitudinal leg. 

On the other hand, the mSTIs are also non-trivial relations for the longitudinal correlation functions in terms of transverse vertices and longitudinal ones. This leads us to the schematic relation, 
\begin{align}
 \label{eq:4pt:mSTI} 
 \Gamma^{(0,n)}_{k,L}=\text{mSTI}^{(0,n)}[\{\Gamma^{(0,m)}_{k,\bot}\}, \{\Gamma_{k,L}^{(0,m)}\}]\,, 
\end{align}
see \cite{Fischer:2008uz} for non-Abelian gauge theories. In consequence, the mSTIs provide no direct information about the transverse correlation functions without further constraint. In the perturbative regime at large momenta, this additional constraint is given by the uniformity of the vertices. In turn, in strongly-correlated regimes such a general constraint is absent. Indeed, one can show that the confinement property in a Yang-Mills theory in a covariant gauge necessitates the absence of uniformity of the vertices at low momenta, for a detailed discussion in non-Abelian gauge theories see \cite{Cyrol:2016tym}.  

Instead, we can simply use \eq{eq:4pt:mSTI} for a given set of transverse correlation functions for constructing a BRST-invariant solution, which signals diffeomorphism invariance. For a given finite set of transverse correlation functions generically such a solution can be found by integrating the flow \eq{eq:FlowSym}. However, it may be \textit{non-local}. The existence of BRST-invariant solutions for a general transverse input emphasises the fact that the derivation of diffeomorphism-consistent solutions is not necessarily the hallmark of a good truncation. However, the comparison of \eq{eq:4pt:mSTI} and \eq{eq:4pt:flowlong} is a further non-trivial constraint on longitudinal correlation functions. Its evaluation is complicated by the fact, that the solutions of two different functional relations for the same set of correlation functions do not agree in general in non-trivial truncations. Furthermore, it is very difficult to provide a measure for the closeness of the solutions. For a related discussion in non-Abelian gauge theories see the recent review \cite{Dupuis:2020fhh} and references therein. 

In summary, the evaluation of diffeomorphism invariance and selfconsistency constitutes an intricate challenge. One has to utilise all the properties and relations discussed above. This holds for all fRG-approaches to quantum gravity and not only to the fluctuation approach: only \textit{local} BRST-invariant solutions should be considered physical, and the evaluation of locality and BRST-invariance or their absence is intricate. 

\subsection{The flat expansion is a curvature expansion}
\label{sec:Expbarg} 	
As briefly mentioned in the introduction of this section, the choice of the background metric is important for the convergence of the vertex expansion. However, an evaluation of the flow equations for the $\Gamma_k^{(0,n)}[\bar g,0]$ for generic metrics is a yet unresolved technical challenge. Even flows for spherically-symmetric backgrounds already pose a formidable technical challenge that has only been solved recently within further approximations that hold for small curvatures \cite{Christiansen:2017bsy, Burger:2019upn}. Therefore in most applications, one resorts to a curvature expansion in powers of the curvature. Such an expansion is tantamount to an expansion about the flat background with vanishing curvature, 
\begin{align}
 \label{eq:flat} 
 \bar g = \mathbb{1}\,, 
\end{align}
the Euclidean analogue of the Minkowski metric. This has been baptised the \textit{flat expansion}. With the flat background \eq{eq:flat}, Fourier transformations can be performed and we are led to correlation functions $\Gamma_k^{(n,m)}( p_1,..,p_n,p_{n+1}...,p_{n+m})$ in momentum space. This gives access to the powerful techniques of standard quantum field theory that allow solving the flow equations for general vertex functions in momentum space. 

This expansion encompasses the standard curvature expansion with the additional benefit that generic covariant momentum dependences are systematically accessible, for a respective brief discussion see \cite{Denz:2016qks}. To understand this statement, we sketch the curvature expansion of the background-field approximation with standard heat-kernel techniques and the flat expansion in momentum space. We shall see that both lead to the same flow equations for the expansion coefficients of diffeomorphism-invariant operators. We expand the full one-field effective action in local curvature invariants and covariant derivatives
\begin{align}
 \label{eq:EHApprox}
 \Gamma_{k}[g_{\mu\nu}]&= \frac{1}{16 \pi G_k} \int\! \mathrm{d}^4 x  
 \sqrt{g} \left(2 \Lambda_k - R\right)
 +O(R^2_{\mu\nu\rho\sigma}, \nabla^2)\,. 
\end{align}
In \eq{eq:EHApprox}, the first term on the right-hand side is the Einstein-Hilbert action with a scale-dependent cosmological constant and Newton coupling. The second term includes all other curvature invariants starting with $R^2,\dots$. Covariant-derivative terms, schematically given by $\int\!\sqrt{g} R\, \nabla^2 R$ and terms with higher-orders in covariant derivatives $\nabla$, kick in at the next order and beyond. Note that the scale-dependent Einstein-Hilbert action without higher-order terms is still a common approximation for the pure gravity sector in particular in many applications to gravity-matter systems. For gravity-matter systems beyond the Einstein-Hilbert truncation see e.g.~\cite{Hamada:2017rvn,Eichhorn:2017eht,Alkofer:2018fxj,deBrito:2019umw,deBrito:2019epw}.

Similarly to \eq{eq:EHApprox}, the flow of the background-effective action can also be expanded in terms of \textit{local} curvature invariants and covariant derivatives. This leads us to 
\begin{align} 
 \frac12 \text{Tr} \,G_k\partial_t R_k\, [g_{\mu\nu}] &=
 \frac{1}{16 \pi} \int\! \mathrm{d}^4 x \sqrt{g} \left( 2 a_{1,k}- a_{R,k} R \right) \notag \\[1ex]
 &\hspace{1.5cm} +O(R^2_{\mu\nu\rho\sigma}, \nabla^2)\,, 
 \label{eq:FlowExpand}
\end{align}
with expansion coefficients $a_{\CO,k}$ of a given operator $\CO$. In particular, we have $a_{1,k}=a_{R^0,k}$. By comparing \eq{eq:EHApprox} and \eq{eq:FlowExpand}, we arrive at the flow equations 
\begin{align}
 \label{eq:EHflow} 
 \partial_t \frac 1{G_k} &= a_{R,k}\,,
 &
 \partial_t \frac{\Lambda_k}{G_k} &= a_{1,k}\,. 
\end{align}
Evidently, any \textit{complete} projection procedure produces the \textit{complete} set of flow equations of all expansion coefficients $a_{\CO,k}$. We emphasise that if the operator basis is overlapping, the flow of the effective action is unique, while the flow of the set of $a_{\CO,k}$ is not. 

The standard procedure for projecting onto the flow of the cosmological constant and the Newton coupling, as well as that of higher-order invariants, is by heat-kernel techniques or explicit summation over the spectrum of the covariant Laplacians, in conjunction with the Euler-Maclaurin formula, see the reviews \cite{Niedermaier:2006wt, Litim:2011cp, Reuter:2012id, Ashtekar:2014kba, Eichhorn:2017egq, Bonanno:2017pkg, Eichhorn:2018yfc, Pereira:2019dbn, Reichert:2020mja, Bonanno:2020bil}. As no other \textit{local} diffeomorphism-invariant operators are present at this order, the flow of $G_k$ and $\Lambda_k$ depends only on the given approximation of $\Gamma_k^{(0,2)}$ on the right-hand side of the flow. As already indicated, at higher orders of the curvature expansion more and more invariants are present and the projections on one single invariant only give unambiguous results if a complete basis of invariants is chosen. In fermionic systems, this is the well-known Fierz ambiguity, see the review \cite{Braun:2011pp} for an extended discussion. Consequently, at higher orders of the expansion one typically has to deal with two truncation artefacts: first we always have to deal with the truncation of $\Gamma_k^{(0,2)}$ and second we have to deal with incomplete bases. We note that only very recently the second order has been mapped out, see \cite{Falls:2020qhj}. This emphasises that we have to deal with an intricate technical challenge. 

Now we derive the flows in \eq{eq:EHflow} within the flat expansion scheme. To that end we note that the only \textit{local} diffeomorphism-invariant term with no derivatives is the volume term $\CV=\int_x\!\sqrt{g}$. Moreover, the only \textit{local} diffeomorphism-invariant term with two derivatives is the curvature scalar term. This implies that we have a unique projection at the flat expansion point \eq{eq:flat}, schematically written as 
\begin{align} 
 - \frac{8 \pi}{\CV} \partial_t \Gamma_k[g]\bigg|_{g\to\mathbbm{1}}&= \partial_t \frac{\Lambda_k }{ G_k}= a_{1,k}\,, \notag \\[1ex]
 16 \pi \partial_{p^2}\partial_t \Gamma_{k,\text{TT}}^{(2)}[g]\bigg|_{g\to\mathbbm{1}} &= \partial_t \frac{1}{G_k}=a_{R,k}\,,  
 \label{eq:EHflat}
\end{align}
where the subscript ${}_\text{TT}$ refers to the projection and normalisation on the traceless-transverse part. More details can be found e.g.\ in \cite{Denz:2016qks}. \Eq{eq:EHflat} simply is \eq{eq:EHflow}, as the flat expansion scheme is a consistent projection scheme. 

This procedure can be extended beyond the set of \textit{local} diffeomorphism-invariant operators: 
\begin{itemize} 	
 \item[(i)] Take general derivatives w.r.t.\ $h_{\mu\nu}(p)$. 
 \item[(ii)] Contract with all possible Lorentz tensor-structures. 
 \item[(iii)] Take derivatives with respect to momenta.
\end{itemize}
In particular, apart from all \textit{local} diffeomorphism-invariant term, the expansion captures general covariant momentum-dependences including potential IR-singular terms as well as topological terms. A diffeomorphism-invariant example for the former is the Polyakov action in two dimensions, 
\begin{align}
 \label{eq:IRSingTerms}
 -\frac{1}{96\pi} \int \!\mathrm d^2 x\sqrt{g}\,R \frac{1}{\Delta} R\,,
\end{align}
see e.g.~\cite{Codello:2010mj, Coriano:2017mux}. IR-singular terms are naturally covered by taking into account full momentum-dependences of full vertices or momentum channels of specific tensor-structures. This has been used extensively in gauge theories such as QCD, not only within the fRG-approach but also in other functional approaches based on Dyson-Schwinger equations or $n$-particle irreducible hierarchies. 

A relevant example for a topological term in gravity is the Gau\ss-Bonnet term with the density 
\begin{align}
 E[g]= \frac{1}{32\pi^2}\!\left( R^2-4 R^{\mu\nu}R_{\mu\nu}+R^{\mu\nu\rho\sigma}R_{\mu\nu\rho\sigma}\right) .
\label{eq:GaussBonnet}
\end{align}
Metric variations of the local density $E[g]$ are non-vanishing to all order of metric derivatives. In turn, its space-time integral $\chi[g]=\int\! \mathrm d^4 x\sqrt{g} \, E[g]$ is the Euler characteristic of the manifold $M$ with $\chi[g]\in \mathbbm{Z}$. Consequently, smooth metric variations of $\chi[g]$ (no change of the geometry) are vanishing. Note, however, that functional derivatives are distributional and do not fall into the class of smooth variations. Moreover, only the combination of the different curvature-squared invariants in \eq{eq:GaussBonnet} add up to the Euler characteristic $\chi[g]$. The single terms have a generic metric-dependence and with appropriate projections, we can capture their running coefficients. This is the manifestation of a more generic feature, which is already used in the extraction of anomalies in perturbation theory and anomalous as well as topological terms beyond perturbation theory, see e.g.~\cite{Pawlowski:1996ch, Reuter:1996be}. 

Below we outline a cautious approach guided by the works \cite{Pawlowski:1996ch, Reuter:1996be} in gauge theories. There, a simply example for a topological invariant is the Pontryagin index in a $U(1)$-theory with the density $1/(32 \pi^2) F_{\mu\nu} \tilde F^{\mu\nu}$ where $\tilde F^{\mu\nu}$ is the dual field strength. This density is quadratic in the field and is discussed in App.~\ref{app:Pontryagin}. Analogously to this example we introduce the Gau\ss-Bonnet term with a local auxiliary field $\theta(x)$, 
\begin{align}
 \chi[g,\theta]= \int \!\mathrm d^4 x\sqrt{g} \, \theta(x) E[g]\,,
 \quad \quad 
 \chi[g,1]\in \mathbb{Z}\,.
\label{eq:EulerChar}
\end{align}
The auxiliary field $\theta(x)= \theta_\text{top}+\Delta \theta(x)$ can be seen as the local coupling of the Gau\ss-Bonnet density. Its constant part $\theta_\text{top}$ with $\nabla \theta_\text{top}=0$ is the topological coupling, while its space-time dependent part $\Delta\theta(x)$ is part of the couplings of the local diffeomorphism-invariants quadratic in the curvature. Applying two derivatives with respect to the metric field in momentum space leads us to
\begin{align}
 \frac{\delta^2 \chi[g,\theta]}{\delta g_{\mu\nu}(p)\delta g_{\rho\sigma}(q)}\bigg|_{g=\delta}
 &= \frac{1}{16\pi^2} \CT^{\mu\nu\rho\sigma}(p,q) \theta(l) \delta(l + p + q) \,.
 \label{eq:derGB}
\end{align}
The tensor-structure $\CT$ is given by
\begin{align}
\CT^{\mu\nu\rho\sigma}
 &= \Pi_0 ( \delta^{\mu\nu} \delta^{\rho\sigma} - \delta^{\mu(\sigma} \delta^{\rho)\nu}) + \Pi_2^{\mu\nu\rho\sigma} \\
&\quad + \delta^{\mu\nu} \Pi_1^{\rho\sigma} + \delta^{\rho\sigma} \Pi_1^{\mu\nu} - \delta^{\mu(\rho} \Pi_1^{\sigma)\nu} - \delta^{\nu(\rho} \Pi_1^{\sigma)\mu} \,,\notag
\end{align}
where we have defined
\begin{align}
\Pi_0 &= p^2 q^2 -(p\cdot q)^2 \,, \notag \\
\Pi_1^{\alpha\beta} &= 2 (p\cdot q) p^{(\alpha} q^{\beta)} - p^\alpha p^\beta q^2 - q^\alpha q^\beta p^2 \,, \notag \\
\Pi_2^{\alpha\beta\gamma\delta} &= p^{\alpha} p^{\beta} q^{\gamma} q^{\delta}+p^{\gamma} p^{\delta} q^{\alpha} q^{\beta} \notag \\
&\quad - p^{\alpha} p^{(\gamma} q^{\delta)} q^{\beta}- p^{\beta} p^{(\gamma} q^{\delta)} q^{\alpha} \,.
\end{align}
The (local) total-derivative property of the Gau\ss-Bonnet density is reflected in the fact that all $\Pi_i$ are vanishing for $l=0$ when momentum conservation implies $p=-q$. Accordingly, with $\theta(x) =\theta_\text{top}$ and $\theta(l)= \theta_\text{top} (2 \pi)^4 \delta(l)$ the right-hand side of \eq{eq:derGB} vanishes. However, by collecting the $\theta$-terms on the left-hand and right-hand side of the flow one can simply project the flow on the running of the coefficient of the topological term. We emphasise that the vanishing of the flow for constant $\theta$ is analogous to the vanishing of the flow $p^2\partial_t Z_\phi$ for $p^2=0$. In conclusion, the present expansion scheme is well-capable and well-suited for describing IR-divergent as well as topological terms. 

In summary, the flat expansion allows projecting the flow equation on the flow of all coefficients $a_{\CO,k}$ for diffeomorphism-invariant operators of the form 
\begin{align}
 \label{eq:GenOps} 
 \CO&= \int_{x}\! \sqrt{g}\,f_{\mu_1\cdots\mu_{4n}}(\nabla_{\!1},\ldots,\nabla_{\!n}) \prod_{i=1}^n R_{\mu_{i_1}\cdots \mu_{i_4}}\,.
\end{align}
Here, $\nabla_{\!i}$ acts only on the $i$th Riemann tensor. In the case of the fluctuation correlation functions $\Gamma_k^{(0,n)}[\bar g]$, no expansion in curvature invariants is possible but an expansion in covariant tensor-structures is possible, though being even more intricate. In case of the flat expansion, this is done with considering all tensor-structures of $\Gamma_k^{(0,n)}(p_{1},\ldots,p_{n})$. How this can be done has been worked out in QCD, see e.g.~\cite{Mitter:2014wpa, Cyrol:2016tym, Cyrol:2016zqb, Cyrol:2017ewj, Corell:2018yil}, respective computational tools are provided e.g.\ by \cite{Huber:2011qr, Cyrol:2016zqb, Huber:2019dkb} or are in preparation.

The findings of the present section can be summarised as follows: 
\begin{itemize} 
 \item[(i)] The flat expansion encompassed the curvature expansion. There is no conceptual difference and both expansions are expansions about the flat background $\bar g=\mathbb{1}$. 
 \item[(ii)] The expansion point of the curvature or flat expansion is not the solution of the EoM, $\bar g_\text{EoM}^\text{fluc}$, and checks of the convergence of the expansion are in high demand. 
 \item[(iii)] The fluctuation approach within the flat vertex expansion resolves the difference between fluctuation and background field. As such it simply improves upon the background-field approximation within the curvature expansion without introducing other approximations: fluctuation approach results benchmark that in the background-field approximation, and provide non-trivial reliability checks for the latter.
\end{itemize}
There is an increasing number of computations that do not rely on the curvature expansion, for example, \cite{Machado:2007ea, Codello:2008vh, Benedetti:2012dx, Dietz:2012ic, Benedetti:2013jk, Dietz:2013sba, Demmel:2014sga, Demmel:2015oqa, Ohta:2015efa, Ohta:2015fcu, Falls:2016msz, Gonzalez-Martin:2017gza, Falls:2017lst, Falls:2018ylp} in the background-field approximation and \cite{Christiansen:2017bsy, Burger:2019upn} in the fluctuation approach. This concludes our discussion of the formal properties of the fluctuation approach.

\subsection{Tensor structure and momentum dependence of vertices}
\label{sec:Approx} 
In the flat expansion, the vertices $\Gamma^{(n)} = \Gamma^{(0,n)}$ are typically rescaled with the wave-function renormalisations $Z_{\phi_a}$ to obtain the RG-invariant vertices $\bar\Gamma^{(n)} = \bar\Gamma^{(0,n)}$
\begin{align}
 \label{eq:vertex-bar}
 \Gamma_k^{(\phi_{a_1} \ldots \phi_{a_n})}(\mathbf{p}) =
 \left(\prod_{i=1}^n Z^{\frac12}_{\phi_{a_i}}(p_i^2)\right)
 \bar \Gamma_k^{(\phi_{a_1} \ldots \phi_{a_n})}(\mathbf{p})\,,
\end{align}
where, $\mathbf{p}=(p_1,\ldots,p_{n})$. The wave-function renormalisations can be fully absorbed by a redefinition of the fields $\bar \phi_a =\sqrt{Z_{\phi_a}}\, \phi_a$. The wave-function renormalisation enter the flow equations only via the anomalous dimensions $\eta_{\phi_a}$ defined by
\begin{align}
 \eta_{\phi_a} (p^2) = - \partial_t \ln Z_{\phi_a} (p^2) \,,
\end{align}
which describe the running of the rescaled fields $\partial_t \bar \phi_a \propto \eta_{\phi_a} \bar \phi_a$. The RG-invariant vertices $\bar \Gamma_k^{(n)}$ are then parameterised with a complete set of tensor-structures $\CT_{j}$ and respective RG-invariant dressings $\CA_{k,j}$ 
\begin{align}
\label{eq:vertex-param}
 \bar \Gamma_k^{(\phi_{a_1} \ldots \phi_{a_n})}(\mathbf{p}) = 
 \CA_{k,j}^{(\phi_{a_1} \ldots \phi_{a_n})}(\mathbf{p}) 
 \CT_j^{(\phi_{a_1} \ldots \phi_{a_n})}(\mathbf{p};\text{couplings}) \,,
\end{align}
where the sum over $j$ is implied. The size of the complete set of tensor-structures increases rapidly for higher-order vertices. The cutoff-dependent dressings $\CA_{k,j}$ capture the overall coupling strength of the respective tensor-structure and its momentum dependence. 

In most applications to gravity, only the Einstein-Hilbert tensor-structures deduced from the curvature scalar and the volume term are taken into account. This leads us to 
\begin{align}
 \label{eq:tensor-structure}
 \CA^{(n)}_k(\mathbf{p}) &= G_n^{\frac n2 -1}(\bar p^2) \,, \notag \\
 \CT^{(\phi_{a_1} \ldots \phi_{a_n})} &= G_\text{N}\,S_\text{EH}^{(\phi_{a_1} \ldots \phi_{a_n})}(\mathbf{p};\Lambda_n)\,, 
\end{align}
with the Einstein-Hilbert action \eq{eq:EH-Action} and the momentum-dependent global dressing $\CA^{(n)}_k$ of the Einstein-Hilbert tensor-structure. The prefactor $G_\text{N}$ in the definition of the tensor-structure leaves the latter independent of $G_\text{N}$. The couplings $G_n$ and $\Lambda_n$ resemble the Newton coupling and the cosmological constant, respectively, for each $n$-point function. They are called \textit{avatars} of the respective coupling. In \eq{eq:tensor-structure}, we have already simplified the momentum dependence of the couplings $G_n$: they only depend on the average momentum $\bar p^2 =(p_1^2+\cdots +p_n^2)/n$. The couplings $G_n$ are extracted from the flow of the $n$-point functions at a momentum-symmetric point. This definition mimics the definition of momentum-dependent couplings in gauge theories. The dimensionless counterparts of $G_n$ and $\Lambda_n$ are denoted by $g_n= G_n\,k^2$ and $\lambda_n=\Lambda_n/k^2$.

For $n=0,1$ we have $\Gamma^{(0,n)}=0$ for a flat background. For $n=2$ we have $G_n^{n/2-1} = G_2^0=1$ and hence there is no Newton coupling $G_2$ for the two-point function. Instead, $\Gamma_k^{(0,2)}$ depends on the graviton mass parameter $\mu=-2 \lambda_2$ and the dimensionless wave-function renormalisation $Z_h(p)$ of the fluctuation graviton. We emphasise that the graviton mass parameter $\mu$ should not be understood as a physical mass. Moreover, the graviton is not directly related to an asymptotic state, for a recent discussion see \cite{Maas:2019eux} and also the review \cite{Bonanno:2020bil}. All dimensionless couplings are shown in \autoref{fig:IR-trajectory} as a function of the cutoff scale on one exemplary UV-IR trajectory.

\begin{figure}[t]
\includegraphics[width=\linewidth]{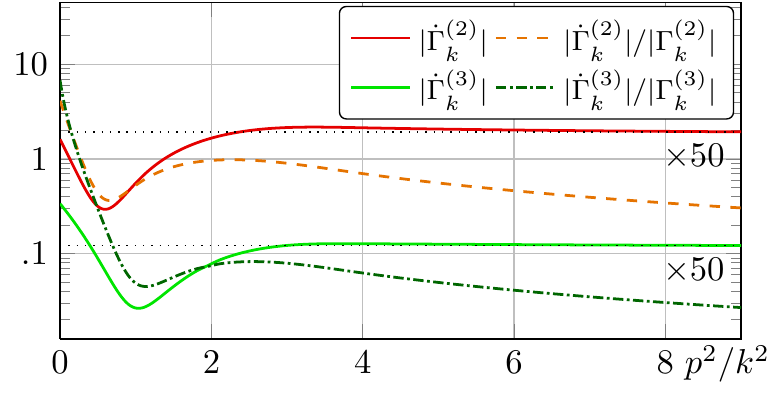}
\caption{
 Displayed are the flows of the traceless transverse parts of the graviton two- and three-point functions, $|\partial_t \Gamma_k^{(2)}|$ and $|\partial_t \Gamma_k^{(3)}|$, as a function of dimensionless momentum. The flows approach constants for large momenta and they do not grow with $p^2$ as expected from a na\"ive counting of momenta. The flows are normalised by the respective $n$-point functions, $|\partial_t \Gamma_k^{(2)}|/|\Gamma_k^{(2)}|$ and $|\partial_t \Gamma_k^{(3)}|/|\Gamma_k^{(3)}|$. These ratios tend towards zero for large momenta which signals momentum locality. The figure is taken from \cite{Christiansen:2015rva}.
}
\label{fig:locality}
\end{figure}

\subsection{Momentum locality}
\label{sec:MomentumLocality}
An important property of a physical coarse-graining procedure is momentum-locality: it ensures that a coarse-graining step at a given cutoff scale $k$ does not influence the physics at \textit{momentum} scales $p \gg k$. In \cite{Christiansen:2015rva}, it was defined by
\begin{align}
\label{eq:mom-locality} 
 \lim_{\frac{p_i^2}{k^2} \to \infty } \frac{|\partial_t \Gamma_k^{(n,m)}(\mathbf{p})|}{|\Gamma_k^{(n,m)}(\mathbf{p})|}&=0\,, 
 &
 \text{with} \quad \mathbf{p}&=(p_1,...,p_{n+m})\,, 
\end{align}
In this definition, all momenta $p_i$ of the correlation function $\Gamma_k^{(n,m)}(\mathbf{p})$ need to be sent to infinity such that there are no trivial cancellations for the momenta of internal propagators. This is for example achieved with a symmetric momentum configuration. The norm of the $n$-point function refers to a normalised tensor projection. 

The condition \eq{eq:mom-locality} is satisfied by all perturbatively renormalisable local quantum field theories of scalars, fermions and vector fields (including gauge fields in linear gauges with linear momentum dependences) by trivial counting of the momenta. In turn, non-renormalisable theories with non-trivial momentum dependences of vertices are easily non-local. For example the scalar field theory with an interaction term of the type $\int_x\!\phi^2 (\partial \phi)^2$ does not fulfil \eq{eq:mom-locality}. Note that this theory has the power-counting of Einstein-Hilbert gravity. 

Thus, a na\"ive momentum counting in gravity leads to the conclusion, that the coarse-graining is not momentum local, neither in Einstein-Hilbert gravity nor in a higher-derivative theory of gravity. One needs non-trivial cancellations between diagrams. In \cite{Christiansen:2014raa} such a cancellation was observed for the first time in the transverse-traceless part of the graviton two-point function with Einstein-Hilbert vertices. In \cite{Christiansen:2015rva} this was extended to the transverse-traceless part of the graviton three-point function. Both cases are displayed in \autoref{fig:locality}. There are three diagrams (plus one ghost diagram) contributing to the flow of the graviton three-point function, see \autoref{fig:diagrams}. The cancellation takes places between all diagrams and holds for all gauge-fixing parameters and all momentum configurations of the three-point function, as long as all external and internal momenta are sent to infinity. 

We close this section with the remark, that the results in \cite{Christiansen:2015rva}, while highly non-trivial, should be considered to be the first step in a fully conclusive analysis. Most notably, the observed locality does not hold for all tensor-structures of the $n$-point functions considered there. In our opinion, this may hint at persistent non-localities introduced by the gauge fixing. If this can be solidified in further investigations, this should allow for selecting gauge fixings that make the coarse-graining procedure for a given regularisation procedure momentum local. Note, that while momentum-locality of a coarse-graining procedure is not a necessary property it certainly improves the convergence of standard approximation schemes which are typically momentum-local. Moreover, if no momentum-local coarse-graining procedure can be found for a given theory, this casts serious doubts on the interpretation of such a theory as a local quantum field theory.

\section{State of the art}
\label{sec:StateArt}
We are now ready to review the state of the art of asymptotically safe quantum gravity within the fluctuation approach. To facilitate accessing the relevance of the different results for the self-consistency of the approach, we start with a brief overview: \smallskip

\textit{UV fixed point} (\autoref{sec:UV-FP}):
The existence of a UV fixed point with a finite-dimensional critical hypersurface ensures the UV finiteness and predictivity of the theory. With the fluctuation approach, this has been investigated for pure gravity in \cite{Christiansen:2012rx, Donkin:2012ud, Christiansen:2014raa, Christiansen:2015rva, Denz:2016qks, Christiansen:2016sjn, Meibohm:2015twa, Christiansen:2017cxa, Christiansen:2017bsy, Eichhorn:2018akn, Eichhorn:2018ydy, Eichhorn:2018nda, Burger:2019upn, Knorr:2017mhu, Knorr:2017fus}. The UV fixed point is comparable with results in the background-field approximation and thus consolidates these results. Three UV attractive directions are found associated with $\sqrt{g}$, $\sqrt{g}R$, and $\sqrt{g}R^2$. First signs for apparent convergence within the vertex expansion were found \cite{Denz:2016qks}.\smallskip

\textit{UV-IR trajectory} (\autoref{sec:UV-IR}): 
A UV-IR trajectory allows us to connect to a classical GR regime and IR-SM physics if matter couplings are included. Classical GR regimes are accessed for $\mu\to 0$ (Gau\ss ian fixed point), $\mu\to\infty$, and $\mu \to -1$, where $\mu=-2 \lambda_2$ as introduced below \eq{eq:tensor-structure}. The case $\mu\to-1$ was investigated in \cite{Christiansen:2012rx, Donkin:2012ud, Christiansen:2014raa, Denz:2016qks}. In the classical regime the modified STIs and modified NIs reduce to standard STIs and NIs, which can be solved for small $k$.\smallskip

\textit{Momentum dependence \& unitarity} (\autoref{sec:mom-dep}):
The full momentum-dependence, in particular of the propagator, opens a path towards a first investigation of unitarity via spectral reconstructions. The truncations already include the momentum dependence of the graviton two-, three,- and four-point functions at the momentum symmetric point \cite{Christiansen:2014raa, Christiansen:2015rva, Denz:2016qks} as well as the momentum dependence of the graviton-matter three-point vertices \cite{Christiansen:2017cxa, Eichhorn:2018akn, Eichhorn:2018nda, Eichhorn:2018ydy}. The momentum dependence has been also used to show the absence of IR-divergences in the IR-regime \cite{Christiansen:2012rx, Christiansen:2014raa}, and to show the absence of $R_{\mu\nu}^2$-contributions at the UV-fixed point \cite{Denz:2016qks}. \smallskip

\textit{Curvature dependence} (\autoref{sec:curv-dep}):
The curvature dependence of the correlation functions allows extending the results from a flat background to a generic background. The full curvature-dependence of the fluctuation correlation functions contain the information of the diffeomorphism-invariant effective action, see \autoref{sec:Hierarchy}. The first steps in this direction in pure gravity and scalar-gravity systems have been done in \cite{Knorr:2017fus, Christiansen:2017bsy, Burger:2019upn}. In \cite{Christiansen:2017bsy, Burger:2019upn}, the difference of the background and quantum EoM due to the mNI was explicitly computed, see \autoref{sec:BackIndep} and \autoref{fig:EoMs}.\smallskip

\textit{Gravity-matter systems} (\autoref{sec:matter-dep}): 
The aim is to incorporate the SM degrees of freedom in asymptotically safe quantum gravity and eventually to retrodict SM parameters and to constrain beyond the SM physics \cite{Shaposhnikov:2009pv, Eichhorn:2017als, Eichhorn:2017ylw, Eichhorn:2017lry, Eichhorn:2017muy, Eichhorn:2018whv, Eichhorn:2019dhg, Kwapisz:2019wrl, Reichert:2019car, Alkofer:2020vtb, Hamada:2020vnf, Eichhorn:2020kca}. Minimally and non-minimally coupled gravity-matter systems have been investigated with (partial) fluctuation-approach techniques in \cite{Folkerts:2011jz, Dona:2013qba, Dona:2015tnf, Meibohm:2015twa, Henz:2016aoh, Eichhorn:2016esv, Christiansen:2017gtg, Eichhorn:2017eht, Christiansen:2017cxa, Eichhorn:2017sok, Eichhorn:2018akn, Eichhorn:2018nda, Burger:2019upn}. A particularly interesting question is for which matter content the UV fixed point exists. First bounds were computed in \cite{Dona:2013qba}, however, a qualitative difference between the results in the background-field approximation and the fluctuation approach were found \cite{Meibohm:2015twa}. It was shown in \cite{Christiansen:2017cxa}, that higher-order curvature terms are needed to fully address this question. For gravity-matter systems with higher-derivative gravity in the background-field approximation see \cite{Hamada:2017rvn, Alkofer:2018fxj, Alkofer:2018baq}. The investigation in \cite{Burger:2019upn} is a first step towards the computational confirmation of the existence of an asymptotically safe fixed point for general gravity-matter in the minimally coupled approximation. This opens a path towards reliable stability investigations of fully coupled gravity-matter systems.\smallskip

\textit{Effective universality} (\autoref{sec:eff-uni}):
Lastly, we discuss the potential \textit{close perturbativeness} of the UV fixed-point regime of asymptotically safe gravity. This leads us to the concept of \textit{effective universality}: so-called \textit{avatars} of the Newton coupling extracted from different correlation functions may agree up to differences that can be inferred from the modified STIs that relates these couplings \cite{Eichhorn:2018akn, Eichhorn:2018ydy}. If present, effective universality may have a dynamical origin. The analysis of this intriguing property is also intricate due to truncation artefacts and RG-scheme dependences.
\smallskip

We close this overview by commenting on the related bi-metric approach and hybrids of the background-field approximation and the fluctuation approach. \smallskip

\textit{Hybrid approaches}: 
In hybrid approaches, one substitutes part of the fluctuation flow equations with background flow equations \cite{Eichhorn:2009ah, Eichhorn:2010tb, Groh:2010ta, Eichhorn:2011pc, Eichhorn:2012va, Codello:2013fpa, Eichhorn:2013ug, Dona:2013qba, Eichhorn:2016vvy, deBrito:2020rwu}. In most cases, this concerns the notoriously difficult pure gravity couplings: the derivation of fluctuation flows of pure gravity vertices such as the three- and four-point functions requires a significant computer-algebraic effort. In advanced truncations, this is accompanied with numerical loop integrations in every flow step as well as interpolations of dressing functions with potentially several momentum and angular dependences. In turn, using the background-field approximation for these vertices reduces this task to computing the flow of a single background coupling, whose flow equation is known analytically. This considerable reduction makes it chiefly important to construct reliable background-field approximation schemes as discussed in \autoref{sec:RegDep}. 

An alternative to the use of the background-field approximation for the pure gravity couplings is their identification with matter-gravity couplings. Such an identification implicitly relies on the concept of \textit{effective universality} discussed in more detail in \autoref{sec:eff-uni}. There it is discussed that while the full system shows effective universality, it is only maintained if using the pure gravity couplings for the matter-gravity couplings. In turn, effective universality, as well as the compatibility with the full system, is lost if using the matter-gravity couplings as pure gravity ones. This hints at a surprisingly complicated interaction structure in gravity-matter systems whose origin is yet to be understood. \smallskip

\textit{Bi-metric approach}: 
The bi-metric approach, developed in \cite{Manrique:2009uh, Manrique:2010mq, Manrique:2010am, Becker:2014qya}, is tantamount to the fluctuation approach reviewed here, as it rests upon the distinction between the background metric and the fluctuation field. Technically, fluctuation and background correlation functions are defined in terms of an expansion of the full metric $g_{\mu\nu}=(1+\epsilon )\bar g_{\mu\nu}$ with the fluctuation field $h_{\mu\nu}= \epsilon \bar g_{\mu\nu}$. This allows one to order the flow and the effective action in powers of $\epsilon$. The power $\epsilon^n$ of the effective action is simply the fluctuation $n$-point function. This reads schematically 
\begin{align}\label{eq:Bimetric} 
\Gamma_k[\bar g,h] = \sum_n \frac{\epsilon^n}{n!}\int \!\Gamma^{(0,h_{\mu_1\nu_1}\cdots h_{\mu_n\nu_n})}_k[\bar g,0]\cdot \bar g_{\mu_1\nu_1} \!\cdots \!\bar g_{\mu_n\nu_n}\,, 
\end{align}  
in analogy to \eqref{eq:vertex-expansion}. The $\Gamma^{(0,n)}_k[\bar g,0]$ have been baptised \textit{level-$n$ vertices} comprising the respective \textit{level-$n$ couplings}. The last and most important step concerns the extraction of the correlation function $\Gamma^{(0,n)}_k[\bar g,0]$ from $\int\Gamma^{(0,n)}_k[\bar g,0] \cdot \bar g^n$, as the computation of the flow requires the knowledge of the correlation function and not their contractions with metrics. This computation is either done by (i) considering an expansion about a specific background such as the flat background, (ii) computing the flow of the effective action for a generic metric $\bar g$, or (iii) assuming a global form of the effective action and simply computing the flow in this closed form. Option (i) is the fluctuation approach reviewed here. It is not built on the metric split with $\epsilon$. Option (ii) asks for advanced computational heat-kernel techniques even within restrictions. These techniques have seen rapid development in the past decade, which may open a path towards their use in (ii). Option (iii) has been considered so far for \textit{level one} couplings. The \textit{level two} correlation functions that are required for the right-hand side of the flow equation then have been obtained within a further background-field approximation. In summary, the bi-metric approach or rather the computational options (ii)-(iii) offer an alternative approach to compute fluctuation correlation functions that may provide important cross-checks for the results discussed here. 
 
\subsection{UV fixed point}
\label{sec:UV-FP}
The UV fixed point in the fluctuation approach has been discussed in \cite{Christiansen:2012rx, Donkin:2012ud, Christiansen:2014raa, Christiansen:2015rva, Denz:2016qks, Christiansen:2016sjn, Meibohm:2015twa, Knorr:2017mhu, Christiansen:2017cxa, Knorr:2017fus, Christiansen:2017bsy, Eichhorn:2018akn, Eichhorn:2018ydy, Eichhorn:2018nda, Burger:2019upn}. This includes work in the vertex expansion about the flat background in pure gravity \cite{Christiansen:2012rx, Christiansen:2014raa, Christiansen:2015rva, Denz:2016qks, Christiansen:2016sjn} and gravity-matter systems \cite{Meibohm:2015twa, Christiansen:2017cxa, Eichhorn:2018akn, Eichhorn:2018ydy, Eichhorn:2018nda} as well as work including curvature dependence \cite{Knorr:2017fus, Christiansen:2017bsy, Burger:2019upn}, a fluctuation potential \cite{Knorr:2017mhu} and in the geometrical approach \cite{Donkin:2012ud}. In \cite{Denz:2016qks}, the tower of fluctuation correlation functions was implemented until the graviton four-point function. All $n$-point functions were evaluated at the momentum-symmetric point with external transverse-traceless projections. A UV fixed point was found at
\begin{align} 
 \label{eq:UV-FP} 
 \left( \mu^*, \lambda_{3}^*, \lambda_{4}^*, g_{3}^*, g_{4}^* \right) 
 &= \left( -0.45,\, 0.12,\,0.028,\, 0.83,\, 0.57 \right) \,,
\end{align}
where $g_n$ and $\lambda_n$ are the dimensionless Newton coupling and the momentum independent part of the graviton $n$-point function, for more details see \cite{Denz:2016qks}. The graviton mass parameter $\mu = - 2 \lambda_2$ is the momentum-independent part of the graviton two-point function. The critical exponents of the fixed point are given by
\begin{align}
 \theta_i &= (4.7,\, 2.0 \pm 3.1 i,\, -2.9,\, -8.0 ) \,,
 \label{eq:crit-exp}
\end{align}
where a positive sign corresponds to a UV-attractive direction. The three UV-attractive directions were associated with the operators $\sqrt{g}$, $\sqrt{g}R$, and $\sqrt{g}R^2$. In contrast, the operator $\sqrt{g}R_{\mu\nu}^2$ is not generated in the present approximation. The latter property was inferred from the momentum dependence of the graviton three- and four-point function, see \autoref{sec:mom-dep}. Importantly, the first signs of apparent convergence were found in \cite{Denz:2016qks}.

\begin{figure}[t]
 \includegraphics[width=\linewidth]{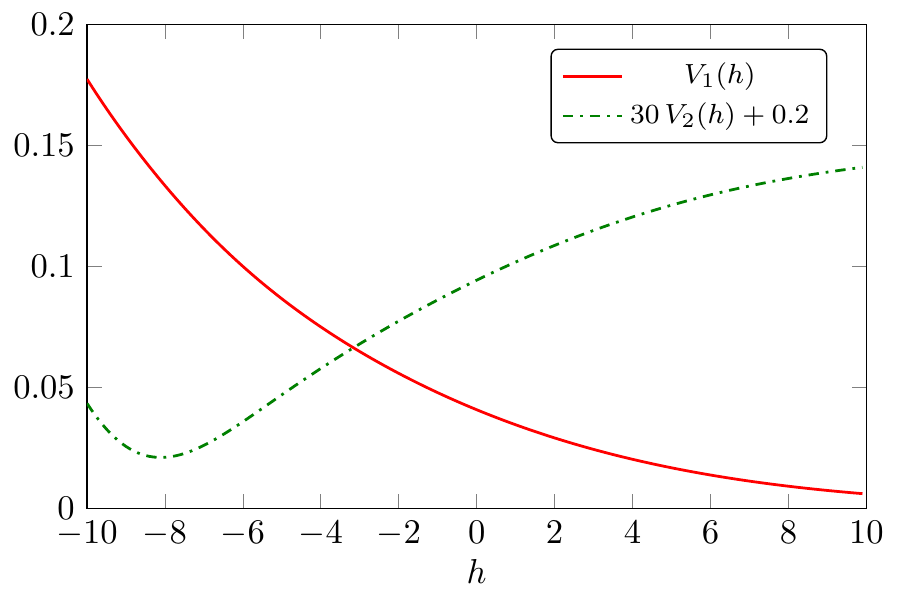}
 \caption{
 Dimensionless fixed-point fluctuation potentials defined via $V = V_1 (h) + \text{Tr}(h^2_\text{TL}) V_2(h)$ where $h$ is the trace part and $h^\text{TL}$ is the traceless part of the fluctuation graviton. Note that we rescaled and shifted $V_2$, i.e., $V_2$ is small compared to $V_1$ and always negative. The results are taken from \cite{Knorr:2017mhu}.
 }
 \label{fig:infinite-order}
\end{figure}

In \autoref{sec:Expbarg} we have shown that the fluctuation approach in the flat expansion improves upon the background field approximation in the curvature expansion, see in particular the discussion at the summary at the end of \autoref{sec:Expbarg}. Accordingly, the fluctuation results for the UV fixed point detailed above extend and corroborate previous findings in the background-field approximation within the curvature expansion. In particular, the results confirm that the latter captures the most important features in pure gravity. For example, the fixed-point value of the cosmological constant in the background-field approximation is typically positive, which is comparable with the negative fixed-point value of $\mu$ in \eq{eq:UV-FP} ($\mu=-2 \lambda_2$). Also, mostly three relevant directions are found in the background-field approximation, see the reviews \cite{Niedermaier:2006wt, Litim:2011cp, Reuter:2012id, Ashtekar:2014kba, Eichhorn:2017egq, Bonanno:2017pkg, Eichhorn:2018yfc, Pereira:2019dbn, Reichert:2020mja, Bonanno:2020bil}, and 
the very recent paper \cite{Falls:2020qhj}. 

A further extension, within the exponential split, has been investigated in \cite{Knorr:2017mhu}. There, the dimensionless fluctuation potential $V$ was approximated with $V = V_1 (h) + \text{Tr}(h^2_\text{TL}) V_2(h)$, where $h$ is the trace part and $h_\text{TL}$ is the traceless part of the fluctuation graviton. The other graviton modes have been dropped. The results for the potentials $V_1$ and $V_2$ are displayed in \autoref{fig:infinite-order}.

\begin{figure}[t]
 \includegraphics[width=\linewidth]{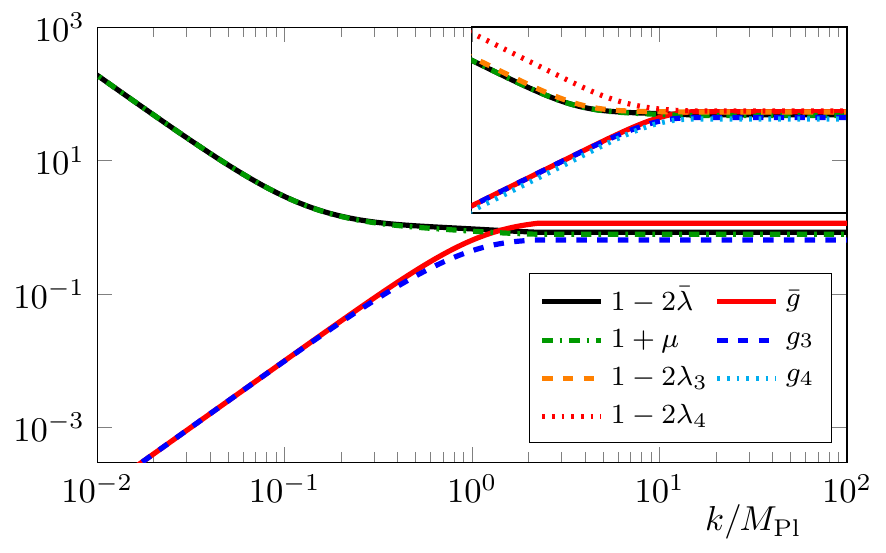}
 \caption{
 Scale-dependence of different fluctuation couplings along a trajectory from the UV-fixed point \eq{eq:UV-FP} to the IR. In the IR the couplings flow according to their canonical running. For small $k$, $\bar g$ and $g_3$ as well as $\bar\lambda$ and $\mu$ are related via the simplified NI \eq{eq:NI-reduced}. The inset shows the complete set of couplings. The results are taken from \cite{Denz:2016qks}.
 }
 \label{fig:IR-trajectory}
\end{figure}

\subsection{UV-IR trajectories}
\label{sec:UV-IR}
UV-IR trajectories in the fluctuation approach and hence the phase structure of quantum gravity have been discussed in \cite{Christiansen:2012rx, Donkin:2012ud, Christiansen:2014raa, Denz:2016qks}. In \autoref{fig:IR-trajectory}, we display a trajectory from the UV fixed point \eq{eq:UV-FP} to the IR where all couplings run classically. In the displayed example, the graviton mass parameter runs to infinity, $\mu \to \infty$. In classical gravity and $\mu>0$ the NIs entail that the cosmological constant is indeed given by $\Lambda= \bar\Lambda= -2 \mu k^2 $ in the limit $k\to 0$ and can take any negative value. This follows from 
\begin{align}
 \label{eq:NI-reduced}
 \frac{\delta \Gamma_k[\bar g,h]}{\delta \bar g} &=
 \frac{\delta \Gamma_k[\bar g, h]}{\delta h} \,,
 &
 \text{for} \quad \lim_{k\to 0} \mu&\to\infty \,.
\end{align}
Moreover the background Newton coupling and (all) the fluctuation Newton coupling agree. This can be seen for the dimensionless versions $\bar\lambda, \lambda_2$ and $\bar g, g_3$ in \autoref{fig:IR-trajectory}. Solving the NIs for the higher couplings corresponds to a fine-tuning problem in terms of choosing an appropriate trajectory. However, a fully diffeomorphism-invariant solution including the higher-order avatars of the couplings has not been fine-tuned yet, see the inlay in \autoref{fig:IR-trajectory}.  

\begin{figure*}[t]
 \includegraphics[width=.49\linewidth]{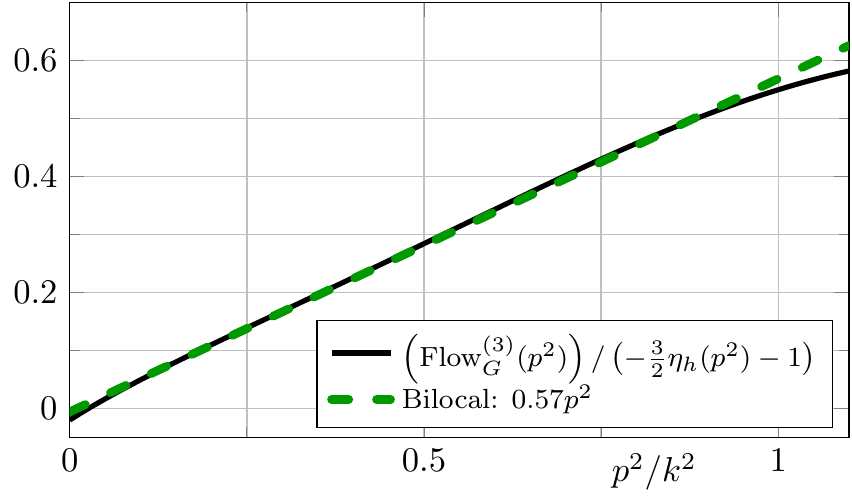} \hfill
 \includegraphics[width=.49\linewidth]{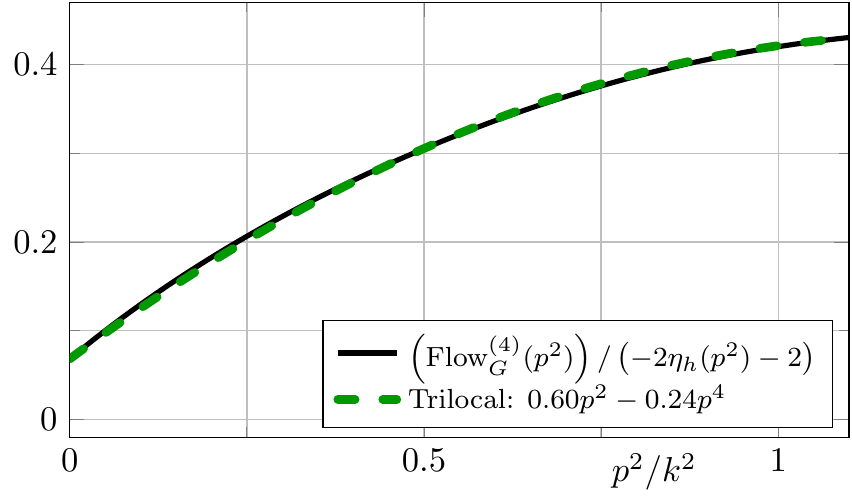}
 \caption{
 Momentum dependence of the transverse-traceless graviton three- and four-point couplings obtained by normalising the vertex flow with $(-\frac n2 \eta_h(p^2)-n+2)$. The graviton three-point coupling (left panel) is well described with a linear $p^2$-function in the momentum range $0\leq p^2 \leq k^2$. This momentum dependence stems from the $R$-tensor structure. The absence of a $p^4$-behaviour implies that the $R_{\mu\nu}^2$-tensor structure is suppressed. On the other hand, the graviton four-point coupling (right panel) shows a clear $p^4$-behaviour, which is associated with $R^2$-tensor structure. The figures are taken from \cite{Denz:2016qks}.
 }
 \label{fig:MomDep}
\end{figure*}

UV-IR trajectories with $\mu \to -1$ in the IR have also been investigated in \cite{Christiansen:2012rx, Donkin:2012ud, Christiansen:2014raa, Denz:2016qks}. Those trajectories are technically challenging since $\mu = -1$ corresponds to a pole in the propagator. We emphasise that the NIs and STIs are in this case non-trivial even for classical gravity: the classical effective action is the convex hull of the classical action, the latter not being convex for $\mu<0$. This entails that $\lambda_2 = -2\mu$ cannot be identified with the cosmological constant $\lambda=\bar \lambda$ even though the sign of the latter must be also positive. Note also that any positive cosmological constant $\Lambda$ can be obtained. The truncation-triggered restriction to $\Lambda=0$ at $k=0$ in the background field approximation is lifted. From the physics point of view, these trajectories are appealing since they correspond to a positive cosmological constant in the IR.

\subsection{Momentum dependence \& unitarity}
\label{sec:mom-dep}
The momentum dependence of correlation functions have been discussed in \cite{Christiansen:2012rx, Christiansen:2014raa, Christiansen:2015rva, Denz:2016qks, Christiansen:2017cxa, Eichhorn:2018akn, Eichhorn:2018nda, Eichhorn:2018ydy}. This momentum dependence encodes the dynamics of the theory and is crucial for the question of unitarity. One of the advantages of the fluctuation approach in the flat vertex expansion is its easy access to the full momentum dependence of fluctuation correlation functions $\Gamma^{(0,n)}_k$ for all cutoff scales $k$. These momentum dependences carry the full dynamics of the underlying theory: all other quantities, ranging from the background correlation functions to diffeomorphism-invariant observables $\CO[g]$ are built from the correlation functions. The latter observables are defined as expectation values $\CO[g]= \CO[\bar g, h=0]$ of diffeomorphism-invariant operators $\hat\CO[\hat g]$ with $\CO[\bar g, \phi]=\langle \hat \CO\rangle$. The $\CO[\bar g, \phi]$ satisfy the flow equation for the expectation values of composite operators derived in \cite{Pawlowski:2005xe}, 
\begin{align}
 \label{eq:compflow} 
 \partial_t \CO_k[g] = -\frac12 \Tr\left[ G_k\,\partial_t R_k\, G_k {\CO}_k^{(0,2)}\right]\![g]\,, 
\end{align}
at vanishing fluctuation field $h=0$. Evidently, the flow \eq{eq:compflow} solely depends on the fluctuation field propagators and $\CO[\bar g, \phi]$. For applications and further investigations of \eq{eq:compflow} see \cite{Pawlowski:2005xe, Igarashi:2009tj, Herbst:2015ona, Pagani:2016pad, Pagani:2016dof, Houthoff:2020zqy, Kurov:2020csd}.

\Eq{eq:compflow} entails in particular that \textit{any} observable inherits its dynamics from that of the full field- and momentum-dependence of the fluctuation two-point function, or rather from the momentum-dependence of the fluctuation correlation functions $\Gamma^{(0,n)}_k$ at a given field expansion point. It is in this sense that the momentum-dependent and RG-invariant vertex dressings $\CA^{(\phi_1\cdots \phi_n)}(\mathbf p)$ encode the dynamics of the theory. In particular, the symmetric-point dressings $G_n(\bar p)$ carry the meaning of momentum-dependent running couplings similar to those in standard quantum gauge theories, and most notably in QCD, for a detailed discussion in the latter case see in particular \cite{Cyrol:2016tym, Cyrol:2017ewj}. We emphasise that while in both cases these couplings are neither observables themselves nor even gauge- or diffeomorphism invariant, they directly encode the dynamics of the theory, and in particular the dominance and/or decoupling of degrees of freedom. If done carefully, they can be also compared to scattering processes related to the respective vertices, for the SM see the comparison of the QCD running (vertex) coupling to scattering experiments at accelerators, see \cite{Tanabashi:2018oca}. 

Moreover, the resolution of the momentum-dependences of $n$-point functions gives at least indirect access to the question of unitarity of asymptotically safe gravity: From the Euclidean data one can reconstruct Minkowski correlation functions and in particular the graviton spectral functions, both that of the fluctuation graviton and that of the background graviton, for more details see \cite{Bonanno:2021squ}. Here we simply comment on the physics content of the graviton spectral functions, see also \cite{Becker:2014jua}. In this context, we will also use the analogy to the gluon in a non-Abelian gauge theory as discussed in \cite{Cyrol:2018xeq}. For a recent discussion of the challenges for unitarity in asymptotically safe gravity see \cite{Bonanno:2020bil}.

To begin with, both the fluctuation graviton and the background graviton two-point functions are not diffeomorphism-invariant. Accordingly, they are not directly related to asymptotic states even though at low energies gravity is weakly-coupled and the theory exhibits a classical momentum- and scale-dependence, see \autoref{fig:IR-trajectory}. The latter property suggests, that, if a K{\"a}ll{\'e}n-Lehmann spectral representation of the graviton propagators exists, the graviton spectral functions may exhibit a particle-like spectrum for low spectral values. In turn, for large spectral values, we enter the UV fixed point regime and the physics content of the spectral functions is unclear.

Note, however, that the same line of arguments would suggest that the gluon spectral function exhibits a particle-type spectral dependence in its perturbative regime for large spectral values. Instead, it can be shown, that, if a K{\"a}ll{\'e}n-Lehmann spectral representation exists, the gluon spectral function is negative for large spectral values and its spectral sum vanishes (Oehme-Zimmermann superconvergence relation). Moreover, it is also negative for small spectral values, see \cite{Cyrol:2018xeq}. These properties hold for both the fluctuation and the background gluon. Since these properties follow directly from the momentum dependence of the Euclidean correlation functions, similar results hold for asymptotically safe gravity, see \cite{Bonanno:2021squ}.

In summary, the spectral properties of diffeomorphism- or gauge-variant correlation functions only indirectly mirror the unitarity of the theory. This situation prohibits any direct conclusion of a lack of unitarity from the occurrence of negative parts of spectral functions including negative poles (ghost states). We also emphasise that the latter statement should not be taken as its converse. Of course, the occurrence of negative parts of spectral function requires a thorough investigation of the physics implications and may well be related to the lack of unitarity of the underlying theory. The example of the non-Abelian theory simply indicates that this is not necessarily the case. Such an investigation requires the analysis of the spectral properties of diffeomorphism-invariant states, for a recent discussion of such a setup see \cite{Maas:2019eux}.

The discussion in this section so far emphasises the importance of the computation of the momentum-dependence of correlation functions both for the dynamics of observables as well as the intricate problem of unitarity. One of the advantages of the fluctuation approach is the direct access to momentum-dependent correlation functions with standard quantum field theory methods:

In \cite{Christiansen:2012rx, Christiansen:2014raa} the full momentum dependence of the graviton and ghost propagator was included via the anomalous dimensions. The computation of the momentum dependence was extended to the graviton three- \cite{Christiansen:2015rva} and four-point function \cite{Denz:2016qks} as well as to the scalar-graviton \cite{Eichhorn:2018akn}, the fermion-graviton \cite{Eichhorn:2018nda}, the gluon-graviton vertex \cite{Christiansen:2017cxa} and the ghost-graviton vertex \cite{Eichhorn:2018ydy}. While only the momenta $0\leq p^2 \lesssim k^2$ contribute to the flow, in all these works the vertices have been computed at the symmetric point for the full momentum range $0\leq p^2 < \infty$. This approximation ignores, in particular, the angular dependence of the vertex dressings. While the angular dependence is important for the discussion of the whole phase space of scattering experiments, it is averaged in the flow diagrams due to the angular loop integrations. The reliability of this approximation has been studied at length in QCD, see \cite{Mitter:2014wpa, Cyrol:2016tym, Cyrol:2017ewj, Corell:2018yil} for details. There, it was shown that the above approximation is very accurate in the absence of resonant interaction channels, and so far no indications have been found for such resonant channels. In conclusion, this analysis provides a non-trivial reliability argument for the approximation described above. Still, for a full reliability check, one has to study extended truncations. 

\begin{figure*}[t]
 \includegraphics[width=\linewidth]{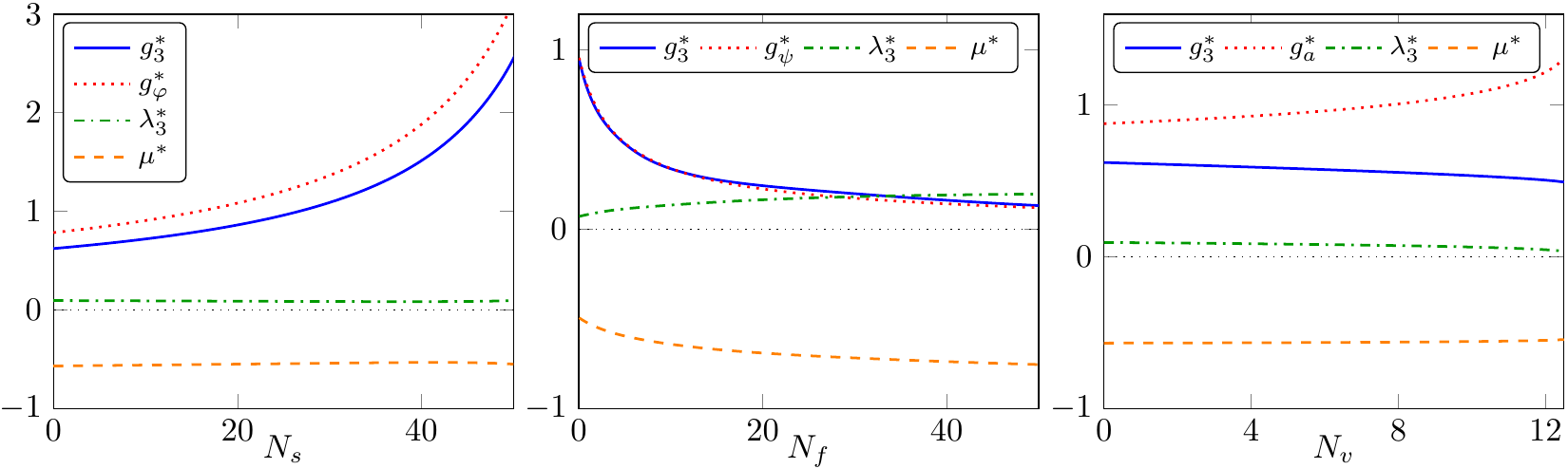}
 \caption{
 Fixed-point values of the fluctuation couplings as a function of the number of scalar (left), fermion (middle), and gauge fields (right). All truncations include the graviton two- and three-point function as well as the respective graviton-matter vertex. In the scalar case, the Newton couplings, $g_3$ and $g_\varphi$, are diverging at $N_s\approx 52 $. The fermionic case is stable for all $N_f$. In the gauge field case, the fixed point is disappearing in the complex plane at $N_v\approx 13$. It was explained in \cite{Christiansen:2017cxa} that the vanishing of the fixed point is an artefact of the truncation and how it can be lifted in the gauge-field case. In \cite{Burger:2019upn}, it was suggested that an expansion about a background that is a solution to the quantum EoM might remove the divergence in the scalar case. The result are taken from \cite{Eichhorn:2018akn} (scalar), \cite{Eichhorn:2018nda} (fermion), and \cite{Christiansen:2017cxa} (gauge).
 }
 \label{fig:matter-dependence}
\end{figure*}

In \cite{Denz:2016qks, Christiansen:2016sjn}, the momentum dependence was used to disentangle contributions from the couplings of the $R^2$ and $R_{\mu\nu}^2$-tensor structures. This was done in \cite{Christiansen:2016sjn} with derivatives at vanishing momentum, while in \cite{Denz:2016qks} the momentum range $0\leq p^2 \leq k^2$ was considered. Importantly, the transverse-traceless graviton three-point function has overlap with $R_{\mu\nu}^2$-tensor structures and not with $R^2$-tensor structures, while the graviton four-point function has overlap with $R_{\mu\nu}^2$- and $R^2$-tensor structures. The momentum-dependence of the couplings is obtained by normalising the vertex flows with $(-\frac n2 \eta_h(p^2)-n+2)$. This is displayed in \autoref{fig:MomDep}. The three-point coupling is well described with a $p^2$-behaviour. Thus, the $R_{\mu\nu}^2$-tensor structure is non-trivially suppressed. The four-point coupling shows a significant $p^4$-behaviour. Due to the absence of a $p^4$-behaviour in the three-point coupling, this suggests that they are related to $R^2$-tensor structures. 

Recently, impressive progress has been made towards momentum-dependent computations in the background-field approximation \cite{Knorr:2018kog, Bosma:2019aiu, Knorr:2019atm, Draper:2020bop, Draper:2020knh}. There, the momentum dependence is captured via form factors $W_k^i$, e.g., $\int_x\! R\,W_k^R(\Delta)R$ and $\int_x\! C_{\mu\nu\rho\sigma} W_k^C(\Delta)C^{\mu\nu\rho\sigma}$. This opens a path towards the comparison of the result of these two approaches. This will allow us to quantify the difference between a background-field approximation and a fluctuation computation.

\subsection{Curvature dependence}
\label{sec:curv-dep}
The curvature dependence of correlation functions in the fluctuation approach has been discussed in \cite{Knorr:2017fus, Christiansen:2017bsy, Burger:2019upn}. Most results in the fluctuation approach were computed on a flat background $\bar g= \mathbb{1}$. Results for generic backgrounds can be obtained from an expansion about the flat background. In \cite{Knorr:2017fus}, this was done with covariant heat-kernel methods up to the first-order curvature couplings. Fixed-point values of all first-order curvature couplings were found and their gauge dependence investigated.

A different approach within the fluctuation approach was taken in \cite{Christiansen:2017bsy, Burger:2019upn} where the fluctuation correlation functions were computed directly on a generic background with constant curvature. The computation reaches up to the graviton three-point function and also includes $N_s$ scalar fields in \cite{Burger:2019upn}. It was found that the curvature dependence of the fluctuation couplings counterbalances the explicit curvature dependence of the respective vertex, making the full vertex approximately curvature independent. This result supports results obtained on a flat background. Furthermore, it was explicitly shown that the background EoM differs from the quantum EoM at the UV fixed point, \eq{eq:FlucEoM=backEoM}. In particular, the background EoM does not have a solution at the UV fixed point, while the quantum EoM has two solutions, a minimum at negative curvature and a maximum at positive curvature, for all $N_s$ that are accessible. This is displayed in \autoref{fig:EoMs}.

\subsection{Gravity-matter systems}
\label{sec:matter-dep}
A theory of quantum gravity necessarily needs to include matter degrees of freedom to describe our Universe. A central question is for which matter content the UV fixed point exists and if certain types of matter field have a stabilising or destabilising effect. Most studies have focused on analysing SM matter fields within the minimally-coupled approximation. In this approximation, the matter fields are considered without self-interaction and only couple to gravity via their kinetic term. There are works in the background-field approximation \cite{Dou:1997fg, Percacci:2002ie, Granda:2005nd, Granda:2005py, Dona:2012am, Biemans:2017zca, Hamada:2017rvn, Alkofer:2018fxj, Alkofer:2018baq, Gies:2018jnv}, in the hybrid approach \cite{Dona:2013qba, Dona:2015tnf}, and in a full fluctuation computation \cite{Meibohm:2015twa, Christiansen:2017cxa, Eichhorn:2018akn, Eichhorn:2018ydy, Eichhorn:2018nda, Burger:2019upn}. For works beyond the minimally-coupled approximation see \cite{Percacci:2003jz, Narain:2009fy, Zanusso:2009bs, Vacca:2010mj, Daum:2010bc, Folkerts:2011jz, Harst:2011zx, Eichhorn:2011pc, Eichhorn:2012va, Henz:2013oxa, Percacci:2015wwa, Labus:2015ska, Oda:2015sma, Eichhorn:2016esv, Henz:2016aoh, Eichhorn:2016vvy, Meibohm:2016mkp, Christiansen:2017gtg, Hamada:2017rvn, Eichhorn:2017sok, Christiansen:2017qca, Eichhorn:2017lry, Pawlowski:2018ixd, deBrito:2019epw, Held:2020kze, Daas:2020dyo}, which also includes scalar-tensor theories and gravitational corrections to the running of matter couplings.

A major keystone in the stability analysis of gravity-matter systems in the minimally coupled approximation was found in \cite{Meibohm:2015twa, Christiansen:2017cxa}. There it was shown that minimally-coupled gravity-matter systems in the Einstein-Hilbert truncation always show a Reuter fixed point as the system can be mapped to a pure gravity system at the level of the path integral. We emphasise that while the explicit computations in \cite{Meibohm:2015twa, Christiansen:2017cxa} are done in the fluctuation approach, the conceptual investigation is general. For a detailed discussion we refer to \cite{Christiansen:2017cxa}. Here, we simply sketch the important steps: In minimally-coupled gravity-matter systems the matter part $S_\text{mat}[\bar g,\phi]$ of the full action $S=S_\text{grav}+S_\text{mat}$ is quadratic (or bi-linear) in the matter fields. To find the Reuter fixed point it is sufficient to discuss the UV limit of graviton correlation functions. Consequently, we consider vanishing matter sources, $J_\text{mat}\equiv 0$. After performing the Gau\ss ian integration over the matter fluctuation fields $\hat{\phi}_\text{mat}$, the path integral of a minimally coupled matter-gravity system takes the schematic form, 
\begin{align}\label{eq:MinCouplGravMat}
Z[J]=\int\!\CD \hat{\phi}_\text{grav}\, e^{- S_\text{grav,eff}[\bar g, \hat\phi_\text{grav}]+\int \! \mathrm d^4 x \sqrt{\bar g} \,J_\text{grav}^a\hat{\phi}_{\text{grav}, a}}.
\end{align}
with 
\begin{align}\label{eq:SgravEff}
S_\text{grav,eff}[\bar g, \hat\phi_\text{grav}] = S_\text{grav}[\bar g, \hat\phi_\text{grav}] + 
\frac12 \text{Tr} \log S_{\text{mat}}^{(2)}[\hat g] \,. 
\end{align}
Here, the full fluctuation field is split into $\phi=(\phi_\text{grav},\phi_\text{mat})$ with $\phi_\text{grav}=( h_{\mu\nu}, c_\mu, \bar{c}_\mu)$, and the hatted field indicate the integration fields. In slight abuse of notation, we wrote $S_\text{mat}^{(2)}[\hat g]$ as the second derivative of the matter action with respect to the matter fields. Its argument is $\hat g=\bar g+\hat h$, the full metric that is integrated over. Hence $S_\text{mat}^{(2)}$ is a covariant operator and the $\text{Tr} \log $-contribution is diffeomorphism-invariant. 

The form of the generating functional in \eq{eq:MinCouplGravMat} is also obtained for UV-complete non-minimally coupled matter theories such as Yang-Mills theories. Then, the $S_\text{grav,eff}[\bar g, \hat h]$ is not of the form \eq{eq:SgravEff} but carries the full non-perturbative metric-dependent part of the effective action of Yang-Mills theories. The UV-completeness within this procedure is required as otherwise the matter path integral cannot be performed. Trivially, minimally-coupled systems are UV-complete. A useful analogue for the study of the UV-stability of minimally-coupled gravity-matter systems is many-flavour QCD. There, the r\^ole of the graviton is taken by the gluon and the quark action is bilinear. 

The representation \eq{eq:MinCouplGravMat} emphasises an intriguing and useful property of the fRG-approach to quantum gravity (and beyond): The phase structure and in particular the fixed-point structure of a generic gravity-matter system can be accessed within pure gravity. In particular, all fixed points are accessible within this setup, if a general fixed point effective action $\Gamma^*_k[\bar g, \phi_\text{grav}]$ is considered. 

This intriguing property also carries an important intricacy of a generic fixed-point analysis: Seemingly the parameterisation \eq{eq:MinCouplGravMat} entails that generic gravity-matter systems are UV-stable if the matter part is UV-complete (with the assumption that the Reuter fixed point exists for pure gravity). This conclusion would apply directly to all minimally-coupled gravity-matter systems. That this argument falls short can be seen at the example of many-flavour QCD. There, an (f)RG-analysis reveals that the QCD $\beta$-function changes its sign for a large enough number of flavours. In the vicinity of this regime, interesting phenomena such as conformal scaling, instabilities and the Caswell-Banks-Zaks fixed point occur, for fRG literature see e.g.\ \cite{Gies:2005as, Terao:2007jm, Braun:2009ns, Braun:2010qs} and references therein. These findings are backed-up by lattice results. The RG-analysis in many-flavour QCD solely relies on the \textit{marginal} operator $\text{tr}\, F_{\mu\nu}^2$. The quantum-corrections from the integrating-out of the quark fluctuations are proportional to 
\begin{align}\label{eq:QCDmarginal}
N_\text{f}\, \text{tr} \,F_{\mu\nu}^2 \, \log \!\frac{F_{\mu\nu}^2}{k^4}\,, 
\end{align}
where $N_\text{f}$ is the number of flavours. The analogous operators in gravity are the curvature-squared operators $R^2$, $R_{\mu\nu}^2$, and $R_{\mu\nu\rho\sigma}^2$. The respective operators including matter quantum fluctuations are 
\begin{align}\label{eq:Gravitymarginal}
N_\text{mat}\, \sqrt{g}\,R^2 \, \log \! \frac{R}{k^2}\,, 
\end{align}
and similar ones for $R_{\mu\nu}^2$, and $R_{\mu\nu\rho\sigma}^2$ and also covariant derivatives. Here, 
$N_\text{mat}$ is the weighted sum over all species and flavours of matter fields.

The logarithmic RG-running of the marginal operator $\text{tr}\, F_{\mu\nu}^2$ in QCD or $R^2$ in gravity necessarily triggers a field-dependence of its coefficient as displayed in \eq{eq:QCDmarginal} and \eq{eq:Gravitymarginal} respectively. In conclusion, the distinctive property of marginal operators is the inherent field-dependence of the quantum corrections. In turn, the coefficients of (local) relevant and irrelevant operators are only scale-dependent. While the latter by definition are not important for a fixed-point analysis, the coefficients of the former ones, if present, can be readily absorbed in the respective pure gauge theory (or gravity) couplings. In the present example of many-flavour QCD relevant operators are indeed absent. In gravity, this applies to the terms in the Einstein-Hilbert action, i.e., the curvature term and the cosmological-constant term. 

In summary, from the perspective of the Yang-Mills system with the generating functional similar to that in \eq{eq:SgravEff}, the marginal operator \eq{eq:QCDmarginal} introduces a new UV-marginal (and hence physical) parameter $N_\text{f}$ that cannot be absorbed in the Yang-Mills coupling. In gravity this applies to the coefficients of the marginal operators $R^2$, $R_{\mu\nu}^2$, and $R_{\mu\nu\rho\sigma}^2$. Thus, also here the flavour number $N_\text{mat}$ of a given matter field is a physical parameter. However, its relevance for the fixed-point analysis originates \textit{solely} from the $N_\text{mat}$-dependent coefficients of the marginal operators $R^2$, $R_{\mu\nu}^2$, and $R_{\mu\nu\rho\sigma}^2$. In contrast, the $N_\text{mat}$ of the relevant operators in the Einstein-Hilbert action is not relevant for the fixed-point analysis. In particular, it cannot trigger instabilities. 

The above properties imply that a fixed-point analysis of a given system within a truncation of the (f)RG-flows, that does not include the flows of the marginal operators, should exhibit the respective fixed-point structure of the pure gravity system in the same truncation. In particular this casts some doubt on any instability findings in the full truncation, if this instability survives in the absence of the marginal operators. 

As an example of this statement, we consider now a minimally-coupled gravity-matter system in the Einstein-Hilbert truncation. Without truncation these systems have the path integral representation \eq{eq:MinCouplGravMat} with \eq{eq:SgravEff}. The Einstein-Hilbert truncation reduces $S_\text{grav,eff}$ in \eq{eq:SgravEff} to 
\begin{align}\nonumber 
S_\text{grav,eff}[\bar g, \hat\phi_\text{grav}] \to &\, S_\text{EH}[g] + S_\text{gf}[\bar g, \hat h] +S_\text{gh}[\bar g, \hat\phi_\text{grav}] \\[1ex] 
&\, + \left. 
\frac12 \text{Tr} \log S_{\text{mat}}^{(2)}[\hat g] \right|_{R,\Lambda}, 
\label{eq:SgravEff-EH}
\end{align}
where the subscript $|_{R,\Lambda}$ stands for the reduction of the full one-loop determinant to its Einstein-Hilbert part with a curvature term and a cosmological-constant term. The respective coefficients can be absorbed in a redefinition of the Newton constant and cosmological constant in $S_\text{EH}[g]$, for more details see \cite{Christiansen:2017cxa}. Hence, \eq{eq:SgravEff-EH} is equivalent to the Einstein-Hilbert truncation of the pure gravity system. The latter shows the Reuter fixed point and so should the minimally-coupled system in this truncation. 

The above result for minimally-coupled systems has the direct consequence, that the Einstein-Hilbert truncation to matter-gravity systems should also exhibit the Reuter fixed point for UV-complete matter systems, as the pure-gravity system does. We add, that this does not exclude the emergence of further fixed points in some $N_\text{mat}$ regime.  

This concludes our discussion of the fixed-point structure and stability properties of gravity-matter systems, its truncation-dependence as well as reliability requirements for truncations. The discussion enables us to formulate relevant properties that have to be considered for a conclusive stability analysis of matter-gravity systems: 
\begin{itemize} 
 \item[(i)] The fixed-point analysis necessarily has to involve all (possibly) relevant operators of the theory under investigation, i.e.\ \eq{eq:QCDmarginal} in many-flavour QCD and \eq{eq:Gravitymarginal} in gravity-matter systems. 
 \item[(ii)] A fixed-point analysis within a given truncation is only fully reliable if it also reproduces the fixed-points of the pure gravity system in the same truncation excluding the marginal operators.  
\end{itemize} 
We now discuss the results in gravity-matter systems given the properties (i) and (ii): In \cite{Meibohm:2015twa}, the first full fluctuation computation for minimally-coupled systems was put forward. On the pure gravity side, the flows of the fluctuation graviton two- and three-point function were included. Importantly, a stabilising mechanism for the fermionic contribution was found for general regulators: the graviton mass parameter is approaching its pole $\mu\to-1$ and thus enhances the graviton contribution, in short: \textit{gravity rules}. This is required from the discussion above. Technically this simply means that the fermion contribution in this setup changes the parameters of the two- and three-point function within the stability regime of the phase diagram of pure gravity in the Einstein-Hilbert truncation. This stabilising mechanism was also found in an extension of the truncation \cite{Eichhorn:2018nda} making the fermion-gravity system a showcase of the mechanism described above. In particular, with the existence of the Reuter fixed point for the minimally-coupled system in the absence of marginal operators in the pure-gravity subsystem, the flow equations of the fermion-gravity system satisfy the requirement (ii). Consequently, a conclusive stability analysis of general fermion-gravity systems can be performed but requires the inclusion of the marginal curvature-squared operators. 

In the same truncation applied to minimally-coupled scalar-gravity systems, it was found within the fluctuation approach in \cite{Meibohm:2015twa, Eichhorn:2018akn}, that the graviton anomalous dimension $\eta_h$ grows with the number of scalars $N_s$ and finally exceeds the value two beyond a critical flavour number $N_{s,\text{stab}}$: $\eta_h> 2$ for $N_s>N_{s,\text{stab}}\approx 20$. For $\eta_h>2$ the overall cutoff scaling of the graviton regulator goes with negative powers of the cutoff scales and effectively the -physical- cutoff decreases. For these large anomalous dimensions, we leave the reliability regime of the approximation. In short, the reliability bound on the truncation makes it impossible to see the stability of the system in this minimally-coupled approximation. From the viewpoint of the pure gravity system, this simply means that the scalar contribution in this setup eventually moves the parameters of the two- and three-point function outside the stability regime of the phase diagram of pure gravity in the Einstein-Hilbert truncation. Consequently, the setup cannot be used for stability investigations in scalar-gravity systems. In \cite{Burger:2019upn}, it was suggested that an expansion about an on-shell background can lift this tension. In summary, at present, there is no conclusive stability analysis for scalar-gravity systems. 

Applying the same truncation to minimally-coupled gauge-gravity systems, it has been shown in \cite{Christiansen:2017cxa} that depending on the regulator the minimally-coupled systems either behave similarly to the fermionic or the scalar system. This suggests that the truncation has to be improved. In summary, a stability analysis of gauge-gravity systems can be performed but the results have to be taken with a grain of salt. A fully conclusive stability analysis for gauge-gravity systems requires an improvement of the truncations used so far in the literature. 

In \autoref{fig:matter-dependence}, we display the state-of-the-art dependence of the fixed-point values on the number of scalar field $N_s$ \cite{Eichhorn:2018akn}, fermion field $N_f$ \cite{Eichhorn:2018nda}, and gauge fields $N_v$ \cite{Christiansen:2017cxa}. The truncations include the flow of the momentum-dependent graviton two- and three-point functions as well as the respective graviton-matter vertex. In the scalar case, the Newton couplings are diverging at $N_s\approx 52$. This is an artefact of the truncation as described in the previous paragraphs and \cite{Christiansen:2017cxa}. The fermion direction is stable for all $N_f$: the graviton mass parameter approaches its pole $\mu\to-1$ and the enhanced graviton contribution counterbalances the matter contribution. In the gauge case, the fixed point is disappearing in the complex plane for $N_v\approx 13$. In \cite{Christiansen:2017cxa}, it was demonstrated that all numbers of gauge fields can be accessed with a different regulator, as discussed in the last paragraph. 

Finally, we speculate on the stability properties of general gravity-matter systems based on the results obtained so far. To that end, we assume that there is a setup such that general minimally-coupled gravity-matter systems in the Einstein-Hilbert truncation show UV stability with a Reuter fixed point similar to the one seen in the fermion-gravity system. This property allows for a consistent truncation as it satisfies (ii). Now, we include tensor-structures from curvature-squared terms, $R^2, R_{\mu\nu}^2$ and $R_{\mu\nu\rho\sigma}^2$. It is convenient to parameterise this complete set of tensor structures in terms of the Ricci squared, the Weyl-tensor squared, and the topological Gau\ss-Bonnet term,  
\begin{align}\label{eq:Gacurve2}
\int \! \mathrm d^4x\sqrt{g}\left( \frac{1}{g_{R^2}} R^2+
\frac{1}{g_{C^2}} C_{\mu\nu\rho\sigma}^2 + \frac{1}{g_E} E\right),
\end{align}
with the dimensionless couplings $g_{R^2}$, $g_{C^2}$ and $g_E$. The Gau\ss-Bonnet density $E$ is defined in \eq{eq:GaussBonnet}, and the Weyl-tensor squared is in four dimensions given by 
\begin{align}
\label{eq:Weyl2}
C_{\mu\nu\rho\sigma}^2=R_{\mu\nu\rho\sigma}^2-2 R_{\mu\nu}^2+\frac13 R^2\,.
\end{align}
We concentrate on the Reuter fixed point with the assumption that it is dominated by the Einstein-Hilbert couplings in contradistinction to the perturbative $R^2$ fixed-point. In \cite{Denz:2016qks}, it has been observed, that $R_{\mu\nu}^2$-contributions, and hence $C^2$-contributions, generated by the Einstein-Hilbert tensor structures at the Reuter fixed point are small. They are sub-leading in comparison to the $R^2$-tensor structure. This has been the topic of \autoref{sec:mom-dep}, see \autoref{fig:MomDep} and the respective discussion. With the assumption of the dominance of the Einstein-Hilbert couplings it implies $1/g^*_{C^2}\approx 0$, and indicates the irrelevance of this operator at the fixed point. 

Moreover, from the quartic term $c_4 \, p^4$ with $c_4\approx -0.24$ in the running of the momentum-dependent coupling of the four-point function displayed in \autoref{fig:MomDep}, we deduce that its contribution $c_\text{EH}\, g_{R^2}^2$ to the $\beta$-function $\beta_{g_{R^2}} =\partial_t g_{R^2}$ of the $R^2$-tensor structure is positive, for a detailed discussion see \cite{Denz:2016qks}. In turn, it is well-known that the $R^2$-coupling itself leads to a negative contribution $-c_{R^2}(g_{R^2})$, which is one-loop universal. We emphasise that both coefficients depend on the full fluctuation propagator. In combination, this leads us to a $\beta$-function 
\begin{align}
\label{eq:R2beta}
\partial_t g_{R^2} = \beta_{g_{R^2} }\simeq  
 c_\text{EH}\, g_{R^2}^2 - c_{R^2}( g_{R^2})\,. 
\end{align}
Switching off the Einstein-Hilbert contribution leads us to the standard Gau\ss ian fixed point for $R^2$-gravity. In turn, at the Reuter fixed point, we assume a small fixed-point value for $1/g_{R^2}$, that may also trigger a small, but non-vanishing fixed-point value for $1/g_{C^2}$. Combining these estimates for $g_{R^2}$ and $g_{C^2}$, we arrive at  
\begin{align}
\label{eq:R^2+Weyl2-ReuterFP}
\frac{1}{g^*_{R^2}}\,,\, \frac{1}{g^*_{C^2}}&\approx 0\,,
&\text{and}&&
 \frac{g^*_{C^2}}{g^*_{R^2}}&\approx 0\,,
\end{align}
in pure gravity. We add, that the relevance analysis in \cite{Denz:2016qks} suggests that the $g_{R^2}$ coupling, while being small, is UV-relevant at the Reuter fixed-point. This finding is corroborated by respective ones in the background approximation. For higher-derivative gravity work in the background-field approximation see e.g.\ \cite{Codello:2006in, Codello:2007bd, Codello:2008vh, Niedermaier:2009zz, Benedetti:2009rx, Ohta:2013uca, Falls:2013bv, Dona:2013qba, Falls:2014tra, Ohta:2015efa, Ohta:2015fcu, Falls:2016wsa, Falls:2016msz, Ohta:2016jvw, Hamada:2017rvn, Falls:2017lst, Ohta:2018sze, Alkofer:2018fxj, Alkofer:2018baq, Falls:2018ylp, Falls:2020qhj}. 

We now proceed to the $R^2$- and $C^2$-contributions from matter fluctuations. Being short of a full fluctuation computation of these terms, we utilise the Nielsen identities in the presence of the cutoff, see \eq{eq:mNIlin} and \eq{eq:backrdep} in \autoref{sec:RegDep} and \autoref{sec:SymId}. The identity \eq{eq:backrdep} comprises the difference between background-metric and fluctuation-field derivatives, while the Nielsen identities \eq{eq:mNIlin} also take into account the difference introduced by the gauge-fixing sector. For the present speculative analysis it suffices to discuss \eq{eq:backrdep}. For example, we find for the $R^2$-contribution,   
\begin{align}
\nonumber 
\left[\frac{\delta \Gamma_k}{\delta \bar g} - \frac{\delta \Gamma_k}{\delta h}\right]_{R^2} \simeq{} & \text{Tr}\,\frac{\delta \sqrt{\bar g} R_k}{\sqrt{\bar g}\,\delta \bar g_{\mu\nu}} G_k\bigg|_{R^2}\\[1ex] 
={}& \Delta g_{R^2}(\vec g) \int \!\mathrm d^4 x\sqrt{\bar g}\, R^2\,, 
\label{eq:rhsmNI}
\end{align}
The right-hand side has a form similar to the flow equation itself, and is UV- and IR-finite. Accordingly, $\Delta g_{R^2}(\vec g)$ is a dimensionless constant, that depends on all couplings taken into account in the computation, summarised as vector $\vec{g}$. This includes the $R^2$- and $C^2$-couplings $g_{R^2}, g_{C^2}$ themselves (or rather avatars thereof), as well as avatars of the dimensionless Newton coupling and the dimensionless cosmological constant, see \autoref{sec:UV-FP}. The scale-derivative of \eq{eq:rhsmNI} vanishes on a fixed point, 
\begin{align}
\partial_t \left[\frac{\delta \Gamma_k}{\delta \bar g} - \frac{\delta \Gamma_k}{\delta h}\right]_{R^2} \simeq \frac{\partial \Delta g_{R^2(\vec{g}^*)} }{\partial g_i} \,\beta_i(\vec{g}^*) =0\,, 
\end{align}
where we have used that the dimensionless coefficient $\Delta g_{R^2}(\vec{g}^*)$ cannot have an explicit $k$-dependence. Hence, \textit{at a fixed point}, this result allows us to identify the matter contribution of the flow for $R^2$-tensor-structures of fluctuation-field vertices with that of the background-field $R^2$-term. The same reasoning also applies to the $C^2$-term. In summary, the above arguments imply that the matter contributions to the curvature-squared couplings should be independent of the background-metric dependence of the regulator, as well as of the shape of the regulator. Moreover, since the ghost contribution to the curvature-squared couplings also does not depend on other scales than the cutoff scale, it should also be regulator-independent. The validity of these general statements can be checked explicitly with the results of \cite{Alkofer:2018fxj, Alkofer:2018baq}. There, different types of regulators have been investigated in $f(r)$-gravity: all couplings except the $R^2$-coupling depend on the Laplacian used in the regulators. The results also confirm a regulator-dependence of the graviton contributions, triggered by the Nielsen identities. As discussed above, this suggests that the pure gravity contributions to the flow should rather be computed within the fluctuation approach. 

The above considerations allow us to discuss the generic structure of gravity-matter flows within the fluctuation approach,  
\begin{align}\nonumber 
\beta_{g_{R^2}} = &\,\left. \beta_{g_{R^2}}\right|_{\text{grav}}  - c_{R^2} N_{R^2} \, g_{R^2}^2\,,\\[1ex]
\beta_{g_{C^2}} =&\,\left. \beta_{g_{C^2}}\right|_{\text{grav}}   
- c_{C^2} N_{C^2} \,g_{C^2}^2 \,,
\label{eq:R^2flow} \end{align} 
where $c_{R^2/C^2}$ are positive coefficients and $N_{R^2/C^2}$ are weighted sums (positive weights) of the numbers of scalars, vectors and fermions. All matter contributions have the same sign, which is the same as that of the gravity-ghost, which is computationally similar. For explicit computations in the background-field approximation see e.g.\ \cite{Dona:2013qba, Hamada:2017rvn, Alkofer:2018fxj, Alkofer:2018baq}.  

The quantitative evaluation of \eq{eq:R^2flow} depends on the full fluctuation flows in pure gravity including flow contributions from curvature-squared invariants. Here, we concentrate on the structure of the $\beta$-function of the  $R^2$ coupling, $\beta_{g_{R^2}}$. The matter contributions are subtracted from the positive Einstein-Hilbert-gravity contribution, see $c_\text{EH}$ in \eq{eq:R2beta}. For a critical number of matter fields, the complete contribution vanishes and we are left with a system that resembles the pure gravity curvature-squared system. This mechanism is very similar leading to the Caswell-Banks-Zaks fixed point in QCD discussed before. Note that in contradistinction to the minimally-coupled system the matter contribution cannot be absorbed in the pure gravity contributions, as they are related to $R^2\log (1+ R/k^2)$-terms. This is visible in the limit of large curvatures, see e.g. \cite{Alkofer:2018baq}. This qualitative analysis has to be sustained with a quantitative computation based on pure gravity flows including higher curvature terms. Such a computation requires improved truncations with the properties (i) and (ii). 

We close this chapter with a brief overview of investigations of gravity-matter systems within the background-field or hybrid approximations. In \cite{Dona:2013qba}, gravity-matter systems in the minimally-coupled approximation were investigated in a hybrid approach: while most contributions to the flow have been computed in the background-field approximation, the matter parts of the anomalous dimensions have been computed in a fluctuation approach setup. Within this approximation destabilising effects for scalars and fermions and stabilising effects for gauge fields were found. The destabilising result for fermions in \cite{Dona:2013qba} is an artefact of the background-field approximation as discussed in \autoref{sec:RegDep}: the background-metric dependence of the regulator influences the (de)stabilising property of minimally-coupled fermions. However, this does not imply that the background-field approximation breaks down for all gravity couplings. The results of \cite{Meibohm:2015twa, Christiansen:2017cxa} showed that in particular the most UV relevant operators have to be taken from a fluctuation computation, i.e., most importantly the graviton mass parameter $\mu$. In turn, the background and the fluctuation Newton coupling behave rather similar under the influence of minimally-coupled matter fields. The sign of leading-order contribution agrees: the scalar and fermionic contribution to the beta function of the Newton coupling at $\CO(g^2)$ is positive, while the gauge contribution is negative.

In summary, the investigations of gravity-matter systems within the fluctuation approach open a systematic path towards reliable stability investigations of fully coupled matter systems as well as that of phenomenological consequences for high energy physics. Still, fully reliable results require a systematic and qualitative improvement of the current truncations. This is the subject of current work in the community.

\begin{figure*}[t]
 \includegraphics[width=\linewidth]{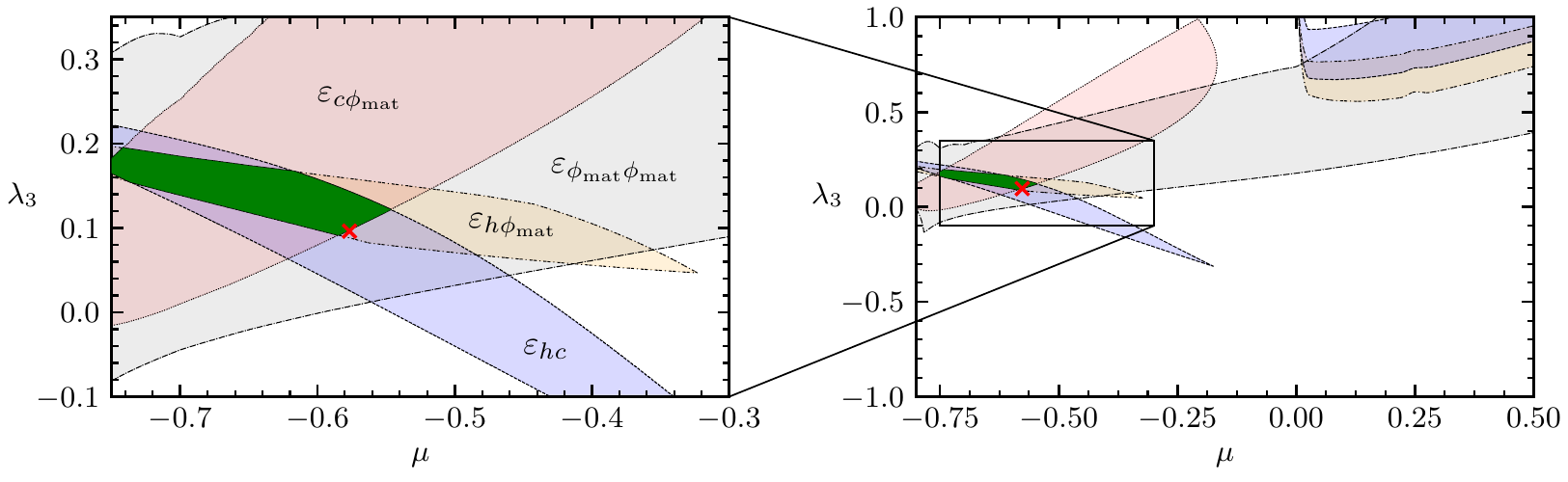}
 \caption{
 Effective universality of the different avatars of the Newton coupling as a function of $\mu$ and $\lambda_3$. The regions of effective universality are defined with $\varepsilon_{ij} < 0.2$ according to \eq{eq:epsilon}. The red cross indicates the UV fixed point, which lies in the region of effective universality. The figure is taken from \cite{Eichhorn:2018ydy}.
 }
 \label{fig:EffUniv}
\end{figure*}

\subsection{Effective universality}
\label{sec:eff-uni}
In the vertex expansion \eq{eq:vertex-expansion}, we have introduced the couplings $g_n$ for each graviton $n$-point function as the running couplings of the Ricci-scalar tensor-structure $(\sqrt{g} R)^{(n)}$, see \eqref{eq:tensor-structure}. In a diffeomorphism-invariant approach these couplings would agree. In turn, in the present gauge-fixed approach these are different avatars of the Newton coupling. While not being identical, $g_i\neq g_j$, they are related by non-trivial mSTIs \eq{eq:mSTI}. 

This is similar in non-Abelian gauge theories, where different avatars of the running strong coupling $\alpha_s=g^2/(4 \pi)$ can be derived from different correlation functions, both from pure glue vertices as well as glue-matter vertices, for a detailed discussion see \cite{Mitter:2014wpa, Cyrol:2016tym, Cyrol:2017ewj} and the recent review \cite{Dupuis:2020fhh}. The $\beta$-functions of all the avatars of the strong coupling are two-loop universal in mass-independent renormalisation schemes. On may also define an RG-scheme with the requirement that $\beta$-functions agree to all orders. However, the standard fRG renormalisation scheme is mass-dependent, so even two-loop universality is not guaranteed. More importantly, identical $\beta$-functions do not necessarily lead to an identical momentum-dependence. Indeed, in non-Abelian gauge theories, the momentum-dependence of different avatars of the running strong coupling differs already at the universal two-loop order, which can be also shown from the STIs. Additionally, in the strongly-correlated IR-regime of a non-Abelian gauge theory, the fRG $\beta$-functions as well as the momentum-dependence of the running couplings differ significantly. Some of them, i.e.\ the three-gluon coupling, even switch sign, while others, i.e.\ the ghost-gluon and four-gluon coupling, stay positive, see~\cite{Cyrol:2016tym}.

In gravity, the situation is even more intricate. To begin with, the Newton coupling is dimensionful and hence the $\beta$-functions of the avatars of the Newton coupling are not universal, leave aside an identical momentum dependence. Additionally, as already mentioned in the context of non-Abelian gauge theories, the standard fRG renormalisation schemes are typically mass-dependent, which adds to the differences, as do truncations. 

\textit{Effective universality} is the concept that in particular at the fixed point, where gravity is in a scaling regime, the quantum theory is dominated by the diffeomorphism invariance of the underlying theory. If this scenario applies, the $\beta$-functions as well as the momentum dependence of different avatars of the Newton coupling should agree or are rather be close to each other on the asymptotically safe UV fixed point. This concept would apply to all couplings, and in particular, the $\lambda_n$ can be understood as avatars of the cosmological constant. Additionally to the Newton couplings from the Ricci-scalar tensor-structure, we have further avatars of the Newton coupling stemming from the gravity-matter correlation functions. 

Given the presence of truncations in explicit computations, the impact of non-trivial mSTIs, and the non-perturbative nature of the UV fixed point, it is left to define a measure for effective universality. In \cite{Eichhorn:2018akn, Eichhorn:2018ydy}, it was quantified how these avatars differ at the UV fixed point using the measure
\begin{align}
 \label{eq:epsilon}
 \varepsilon_{ij}(g, \mu, \lambda_3)=
 \left|\frac{\Delta\beta_{g_i}-\Delta\beta_{g_j}}{\Delta\beta_{g_i}+
 \Delta\beta_{g_j}}\right|_{g_i=g_j=g}\,,
\end{align}
where $\Delta \beta_{g_{i}}$ is the anomalous part of the $\beta$-function $\beta_{g_{i}}$ obtained by subtracting the canonical running
\begin{align}
 \Delta\beta_{g_i} = \beta_{g_i} - 2\,g_i\,.
\end{align}
In \cite{Eichhorn:2018ydy}, five avatars of the Newton coupling were included stemming from the three-point functions, $\Gamma_k^{(hhh)}$, $\Gamma_k^{( c\bar c h)}$, $\Gamma_k^{(\varphi \varphi h)}$, $\Gamma_k^{(\psi\psi h)}$, and $\Gamma_k^{(AAh)}$. Thus the set of $g_i$ is given by $i \in \{h,c,\varphi,\psi,A\}$ where $g_h=g_3$ in the previous notation. In \eq{eq:epsilon}, the $\beta$-functions are identical for $\varepsilon_{ij}=0$ and we have full universality. A small value of $\varepsilon_{ij}$ indicates almost identical $\beta$-functions and thus 'effective universality'. In \cite{Eichhorn:2018ydy}, these small values were estimated to be $\varepsilon_{ij}<0.2$. This estimate is based on a systematic error estimate of the used truncations as well as the impact of the mSTIs. In turn, a larger value of $\varepsilon_{ij}$ shows that universality is strongly broken and that the mSTIs are highly non-trivial.

The universality measures $\varepsilon_{ij}$ are functions of all couplings and we display them in \autoref{fig:EffUniv} for $g_i=g_h^*$ as functions of $\mu$ and $\lambda_3$. Remarkably, the UV-fixed point lies in the green area, which signals $\varepsilon<0.2$ and thus effective universality holds. As discussed above, this statement is non-trivial since the mSTIs can introduce large differences between the avatars, in particular, if the fixed point is highly non-perturbative. In turn, this result gives a strong hint that the UV fixed point is in the semi-perturbative region. Interestingly, a semi-perturbative behaviour was also found in large-order Ricci-scalar expansions of the effective action in the background-field approximation \cite{Falls:2013bv, Falls:2014tra, Falls:2016wsa, Falls:2017lst, Falls:2018ylp}. There it was found that the critical exponents of the high-order curvature invariants are close to their canonical values. 

We emphasise that the observed effective universality is a highly non-trivial result. If it can be sustained in further analyses, it is presumably dynamical. This conjecture is supported by the following observation: for a marginal, universal, coupling one may simply compute one avatar of the coupling and identify the other avatars with the computed one. In turn, in a theory like gravity, where the effective universality is potentially generated dynamically, this may only work in specific RG-schemes. One may even define a natural RG-scheme by $\epsilon_{ij} \equiv 0$. This entails, that in other RG-schemes only a subset of the couplings will have the natural $\beta$-functions. Note, that the latter property is additionally triggered by the inherent truncations of explicit computations. 

In any case, within a given RG-scheme some of the $\beta$-functions may satisfy $\epsilon_{ij} \equiv 0$, while others may not. The identification of all avatars of the given coupling with a specific one will only work if the latter coupling is chosen from the natural subset. Such an identification is an implicit way of enforcing the natural RG-scheme. In turn, if all couplings are identified with an avatar which is not in the natural subset, the system may be corrupted. This can even lead to a loss of the fixed point.

In gravity-matter systems, we indeed observe, in given truncations, such a behaviour: if all avatars of the Newton coupling are identified with the three-graviton coupling $g_h$, that is $g_i=g_h$, the results are close to the full ones with multiple avatars of the Newton coupling. In turn, identifying all avatars of the Newton coupling with a gravity-matter avatar fails. In summary, this hints at a surprisingly complicated interaction structure in gravity-matter systems. Its origin is yet to be understood and may give us further valuable insights into the dynamics of these systems.

In short, these investigations of effective universality indicate a \textit{close perturbativeness} of the UV fixed-point regime of asymptotically safe gravity. 

\section{Summary \& outlook}
\label{sec:Summary}
In this contribution, we have reviewed the state of the art of the \textit{fluctuation approach} to quantum gravity. This approach is based upon the computation of the correlation functions of the dynamical graviton fluctuation field $h_{\mu\nu}$ within a systematic vertex expansion. This can be done within general parameterisations of the full metric, but most results have been achieved in the linear split, $g_{\mu\nu}=\bar g_{\mu\nu} +h_{\mu\nu}$. While the correlation functions of the fluctuation field are not observables by themselves and carry a gauge-dependence, the computation of observables in quantum gravity requires the knowledge of the fluctuation correlation functions, and they indeed encode the dynamics of quantum gravity. 

By now the fluctuation approach has matured, see the overview of the results in \autoref{sec:StateArt}. We see signs of \textit{apparent convergence} of the results in pure gravity. Moreover, by now we can reliably evaluate the stability of general gravity-matter systems. In combination, the fluctuation approach now allows for reliable physics predictions for the UV regime of asymptotically safe gravity including its unitarity. The approach also allows for reliable physics predictions for the 
'IR' particle physics within the asymptotically safe standard model. \\ 

{\bf Acknowledgements}\\[-2ex]

\noindent We thank A.~Bonanno, B.~B\"urger, N.~Christiansen, T.~Denz, A.~Eichhorn, K.~Falls, H.~Gies, A.~Held, B.~Knorr, P.~Labus, S.~Lippoldt, J.~Meibohm, D.~Litim, A.~Pereira, A.~Rodigast, B.-J.~Schaefer, M.~Schiffer,  F.~Saueressig, J.~Smirnov, M.~Yamada, and C.~Wetterich for discussions and work on the subjects reported on. This work is supported by the DFG, Project-ID 273811115, SFB 1225 (ISOQUANT), as well as by the DFG under Germany's Excellence Strategy EXC - 2181/1 - 390900948 (the Heidelberg Excellence Cluster STRUCTURES), by the Danish National Research Foundation under Grant No.~DNRF:90, and by the Science and Technology Research Council (STFC) under the Consolidated Grant ST/T00102X/1. 

\appendix

\section{Notation}
\label{app:Notation}
Our convention for functional derivatives are given by  
\begin{align}
 \label{eq:funDerGen} 
 \frac{\delta J_{\mu_1\cdots \mu_n}(x)}{\delta J_{\nu_1\cdots \nu_n}(y)} = \frac{1}{\sqrt{\bar g}} \delta(x-y)\, \delta^{(\nu_1}_{\mu_1} \cdots \delta^{\nu_n)}_{\mu_n} \,, 
\end{align}
where the parenthesis in the superscript of the Kronecker-$\delta$'s stands for the symmetrisation of the indices including a normalisation factor $1/n!$. For example we have 
\begin{align} \label{eq:funDerh} 
 \frac{\delta h_{\mu_1\mu_2}(x)}{\delta h_{\nu_1\nu_2}(y)} = \frac{1}{\sqrt{\bar g}} \delta(x-y)\, \frac{1}{2} \left(\delta^{\nu_1}_{\mu_1}\delta^{\nu_2}_{\mu_2}+ \delta^{\nu_2}_{\mu_1}\delta^{\nu_1}_{\mu_2}\right)\,.  
\end{align}
This leads to the correlation functions of the fluctuation fields as given in \eq{eq:Zh}.

The metric $\gamma^{ab}$ in field space is diagonal for bosons $\varphi$, and is symplectic for fermions $\psi,\bar\psi$, 
\begin{align}
\label{eq:FieldMetric}
(\gamma_\varphi^{ab})&= \mathbb{1}\,,
&
 (\gamma_\psi^{ab}) &= \left(\begin{array}{cc} 0 & 1 \\  -1 & 0\end{array}\right)\,,
\end{align}
with the Northwest-Southeast convention 
\begin{align}\label{eq:Co+Contra}
\phi^a &=\gamma^{ab}\phi_b\,,
&
\phi_a &=\phi^b \gamma_{ba}\,. 
\end{align}
These definitions entail 
\begin{align}
\gamma_b{}^a &=\delta^a_b\,,
&\text{and}&
&
(\gamma_\varphi)^a{}_b &=\delta^a_b\,, 
&
(\gamma_\psi)^a{}_b &=-\delta^a_b\,,
\end{align}
more details can be found in \cite{Pawlowski:2005xe}. 

\section{Pontryagin index in \texorpdfstring{$U(1)$}{U(1)} gauge theories}
\label{app:Pontryagin}
The Pontryagin index $P$ of a four-dimensional $U(1)$-gauge theory in flat space is a simply example for a topological index in quantum field theory. For general field configurations it is a non-vanishing integer on manifolds such as $\mathbbm{T}^4$, the four-dimensional torus, e.g.\ underlying standard lattice simulation. We write in general  
\begin{align}
\label{eq:Pontryagin}
P[A,\theta] = \frac{1}{32\pi^2} \int_x \! \theta(x) F_{\mu\nu} \tilde  F^{\mu\nu}\,, \qquad 
P[A,1] \in\mathbbm{Z}\,,  
\end{align}
with the Pontryagin index $P[A]=P[A,\theta=1]$. The (dual) field strength, $F_{\mu\nu}$ and $\tilde F_{\mu\nu}$, are given by   
\begin{align}
\label{eq:FdualF}
F_{\mu\nu}&=\partial_\mu A_\nu -\partial_\nu A_\mu \,,
&
\tilde F^{\mu\nu}&=\frac{\epsilon^{\mu\nu\rho\sigma}}{2} F_{\rho\sigma}\,. 
\end{align}
In momentum space $P[A,\theta]$ reads 
\begin{align}\label{eq:PontryaginMom}
P[A,\theta] = \frac{\epsilon^{\mu\nu\rho\sigma}}{16\pi^2} \int_{p,q} \! \theta(-(p+q))  \,p_{\mu} A_{\nu}(p) \, q_{\sigma} A_{\rho}(q) \,,
\end{align}
The flow of $\theta$ has been studied in \cite{Reuter:1996be} for the topological charge in Yang-Mills theories. Two derivatives with respect to the gauge field in momentum space lead us from \eqref{eq:PontryaginMom} to
\begin{align}
\label{eq:derTop}
\frac{\delta P[A,\theta]}{\delta A_\alpha(p) \delta A_\beta(q) }= \frac{\epsilon^{\alpha\beta\rho\sigma}}{8\pi^2}  p_\rho q_\sigma \,\theta(-(p+q)) \,. 
\end{align}
For a topological term with constant $\theta=\theta_\text{top}$ we have $\theta(l) =\theta_\text{top} (2 \pi)^4 \delta(l)$. Inserting this choice into \eqref{eq:derTop}, the term vanishes with $\epsilon^{\alpha\beta\rho\sigma} p_\rho p_\sigma=0$. 

\bibliography{GravityStatus}

\end{document}